\documentstyle[amstex,11pt,a4]{report}
\input{epsf}

\newcommand{\HASHifHelv}[2]{#2}
\newcommand{\HASHreport}[1]{}

\renewcommand{\numberline}[2]{\makebox[1cm][r]{#1} #2}

\newcommand{\mysection}{\section*}

\newenvironment{Itemize}
{\begin{list}{\labelitemi}{\setlength{\leftmargin}{9mm}}}
{\end{list}}

\newtheorem{thm}{Theorem}[chapter]

\newtheorem{lemma}[thm]{Lemma}
\newtheorem{defn}[thm]{Definition}

\newcommand{\Tr}{\operatorname{Tr}}

\newcommand{\Real}{\operatorname{Re}}

\newcommand{\eqref}[1]{(\ref{#1})} 

\newcommand{\deriv}[2]{\frac{d#1}{d#2}}
\newcommand{\pderiv}[2]{\frac{\partial#1}{\partial#2}}

\newcommand{\proofbegin}
{\HASHifHelv{ {\noindent{\bf Proof}\hspace{1.5ex}} }
           { {

            \noindent{\bf Proof}\hspace{1.5ex}}  }
}
\newenvironment{proof}{\proofbegin}{\proofend}
\newcommand{\proofend}
{\HASHifHelv{ \nopagebreak\hspace*{\fill}\nopagebreak\mbox{$\Box$} }
            { \nopagebreak\hspace*{\fill}\nopagebreak\mbox{$\Box\!\!\!\!$} } }

\newenvironment{example}{\examplebegin}{\exampleend}
\newcommand{\examplebegin}{\paragraph*{Example}}
\newcommand{\exampleend}{\proofend}

\newcounter{myfootnote}

\newcommand{\N}{{\Bbb N}}
\newcommand{\R}{{\Bbb R}}
\newcommand{\C}{{\Bbb C}}

\newcommand{\bra}{\left(\rule{0mm}{2.5ex}\right.}
\newcommand{\ket}{\left.\rule{0mm}{2.5ex}\right)}

\newcommand{\cbra}{\left\{\rule{0mm}{2.5ex}\right.}
\newcommand{\cket}{\left.\rule{0mm}{2.5ex}\right\}}

\newcommand{\half}{{\textstyle \frac{1}{2}}}

\newcommand{\mthsbldltr}[1]{{\bold #1}}

\newcommand{\Bj}{\mthsbldltr{j}}

\newcommand{\Br}{\mthsbldltr{r}}

\newcommand{\Bx}{\mthsbldltr{x}}

\newcommand{\Bp}{\mthsbldltr{p}}       
\newcommand{\Bq}{\mthsbldltr{q}}       
\newcommand{\BS}{\mthsbldltr{S}}       

\newcommand{\CS}{\mbox{$\cal S$}}    

\newcommand{\FH}{{\cal H}}           

\newcommand{\x}{\times}

\newcommand{\M}[1]{\!#1\!}

\newcommand{\Min}{\!\in\!}

\newcommand{\Mneq}{\!\neq\!}

\newlength{\framemarg}
\newlength{\textlessmarg}
\newcommand{\dbra}{\langle}
\newcommand{\dket}{\rangle}
\newcommand{\bib}{\subsubsection*{Bibliography}}

\newlength{\quotemarg}
\newlength{\quotewidth}
\newcommand{\aquote}[2]{\hspace*{\quotemarg}%
\parbox{\quotewidth}{\begin{center}{\it #1}
\\ \ \\ \hspace*{\fill}#2\end{center}}}

\newcommand{\newpart}[3]{\newpage%
\thispagestyle{plain}
{\Huge Part #1}\\ \ \\ \ \\ {\Huge #2}
\vspace*{\fill}\\{\large\bf\it #3}\vspace*{\fill}\vspace*{\fill}%
\addcontentsline{toc}{chapter}
{\protect\numberline{{\large #1}}{\protect\underline{\large\sl #2}}}%
}

\newcommand{\mychapter}[1]
{\chapter{#1}%
}

\begin{document}%
\thispagestyle{empty}
%
%
{\Large
\begin{center}
University of London\\
Imperial College of Science, Technology and Medicine\\
Department of Physics
\end{center}
}

\vspace*{\fill}

\begin{center}
{\huge\bf The Nine Lives of\\[3mm]Schr\"odinger's Cat}\\
\mbox{} \\
{\large\it\bf
On the interpretation of\\
non-relativistic quantum mechanics}\\
\mbox{} \\
\mbox{} \\
\Large Zvi Schreiber \\
\ \\
\large Rakach Institute of Physics\\
\large The Herbrew University\\
\large Givaat Ram\\
\large Jerusalem 91904\\
\large e-mail \verb+zvi1@shum.cc.huji.ac.il+\\
\end{center}

\vspace*{\fill}

{\Large
\begin{center}
A thesis submitted in partial fulfilment\\
of the requirements for the degree of\\
Master of Science of the University of London \\
\mbox{} \\
October 1994
\end{center}
}
\newpage\thispagestyle{empty}
\tableofcontents%
\setcounter{page}{0}
%
%
\newpart{I}{Introduction}
{
  The most remarkable aspect of quantum mechanics is that it needs
  interpretation.  Never before has there been a situation in which
  a physical theory provides correct predictions for the behaviour of
  a system without providing a clear model of the system.

  By way of introducing the interpretational issues, Part I briefly
  sketches the historical development of the formalism and concepts
  of quantum mechanics.

  (A detailed history may be found in \cite{Jammer:89}.)
}
\pagestyle{headings}%
\renewcommand{\thechapter}{I.\arabic{chapter}}
\setcounter{chapter}{0}
\mychapter{Introduction}

\section{A brief history of quantum mechanics}

In classical physics, some aspects of nature are described using
particles, others using waves.
Light, in particular, was described by waves.

In 1900, Max Planck showed that the spectrum of blackbody radiation,
which could not be explained in terms of waves, could be explained
by the assumption that light is emitted in discrete quanta of energy
\cite{Planck:90,Planck:91}.
Later, in 1905, Albert Einstein showed that the photoelectric effect,
which also did not fit the wave model, could be explained by the
assumption that light is {\em always\/} quantised \cite{Einstein:05}.
This controversial suggestion was not accepted until after the famous
experiments of Arthur Holly Compton in 1923
\cite{Compton:23a,Compton:23b}.

In 1923, Louis de Broglie suggested that, conversely,
`particles' might display wave-like properties
\cite{deBroglie:23a,deBroglie:23b,deBroglie:23c,deBroglie:24}.
This was later confirmed, most strikingly
by the Davisson-Germer experiment of 1927
\cite{DavissonGermer:27}.

In 1925, Werner Heisenberg introduced matrix mechanics
\cite{Heisenberg:25}.  This formalism
was able to predict the energy levels of quantum systems.
It was cast in the form of a {\it wave function\/} and
differential equation by Erwin Schr\"odinger
in the following year \cite{Schrodinger:26}.
The efforts of Paul Dirac \cite{Dirac:26}, Pascual Jordan \cite{Jordan:27}
and others at synthesising the two approaches (the so called
{\it transformation theory\/}) led to
the more general formalism of {\it quantum mechanics\/}
\cite{vonNeumann:55}.

These developments marked a turning point in physics.
For the first time, physical results were being derived, not from
a model of the universe, but from abstract mathematical constructs
such as the Schr\"odinger wave function.  There was an unprecedented need
for {\it interpretation}.

In 1926, Max Born suggested that the values in a particle's
wave function gave the {\em probabilities\/} of finding the particle in a
given place \cite{Born:26}.
This really set the cat amongst the pigeons.
The suggestion that the laws of physics were non-deterministic  flew
in the face of everything achieved since Newton.

In retrospect, the work of Born precipitated an even more radical
revolution: a switch in physics from ontology (discussion of {\it what is\/})
to epistemology (discussion of {\it what is known\/}).
According to Born, an experiment to determine the position of a quantum
system would give results with probabilities dictated by
the wave function.  This statement talks about what {\it results\/}
are obtained instead of modelling the {\it process}.
It involves an artificial division of the world into {\it system\/}
and {\it observer}.

The following interpretational questions were raised.
\begin{itemize}
\item
  Accepting quantum mechanics as an epistemological theory, is there
  a specific place at which the observer-system cut should be located?
  In more concrete terms: are there certain systems which may be
  described by a wave function and others which may not?

  If there is no such fixed observer-system cut,
  is it consistent to allow this cut to be
  made at different places according to convenience?  In other words,
  may we always {\it choose\/} how much of a system to describe by
  a wave function?

\item
  Is there an ontological theory underlying the epistemological formulation
  of quantum mechanics?  In other words, is there a model of the
  universe from which we can derive the fact that certain systems
  may be treated using quantum mechanics?

  If such a model is proposed, does it explain the non-determinism
  of quantum mechanics in terms of an underlying deterministic mechanism?
\end{itemize}

In response to these questions, two main schools of thought arose.
Bohr and his followers (notably Heisenberg)
embraced the epistemological viewpoint and
argued that it cannot be possible to find an underlying ontological
theory.

Einstein and his followers (notably Schr\"odinger) felt that
a deterministic ontological physics must underly the
non-deterministic epistemological predictions of quantum theory.
The states in such underlying theories were dubbed
{\it hidden variables}.

A critical twist came in 1935 when Einstein, Podolsky and Rosen (EPR)
showed that the formalism of quantum mechanics implied nonlocality:
an action at one point may have immediate consequences at a distant
point without any apparent intervening mechanism
\cite{EinsteinPodolskyRosen:35}.
They inferred that quantum mechanics cannot be complete.
There must be an underlying theory which is local.

This highlighted a division amongst those who accepted quantum mechanics
as a fundamental theory.

Bohr himself was a positivist.  He insisted on describing things through
macroscopic observations only and rejected the idea that the
Schr\"odinger wave function  represented the state of a particle.
At best, the wave function was a useful way of predicting the
macroscopic results of an experiment on the particle.
As such, it was essentially physically meaningless to discuss whether
the wave function behaved nonlocally.

John von Neumann and Paul Dirac, on the other hand, were perfectly
happy to talk about the state of a particle.  However, rather than
accepting EPR's argument for an underlying theory, they simply
accepted that the world is indeed nonlocal.

Later, in 1964, the controversy was to take another turn when John Bell
showed that it was the {\it predictions\/} of quantum theory
and not merely the formalism that implied nonlocality \cite{Bell:64}.
Even if an underlying deterministic theory were found, it would
not be local!

In the mean time, there had been two important developments in the
interpretation of quantum mechanics.
In 1952, David Bohm proposed a hidden variables
theory \cite{Bohm:52}.
(The core of Bohm's idea had earlier been proposed by de Broglie
\cite{deBroglie:27} and abandoned \cite{deBroglie:30}.)
Although it had some strange features including, of course,
nonlocality, this was the first real candidate for a deterministic
theory underlying quantum mechanics.

In 1957, Hugh Everett III proposed a radical interpretation.
He suggested that the Schr\"odinger wave function describes not one
world but an infinite and growing collection of realities
\cite{Everett:57a}.
When the position of a particle is measured, rather than saying it
may turn out to be here or there, Everett suggested that it will be
both here and there, in parallel realities.

As well as being bizarre, Everett's interpretation was rather vague.
What constitutes a measurement for the purposes of causing reality to split?

Considerable progress was made on this question through study of
the phenomenon of {\it decoherence}
\cite{FeynmanVernon:63,Zurek:81,Zurek:82,CaldeiraLegget:83,JoosZeh:85,%
Zurek:86}.
This study abandoned simplified models of isolated lab equipment and
started to consider the effects of the environment.
This led to a number of ``post-Everett'' interpretations, some building
closely on Everett's ideas, others not, but all a little more
sophisticated and a little less vague.  {\it Histories\/} have emerged
as the favourite formalism for this work.

\section{The interpretational issues}

Quantum mechanics is formulated in terms of a system and an observer.
It describes the state of the system using the Schr\"odinger equation,
or, more abstractly, using a mathematical construct called
{\it Hilbert space}.
The theory predicts the possible results for any experiment
performed by the observer, as well as the associated probabilities.

The main interpretational issues are as follows.

\subsection*{The measurement problem}

The central interpretational issue will emerge in Chapter~\ref{c:qstate}.
It is the fact that measurement, when modelled within the quantum formalism,
gives results different to those predicted for external measurements,
ie.\ measurements on the system.
This is called the {\it measurement problem}.

The problem is exemplified by Schr\"odinger's maltreated cat
\cite{Schrodinger:35}, after which this work is named.
Schr\"odinger's cat is placed in a sealed box with a bomb.
The bomb is triggered by the decay of a particle.

According to quantum mechanics, the particle quickly takes on a
state in which it is in a superposition of being
decayed or intact.
If the bomb checks the particle after one minute,  at that point the particle
will take on a definite state of decayed or intact, with particular
probabilities of each, and the bomb will explode or not as appropriate.
The cat is either alive or die.

This however assumes that the particle is a quantum system and the
bomb is an external observer
(alternatively that the cat is an external observer).
Now suppose the entire box is treated as a quantum system while
Schr\"odinger\footnote
{
  Apologies for the misinterpretation of ``Schr\"odinger's cat''
  as the cat belonging to Schr\"odinger as opposed to the cat experiment
  conceived by Schr\"odinger.
}
is the observer.  Then quantum mechanics predicts
that the entire cat is in a superposed state of alive and dead until
Schr\"odinger comes along.
(In principle,
Schr\"odinger could actually check whether this superposition exists therefore
establishing the location of the system-observer cut, though not in practice.)

This problem means that one cannot move around the observer-system
cut at will.  One can also not do away with it and treat the whole
universe as a quantum system because in such a treatment the cat will
always be in a superposition of dead and alive, an absurd result.

In summary, one does not know what constitutes a quantum system and
what does not.  One only knows that the whole universe (or any other
{\it closed system}, ie.\ system without an observer) cannot sensibly
be treated as a quantum system.  One therefore cannot use quantum
mechanics to describe the whole world or to recover classical physics.

\subsection*{Non-determinism}

Quantum mechanics stipulates that the result of an external observation
is non-deterministic.  It predicts the probabilities.
But is the world truly non-deterministic?
Or does the uncertainty result from our ignorance of some details
of the state of the system and/or apparatus?

\subsection*{Locality}

The conventional interpretation of quantum mechanics is {\it nonlocal}.
Actions at one point may have consequences at a distant point without
any apparent intervening mechanism.
Specifically,
actions at one point can cause an observable at a distant location to
take on a definite value.

Is this really what is happening and if so can it be used to
communicate instantly with a remote site ({\it signal nonlocality\/})?
Or perhaps the effect is an illusion.  Perhaps the remote
observable always had a definite value and it is simply being revealed
by the local action.

\section{Outline of thesis}

Part II describes the rules of quantum mechanics and formal
aspects of quantum theory which are relevant to interpretation.

Part III concerns interpretation.
Nine interpretations are considered.

In Part IV, a brief stock-taking takes place.  Has a favourite
interpretation emerged?  Did curiousity kill the cat?
%
%
\newpart{II}{Quantum mechanics}
{
  The orthodox formulation of quantum mechanics is presented as
  a series of rules.  Some alternative formulations ---
  the Heisenberg picture, the Schr\"odinger wave function,
  the density matrix and the logico-algebraic approach --- are also
  presented.  The case of a spin half particle is briefly described.

  Then, in the second chapter, the formalism of histories is
  developed.

  Issues of interpretation are not considered until Part~III.

  Note that these chapters assume familiarity with the mathematics
  of Hilbert space and self-adjoint operators including the spectral
  theorem in the projection-valued-measure (p.v.m.) form.  The concepts of
  a boolean lattice (= boolean algebra), algebraic congruence and
  Borel set are also used.
  (A useful text for this type of maths is \cite{ReedSimon:80}).
}
\renewcommand{\thechapter}{II.\arabic{chapter}}
\setcounter{chapter}{0}
\mychapter{Formulation of quantum mechanics} \label{c:msrmnt}

\section{Dirac notation}

Quantum mechanics makes extensive use of the Hilbert space construction.
Usually the notation introduced by Paul Dirac \cite{Dirac:47} is used.

In Dirac's notation every state in Hilbert space is called a {\it ket\/}
and written
$|\psi\rangle$ where $\psi$ identifies the state (eg.\ an eigenstate
is often identified by its eigenvalues when the relevant operator is
made clear by context).

The inner product of $|\phi\rangle$ with $|\psi\rangle$ is written
$\langle\phi|\psi\rangle$ and called a {\it bracket}.
This suggests that $\langle\phi|$ should be
called a {\it bra\/} (in fact, the bra associated with $|\phi\rangle$).
It may be seen as a linear map from the Hilbert space to $\C$.

Note that
$\langle \hat{A}\psi,\phi\rangle=\langle\psi|\hat{A}^{\dagger}|\phi\rangle$
so the bra associated with the ket $\hat{A}|\psi\rangle$ is written
$\langle\psi|\hat{A}^{\dagger}$.

For $|\phi\dket\Min\FH_1$, $|\psi\dket\Min\FH_2$,
the tensor product $|\phi\dket\otimes |\psi\dket$ is simply written
$|\phi\dket|\psi\dket$.
It follows that $|\phi\rangle\langle\phi|$ projects onto the subspace
spanned by $|\phi\rangle$.


The notation $|\phi\rangle\langle\psi|$ is used for the linear map which
takes $|\chi\rangle$ to $|\phi\rangle\langle\psi|\chi\rangle$, ie.\ to
$(\langle\psi|\chi\rangle)|\phi\rangle$.

\section{Systems and observables}

The following slightly vague definitions of a system and an observable
will be used.

\begin{defn}
  A physical {\bf system\/} is any subset of the matter in the universe
  which does not interact with the other matter except when measured,
  ie.\ such interaction which occurs is carefully controlled.
\end{defn}

It is now necessary to define properties such as position and momentum
of a system.  Following Dirac, these are called {\it observables}.

\begin{defn}
  A {\bf physical observable\/} of a system is a type of measurement
  which may be performed on the system.
\end{defn}

For example, if the system comprises a particle and this particle is allowed
to hit a photographic plate, the position of the dot on the photographic plate
may be called the {\it position\/} observable.

Note that this definition defines the position
of the particle in terms of something that can be seen, namely the dot on
the photographic plate.

However, for now it is assumed that the position is an
{\em inherent property\/}\footnote
{
  This assumption is important for now although it is rejected
  in Bohr's work, see Chapter~\ref{c:bohr}.
}
of the system which is {\em measured\/}
by the photographic plate.  The importance of this assumption is
in that it allows one to say that two {\em different\/} apparatus
measure the {\em same observable}.
Although inherent, the value of an observable may not always be
defined.

It will generally be assumed that observables are described by real
numbers since this can always be arranged.
It will not be assumed that a measurement yields a sharp value;
after all, a photographic plate does not have perfect resolution.
Instead, it is assumed that an experiment narrows down
the value of the observable to some Borel set of real numbers.

\section{Formulation} \label{s:rules}

The conventional formulation of quantum mechanics is now given.
It is loosely based on formulations such as \cite[Ch.3]{dEspagnat:71},
\cite{dEspagnat:89}
although there are several differences.

\subsection*{Postulates}

\begin{itemize}
\item
  Every type of physical system may be associated with a separable
  Hilbert space $\FH$ and each time
  $t\Min\R$ with a self-adjoint operator $\hat{H}(t)$
  called the Hamiltonian, densely defined on $\FH$
\item
  a certain subset of instances of a given system may be associated
  with a non-zero vector in the associated Hilbert space
  $\FH$ (giving the system's state)
\item
  every physical observable on a system may be associated with
  a self-adjoint operator densely defined on the associated
  Hilbert space
\end{itemize}
in such a way that the following rules hold.

\begin{description}
\item[Composite system]
  If system $S_1$ is associated with Hilbert space  $\FH_1$
  and Hamiltonian $\hat{H}_1(t)$
  and system $S_2$ is associated with Hilbert space  $\FH_2$
  and Hamiltonian $\hat{H}_2(t)$
  then $S_1\M+S_2$, the union of the two systems, is associated
  with $\FH_1\otimes\FH_2$.
  The Hamiltonian of the joint system
  if $\hat{H}_1(t)\otimes\hat{1}+\hat{1}\otimes\hat{H}_2(t)$.

  If $S_1$ is in a state which may be described by vector $|\phi\rangle$
  and $S_2$ is in a state which may be described by vector $|\psi\rangle$
  then $S_1\M+S_2$ is in a state which may be described by the vector
  $|\phi\rangle|\psi\rangle$.

\item[Time evolution]
  While no measurements are performed on the system the state evolves
  in time according to the Schr\"odinger
  equation
  \begin{equation*} 
   \boxed{
    i\hbar \deriv{|\psi_t\rangle}{t} = \hat{H}(t) |\psi_t\rangle
    \mbox{ .}
   }
  \end{equation*}
  $\hbar$ is a constant the value of which depends on the units used for
  distance, time and mass.

  For a constant Hamiltonian this is solved
  by the unitary transformation
  \[
   \boxed{
    |\psi_t\rangle = e^{-i\hat{H}t/\hbar}|\psi_0\rangle
    \mbox{ .}
   }
  \]

\item[Statistical formula]
  The result of measuring a real-valued physical observable
  $A$ is inherently non-deterministic.
  When the system is in state $|\psi\rangle$,
  the probability of obtaining a result in the Borel set $\Omega$ is
  \begin{equation*} 
    \boxed{
      \frac{\| \hat{P}_{\Omega}|\psi\rangle \|^2}{\||\psi\rangle\|^2}
      \mbox{ \ ie.\ \ }
      \frac{ \langle \psi| \hat{P}_{\Omega}|\psi \rangle}
           { \langle \psi| \psi \rangle}
    }
  \end{equation*}
  where $\{\hat{P}_{\Omega}\}$ is the p.v.m.\ associated with
  $\hat{A}$.

\item[Ideal measurement]
  For every measuring apparatus it is in principle possible to construct
  an apparatus to measure the same observable in an {\it ideal\/} way.
  This means that if the measurement is repeated twice rapidly it will
  yield the same value on both occasions.
  (Such a measurement is sometimes called {\it measurement of the first
  kind\/} following Pauli.  Note that in the case where the
  system was described by a state vector this rule amounts to
  `wave function collapse'.)

\item[Commuting observables]
  Suppose $\hat{A}$ and $\hat{B}$ commute.  If $A$, $B$, $A$ are
  measured ideally in rapid succession then the value of $A$ will
  be the same on both occasions.

\item[Realisation of states]
  Every state in the Hilbert space of a system is in principle
  realisable unless
  excluded by a {\it superselection\/} rule.  (These rules are peculiar
  to certain systems and are not described here.  Note that this
  postulate generalises what is sometimes called the
  {\it linear superposition rule},
  namely that the linear combination of states is a state.)

\item[Existence of systems with a state vector]
  If a set $\hat{A}_1,\ldots,\hat{A}_n$ of commuting observables
  with p.v.m.s $\{\hat{P}^1_{\Omega}\},\ldots,\{\hat{P}^n_{\Omega}\}$
  are measured yielding results in Borel sets $\Omega_1,\ldots,\Omega_n$
  respectively, and if
  \(
    \hat{P}^1_{\Omega_1}\hat{P}^2_{\Omega_2}\cdots\hat{P}^n_{\Omega_n}
  \)
  projects onto a space spanned by the single vector $|\phi\rangle$ then
  the system is left in a state described by $|\phi\rangle$.
  (Note that without this rule there is no guarantee that any
  system could ever be described by a state vector.)

\item[Persistence of a state vector]
  If a state may be described by a state vector then it may be described
  by a state vector after an ideal measurement is performed.

%
\end{description}

\subsection*{Notes}

\subsubsection*{What is a rule?}

Many formulations include the facts that a system may be associated with
a Hilbert space and an observable with an operator etc.\ as rules.
However, these are devoid of physics content.
These are merely the mathematical models of choice and it is the
statements about these constructions that gives the physics.
In the above formulation, these have therefore been separated.

The situation is analogous to modern algebra where the list of
operations on the underlying set is separated from the list of
axioms obeyed by the operations.

\subsubsection*{What is a state?}

It is easily checked that state $|\phi\rangle$ and $c|\phi\rangle$
are physically equivalent for $c\in\C\setminus\{0\}$.
The choice $\langle \phi|\phi\rangle=1$ is often imposed although here
this convention is not assumed.

%
%
%
%
%
%
\subsubsection*{Special forms of rules}

The postulates above are given in a very general form.
More convenient forms may be derived for special cases.
It is these special forms which are most useful in practice.
\begin{itemize}
\item
  The statistical formula may be simplified for the case of a discrete
  spectrum.

  If $\hat{A}$ has eigenvalue
  $\lambda$ with associated eigenstates forming a space $V$ then the
  probability that measuring $A$ will give a value $\lambda$ is
  \begin{equation*} 
    \frac{\| \hat{P}|\psi\rangle \|^2}{\||\psi\rangle\|^2}
  \end{equation*}
  where $\hat{P}$ projects onto $V$.

  If in fact the eigenvalue $\lambda$ is non-degenerate,
  ie.\ is associated with the single
  normalised eigenvector $\phi$,
  then the
  probability that measuring $A$ will give a value $\lambda$ is
  \[
    \frac{|\langle\psi|\phi\rangle|^2}{\||\psi\rangle\|^2}
    \mbox{ .}
  \]

\item
  An ideal measurement may be defined in terms of its affect on the
  state vector when such a vector exists.
  If an ideal measurement of observable $A$ is made at time $t$
  in such a way that its value
  is determined to be in the Borel set $\Omega$ then
  the system state changes discontinuously at time $t$
  as a result of the measurement to some vector in the space onto which
  $\hat{P}_{\Omega}$ projects,
  where $\{\hat{P}_{\Omega}\}$ is the p.v.m.\ associated with
  $\hat{A}$.

  There is a case in which the wave function collapse is completely
  determined, namely if observable $A$ is measured yielding eigenvalue
  $\lambda$ which is associated with a single eigenstate $|\phi\rangle$.
  Then the state collapses to $|\phi\rangle$.

%
%
  In the more general case, a completely determined collapse is possible
  if it is assumed that the measurement gives no information other
  than $A$ lying in set $\Omega$.  In this case the measurement is
  sometimes called `moral' and the system is left in state
  \begin{equation*} 
      \frac{\hat{P}_{\Omega}\psi}
               {\|\hat{P}_{\Omega}\psi\|}
      \mbox{ .}
  \end{equation*}
\end{itemize}

\subsubsection*{The composite system}

The composite system rule is particularly important in the interpretation
of quantum mechanics.  One might have expected that the state of the
composite system would always take the form
$|\phi\rangle|\psi\rangle$
where $|\phi\rangle$ and $|\psi\rangle$ are respectively states of
$S_1$ and $S_2$.  Thus, $S_1+S_2$ would be associated with the
Hilbert space $\FH_1 \x \FH_2$.

Instead, it is associated with the tensor product space $\FH_1 \otimes \FH_2$
in which there are vectors of the form
\[
  |\phi_1\rangle|\psi_1\rangle + |\phi_2\rangle|\psi_2\rangle
\]
which cannot be written in the simple form
\(
  |\phi\rangle|\psi\rangle
  \mbox{ .}
\)
In other words, {\it in general it is not possible to associate
a state with each subsystem of a composite system}.

This bizarre result is sometimes called the {\it non-separability}
of quantum mechanics.  It leads to nonlocality.

It is interesting to note that this result follows
without use of the composite system rule.

Suppose $|\phi_1\rangle$ and $|\phi_2\rangle$ are states of
$S_1$ corresponding to different values of an observable $P_1$.
It follows that $|\phi_1\rangle$ and $|\phi_2\rangle$ are orthogonal\footnote
{
  $|\phi_1\rangle$ and $|\phi_2\rangle$ are in spaces onto which
  $\hat{P}_{\Omega_1}$ and $\hat{P}_{\Omega_2}$ respectively project, with
  $\Omega_1$ and $\Omega_2$ disjoint.  Here $\{\hat{P}_{\Omega}\}$
  is the p.v.m.\ associated with $\hat{P}_1$.  It follows from the
  definition of a p.v.m.\ that $|\phi_1\rangle$ and $|\phi_2\rangle$
  are orthogonal.
}.
Suppose also that $|\psi_1\rangle$ and $|\psi_2\rangle$ are states of
$S_2$ corresponding to different values of an observable $P_2$.

Now it is at least implicit in the rules of quantum mechanics
that a composite system {\em may\/} be in a state
$|\phi\rangle|\psi\rangle$ for any $|\phi\rangle$ and $|\psi\rangle$,
otherwise one could {\it never\/} consider the state of a subsystem.

So $|\phi_1\rangle|\psi_1\rangle$ and $|\phi_2\rangle|\psi_2\rangle$
are valid states of $S_1+S_2$.
But the linear superposition rule implies that
\[
  |\phi_1\rangle|\psi_1\rangle + |\phi_2\rangle|\psi_2\rangle
\]
is a valid state (assuming that no superselection rule excludes this).
It may be checked that this state cannot possibly
be expressed in the form\footnote
{
  In this state one can measure $P_1$ and $P_2$ and the
  rules imply that the results will be correlated, ie.\ will
  yield values corresponding to $|\phi_1\rangle$ and $|\psi_1\rangle$
  or $|\phi_2\rangle$ and $|\psi_2\rangle$.  Such a correlation
  could not occur in a state of the form $|\phi\rangle|\psi\rangle$.

}
$|\phi\rangle|\psi\rangle$.

Therefore the non-separability of quantum mechanics follows immediately
from the association of separable Hilbert spaces with systems
and self-adjoint operators with observables.

\subsubsection*{May the rules be derived from physical axioms?}

The idea that a system may be described by a Hilbert space is
highly abstract and, as was just shown, has surprising consequences.
Can this abstract idea be derived from postulates of a more physical
nature?

Von Neumann suggested that such a programme should be attempted and
made several important contributions in this direction
\cite{vonNeumann:55} (and later with Birkhoff \cite{BirkhoffvonNeumann:36},
see \S\ref{s:logicoalgebraic}).

There have been many further contributions (eg.\ \cite{Zieler:61,Piron:64}).
However, all these attempts either involve abstract assumptions after
all or fail to fully reproduce quantum mechanics.
\subsubsection*{Other rules}

Von Neumann put forward the following two rules.
\begin{itemize}
\item
  Every vector in Hilbert space is, in principle, realisable.
\item
  Every densely defined self-adjoint operator is associated with some
  physical observable.
\end{itemize}

These assumptions are now known to be false.
In some quantum mechanical systems, there are valid states
$|\phi\rangle$, $|\psi\rangle$ such that the linear combination
$|\phi\rangle+|\psi\rangle$ is not realisable
\cite{WickWightmanWigner:52}.
These restrictions are called {\it superselection rules}.

Further, the projection onto $|\phi\rangle+|\psi\rangle$, although
self-adjoint, is not a physical observable\footnote
{
  If it was, and if it was measured when the system was in state
  $|\phi\rangle$, there would be a finite probability of obtaining the
  value $1$ leaving the system in state
  $(|\phi\rangle\M+|\psi\rangle)/\sqrt{2}$, a contradiction.
}.

Even in those particular quantum systems where all states {\em are\/}
realisable,
it is not clear that all densely defined self-adjoint operators
correspond to physical observables.
In particular, if $\hat{P}$ corresponds to a physical observable,
a function $f(\hat{P})$ of $\hat{P}$ will typically not be directly
observable.

Nevertheless, these assumptions are of historical importance and feature
in some of the older work on the interpretation of quantum mechanics.

\section{The Heisenberg picture}

In the above presentation, called the Schr\"odinger picture,
the state of the system evolves with time and the operators giving
observables are fixed.

Now suppose one wanted to calculate probabilities associated with a
measurement of $A$ at time $t$ for 1,000 systems for which the
initial state is known.
Is it necessary to apply the Schr\"odinger equation to all 1,000 systems
in order to work out their states at time $t$?

In fact, it is not.  It turns out that there is an operator $\hat{A}_H(t)$
which may be measured at time $0$ giving exactly the same results as
measuring $\hat{A}$ at time $t$.  Therefore, even if the measurement
is really performed at time $t$, the results may be predicted
equally well by analysing the measurement of $\hat{A}_H(t)$ at time $0$.
(The fact that such an operator exists is not surprising.
After all, the states at time $t$ are related to those at time $0$
by a unitary transformation, ie.\ an isomorphism.)

In this way, it is never necessary to apply the Schr\"odinger equation.
Instead, the operators associated with observables may be said to
evolve with time.  This is the {\it Heisenberg picture}.

It only remains to derive the form of $\hat{A}_H(T)$.
This will be done for the case of a fixed Hamiltonian.
Applying $\hat{A}$ to $|\psi_t\rangle=\exp(-i\hat{H}t/\hbar)|\psi_0\rangle$
is in some sense similar to applying
\ $\exp(i\hat{H}t/\hbar)\hat{A}\exp(-i\hat{H}t/\hbar)$ \ to $|\psi_0\rangle$
so a good candidate for $\hat{A}_H(t)$ is
\ $\exp(i\hat{H}t/\hbar)\hat{A}\exp(-i\hat{H}t/\hbar)$.

To check that this is the correct time dependence,
let $\{\hat{P}^t_{\Omega}\}$ be the p.v.m.\ associated with $\hat{A}(t)$.
It is easy to check that the p.v.m.\ associated with
$\hat{A}_H(t)$ is
\[
  \{\exp(i\hat{H}t/\hbar)\hat{P}^t_{\Omega}\exp(-i\hat{H}t/\hbar)\}
  \mbox{ .}
\]

Now consider the probability that measuring $A$ at time $t$ gives
a value in the Borel set $\Omega$.
In the Schr\"odinger picture the predicted probability is
\[
  \dbra\psi_0|
   [\exp(-i\hat{H}t/\hbar)(t)]^{\dagger}\hat{P}^t_{\Omega}
   [\exp(-i\hat{H}t/\hbar)]
   |\psi_0\dket
\]
whereas in the Heisenberg picture the predicted probability is
\[
  \dbra \psi_0|
    [\exp(i\hat{H}t/\hbar)(t)\hat{P}^t_{\Omega}
    \exp(-i\hat{H}t/\hbar)(t)]
  |\psi_0\dket
  \mbox{ .}
\]
These are equal.

Further, if in the Schr\"odinger picture the state collapses to
a vector in the space onto which $\hat{P}_{\Omega}$ projects, say $V$,
in the Heisenberg picture the state collapses to a vector in
the space onto which
$\{\exp(i\hat{H}t/\hbar)\hat{P}^t_{\Omega}\exp(-i\hat{H}t/\hbar)\}$
projects and this is just \ $exp(i\hat{H}t/\hbar)V$.
Therefore the Schr\"odinger picture state is still obtained from the
Heisenberg picture state by applying \ $\exp(-i\hat{H}t/\hbar)$.

Hence
\[
 \boxed{
  \hat{A}_H(t)=\exp(i\hat{H}t/\hbar)\hat{A}\exp(-i\hat{H}t/\hbar)
 }
\]
is the required relationship.

This may be given compactly by the differential equation
\begin{equation}
 \boxed{
  \deriv{\hat{A}_H}{t}
    =\frac{i}{\hbar} [\hat{H},\hat{A}_H(t)]
  \mbox{ .}
 }
\end{equation}
This is the {\it Heisenberg equation of motion}.
It is valid even when the Hamiltonian is not fixed.

\section{Quantisation and the Schr\"odinger representation}

\subsection*{Quantisation}

Consider a mechanical system with $n$ degrees of freedom.
Classically its state is described by $n$ generalised coordinates
$q_i,\ldots,q_n$, and $n$ generalised momenta $p_1,\ldots,p_n$.
The dynamics are given by the Hamiltonian function
\[
  H(q_1,\ldots,q_n,p_1,\ldots,p_n)
  \mbox{ .}
\]

It is found that the correct quantum description of such a system
requires operators $\hat{q}_i$ and $\hat{p}_i$ such that
\begin{equation} \label{e:pqcomm}
  [\hat{q}_i,\hat{p}_j] = i\hbar\delta_{ij}\hat{1}
  \quad\quad
  [\hat{q}_i,\hat{q}_j] = \hat{0}
  \quad\quad
  [\hat{p}_i,\hat{p}_j] = \hat{0}
  \mbox{ .}
\end{equation}
The commutator $[\hat{q}_j,\hat{p}_j] = i\hbar\hat{1}$ captures
the wave nature of matter while the others establish the freedom to
perform simultaneous measurements on different degrees of freedom.
Any operators satisfying these conditions may be used to represent
the observables of generalised coordinates and momenta.

It should be noted at this point that these relationships imply an
infinite-dimensional Hilbert space because in a finite-dimensional space
operators may be represented by matrices and there are no matrices
for which\footnote
{
  In fact, the diagonal in $QP\M-PQ$ always vanishes.
}
$QP\M-PQ=i\hbar I$.

In finding the quantum Hamiltonian one is guided by the
{\it correspondence principle\/} which states that in certain
limits the quantum system should behave classically.
It turns out that this implies that
the Hamiltonian in the quantum system should bare the same relationship
to the operators $\hat{q}_i$ and $\hat{p}_i$ as the classical Hamiltonian
does to the values $q_i$ and $p_i$.
This can be ambiguous but here only the simplest case is of interest
and in this case there is no ambiguity.

The case of interest is that of particles described by Cartesian coordinates
$q_1,\ldots,q_n$ in a potential $V$.

The classical Hamiltonian is
\[
  H(q_1,\ldots,q_n,p_1,\ldots,p_n) = \Bp^2/2m + V(\Bq)
  \mbox{ .}
\]
The quantum Hamiltonian is therefore
\begin{equation} \label{e:VHamiltonian}
  \hat{H}=
  \mbox{ ``}
  H(\hat{q}_1,\ldots,\hat{q}_n,\hat{p}_1,\ldots,\hat{p}_n)
  \mbox{'' }
  =
  \hat{\Bp}^2/2m + V(\hat{\Bq})
  \mbox{ .}
\end{equation}

\subsection*{The Schr\"odinger representation}

Still considering the case of particles in a potential with Cartesian
coordinates, one of course may choose any infinite dimensional
separable Hilbert space to work with since these are all isomorphic.
A particularly useful Hilbert space, the space of Lesbegue measurable maps
from $R^n$ to $\C$ with inner product
$\langle \phi,\psi\rangle = \int_{\R^n} \phi^*\psi$
[written $L^2(\R^n)$], was used
in Schr\"odinger's formulation of quantum mechanics.

In this separable Hilbert space, the operators defined by
$(\hat{q}_i f)(\Bq)=q_if(\Bq)$ and
$(\hat{p}_i f)(\Bq)=-i\hbar (\partial f/\partial q_i) (\Bq)$
(or strictly the self-adjoint extensions of these operators)
happen to satisfy the commutation relations \eqref{e:pqcomm}.
These will be chosen to represent the coordinates and momenta.


The useful thing is that the p.v.m.\ belonging to the operator $\hat{q}_i$
is simply
\[
  (\hat{P}_{\Omega}\psi)(\Bq) =
  \left\{
    \begin{array}{ll}
      \psi(\Bq) & (q_i\Min\Omega) \\
      0         & (q_i\M{\not\in}\Omega)
    \mbox{ .}
    \end{array}
  \right.
\]

Therefore, when the position $\Bq$ is measured in state
$\psi$, the probability of
finding it in Borel set $\Omega\subseteq\R^n$ is simply
\[
  \int_{\Omega} |\psi(\Bq)|^2 d\Bq
\]
provided that $\psi$ is normalised so that
$\int_{\R^n} |\psi(\Bq)|^2 d\Bq=1$.
Thus $|\psi(\Bq)|^2$ gives the probability density for finding
the particle near $\Bq$!

$\psi$ is called the {\it Schr\"odinger wave function}.

Note that in this `Schr\"odinger representation' the Hamiltonian
\eqref{e:VHamiltonian} becomes
\begin{equation} \label{e:SchVHamiltonian}
  \hat{H}=\frac{-\hbar^2\nabla^2}{2m}+V(x)\hat{1}
  \mbox{ .}
\end{equation}

\subsection*{The probability density current}

  Let $\psi$ be a wave function.
  A probability density current $\Bj$ for $\psi$ is a function satisfying
  \begin{equation*} 
   \boxed{
    \pderiv{}{t} \int_V |\psi(t,\Bx)|^2d\Bx
    =
    \int_S \Bj(t,\Bx).d\BS
   }
  \end{equation*}
  for every volume $V$ bounded by a surface $S$
  (with suitable smoothness conditions).

  A probability density current exists for the standard Hamiltonian,
  \eqref{e:SchVHamiltonian}.
  This may be derived as follows:
  \begin{equation*} 
  \begin{split}
    \pderiv{}{t}\int_{V}(\psi^*\psi)dV
    &=
    \int_{V}\pderiv{}{t}(\psi^*\psi)dV\\
    &=
    \int_V \left( \pderiv{\psi}{t}\psi^* + \pderiv{\psi^*}{t}\psi \right)dV
    \\&=
    \frac{1}{i\hbar}\int_V \left(\psi^*\hat{H}\psi-\psi\hat{H}\psi^*\right)dV
    \\&=
    \frac{-\hbar}{2im}\int_V
      \left(\psi^*\nabla^2\psi-\psi\nabla^2\psi^*\right)dV
    \\&=
    \frac{-\hbar}{2im}\int_V
      \nabla.\left(\psi^*\nabla\psi-\psi\nabla\psi^*\right)dV
    \\&=
\frac{-\hbar}{2im}\int_{S}\left(\psi^*\nabla\psi-\psi\nabla\psi^*\right).{\bf
dS}
    \mbox{ .}
    \end{split}
  \end{equation*}
  The last step uses the divergence theorem.
  So the probability density current is given by
  \[
    \Bj
    =
    \frac{\hbar}{2im}(\psi^*\nabla\psi-\psi\nabla\psi^*)
    =
    \frac{\hbar}{m}\mbox{ Im}(\psi^*\nabla\psi)
    \mbox{ .}
  \]

\section{Mixed states and the density matrix} \label{sec:dnstymtrx}

It is sometimes useful to derive the probabilities and expectation values
for a physical observable of a quantum mechanical system when
knowledge of the system's state is not complete.
Suppose that we know that the system is in normalised state $|\psi_i\rangle$
with probability $w_i$.

It turns out that the properties of such a {\it mixed state\/} are
not in general equal to the properties of any one {\it pure state}.
But not all the details of the probabilities need to be known;
it is sufficient to know the operator
\[
  \hat{\rho} = \sum_i w_i\hat{P}_{|\psi_i\rangle}
  = \sum_i w_i|\psi_i\rangle\langle\psi_i|
  \mbox{ .}
\]
This is called the {\it density matrix\/} or {\it statistical operator}.
Two different probability distributions may lead to the same
density matrix, and it is therefore necessary to show that
the density matrix contains all necessary information.
Namely, the following must be expressed in terms of the density matrix:
\begin{itemize}
\item
  the probability of measuring a given value $a$ of an observable $A$
\item
  the collapsed density matrix after a measurement is performed
\item
  the density matrix at a later time.
\end{itemize}

\paragraph*{Probabilities}

Let $A$ be an observable with corresponding operator $\hat{A}$.
Suppose that $\hat{A}$ has eigenvalue $a$ and that the projection
onto the corresponding space of eigenvectors is $\hat{P}_a$.
Then
\[
  \Pr(A\M=a;\hat{\rho}) = \sum_i w_i\Pr(A\M=a;|\psi_i\rangle)
  = \sum_iw_i\Tr(\hat{P}_{|\psi_i\rangle}\hat{P}_a)
  = \Tr\left(\left[\sum_iw_i\hat{P}_{|\psi_i\rangle}\right]\hat{P}_a\right)
  \mbox{ .}
\]
Therefore
\begin{equation*} 
  \boxed{
    \Pr(A\M=a;\hat{\rho}) = \Tr(\hat{\rho}\hat{P}_a)
    \mbox{ .}
  }
\end{equation*}

In the more general case where $\hat{A}$ has a continuous spectrum, if
its p.v.m.\ is $\{\hat{P}_{\Omega}\}$ then
\begin{equation*} 
  \boxed{
    \Pr(A\Min\Omega;\hat{\rho})
    = \Tr(\hat{\rho}\hat{P}_{\Omega})
    \mbox{ .}
  }
\end{equation*}


\paragraph*{The collapsed density matrix}

The collapse of a state is not unique.  Here the particular case
of a moral measurement is considered.

It is shown that in the case of a moral measurement the density
matrix after the measurement may be expressed in terms of the density
matrix before the measurement and the result.  Presumably this result
generalises to other types of measurement.

Suppose a system has density matrix
$\hat{\rho} = \sum_i w_i |\psi_i\rangle\langle\psi_i|$.
And that $A$ is then measured with the result $a$.
Let $\hat{A}$ be the (positive definite) operator corresponding to $A$
and let $\hat{P}_a$ project onto the space of eigenstates of $\hat{A}$
with eigenvalue $a$.

If the system was in fact in state $|\psi_i\rangle$ before the
measurement, it will now be in state $\hat{P}_a|\psi_i\rangle$.
But note that the probability associated with this eventuality is not
simply $w_i$ but rather a conditional probability which takes account of
the new available information, namely that measuring $A$ gave $a$.
Write $\psi_i$ for the assertion that the original state was $|\psi_i\rangle$
and $A\M=a$ for the assertion that $A$ measured $a$.
Then
\begin{equation*}
 \begin{split}
    w_i' &= \Pr(\psi_i \:|\: A\M=a) \\
         &= \frac{\Pr(\psi_i\mbox{ and }A\M=a)}{\Pr(A\M=a;\hat{\rho})} \\
         &= \frac{\Pr(\psi_i)\Pr(A\M=a;|\psi_i\rangle)}{\Pr(A\M=a;\hat{\rho})}
\\
         &= \frac{w_i\Tr(\hat{P}_{|\psi_i\rangle}\hat{P}_a)}
                {\Tr(\hat{\rho}\hat{P}_a)}
    \mbox{ .}
 \end{split}
\end{equation*}
Then the density matrix after the collapse is
\[
  \hat{\rho}' = \sum_i w_i' \hat{P}_{\hat{P}_a|\psi_i\rangle}
         = \sum_iw_i\frac{\Tr(\hat{P}_{|\psi_i\rangle}\hat{P}_a)}
                {\Tr(\hat{\rho}\hat{P}_a)}
                \hat{P}_{\hat{P}_a|\psi_i\rangle}
  \mbox{ .}
\]

But in general
\[
  \hat{P}_{|\phi\rangle}
  = \frac{|\phi\rangle\langle\phi|}{\langle\phi|\phi\rangle}
\]
so
\begin{equation} \label{eqn:qm:Pustt}
  \hat{P}_{\hat{P}_a|\psi_i\rangle} =
  \frac{\hat{P}_a|\psi_i\rangle\langle\psi_i|\hat{P}_a^{\dagger}}
       {\langle\psi_i|\hat{P}_a^{\dagger}\hat{P}_a|\psi_i\rangle}
  =
  \frac{\hat{P}_a|\psi_i\rangle\langle\psi_i|\hat{P}_a}
       {\langle\psi_i|\hat{P}_a|\psi_i\rangle}
  =
  \frac{\hat{P}_a\hat{P}_{|\psi_i\rangle}\hat{P}_a}
       {\langle\psi_i|\hat{P}_a|\psi_i\rangle}
  =
  \frac{\hat{P}_a\hat{P}_{|\psi_i\rangle}\hat{P}_a}
       {\Tr(\hat{P}_{|\psi_i\rangle}\hat{P}_a)}
  \mbox{ .}
\end{equation}
(The last step is verified by completing $\psi_i$ into an orthonormal
complete set and using this set to compute the trace.)
Therefore
\begin{equation} \label{e:qm:dnstycollps}
 \boxed{
  \hat{\rho}' = \sum_iw_i\frac{\hat{P}_a\hat{P}_{|\psi_i\rangle}\hat{P}_a}
                              {\Tr(\hat{\rho}\hat{P}_a)}
              =\frac{\hat{P}_a\hat{\rho}\hat{P}_a}{\Tr(\hat{\rho}\hat{P}_a)}
  \mbox{ .}
 }
\end{equation}

\paragraph*{Time evolution}

First consider the time evolution of the projection
$\hat{P}_{|\psi_i(t)\rangle}$ associated with the state
$|\psi_i(t)\rangle$.
It is possible to write
$|\psi_i(t)\rangle\M=\hat{U}(t)|\psi_i(0)\rangle$ where $\hat{U}(t)$
is unitary and satisfies equations
\begin{equation*} 
  i\hbar \deriv{\hat{U}}{t} = \hat{H}(t)\hat{U}(t)
\end{equation*}
and
\begin{equation*} 
  -i\hbar \deriv{\hat{U}^{\dagger}}{t} = \hat{U}^{\dagger}(t)\hat{H}(t)
  \mbox{ .}
\end{equation*}

Now rather like in \eqref{eqn:qm:Pustt} it is possible to write
\[
  \hat{P}_{\hat{U}|\psi_i\rangle} =
  \frac{\hat{U}|\psi_i\rangle\langle\psi_i|\hat{U}^{\dagger}}
       {\langle\psi_i|\hat{U}^{\dagger}\hat{U}|\psi_i\rangle}
  =
  \frac{\hat{U}|\psi_i\rangle\langle\psi_i|\hat{U}^{\dagger}}
       {\langle\psi_i|\psi_i\rangle}
  =
  \hat{U}\hat{P}_{|\psi_i\rangle}\hat{U}^{\dagger}
  \mbox{ .}
\]
Therefore
\[
  i\hbar \deriv{}{t} \hat{P}_{\hat{U}(t)|\psi_i\rangle}
  =
  \hat{H}(t)\hat{U}(t)\hat{P}_{|\psi_i\rangle}\hat{U}^{\dagger}(t)
  - \hat{U}(t)\hat{P}_{|\psi_i\rangle}\hat{U}^{\dagger}(t)\hat{H}(t)
  \mbox{ .}
\]
So
\[
  i\hbar \deriv{}{t} \hat{P}_{\hat{U}(t)|\psi_i\rangle}
  =
  [\hat{H}(t),\hat{P}_{\hat{U}(t)|\psi_i\rangle}]
  \mbox{ .}
\]

Now $\rho(t)$ is just $\sum_i w_i \hat{P}_{\hat{U}(t)|\psi_i\rangle}$
and therefore
\[
  \boxed{
    i\hbar \deriv{}{t} \hat{\rho}(t)
    =
    [\hat{H}(t),\hat{\rho}(t)]
    \mbox{ .}
  }
\]

If $\hat{H}$ is constant, this is solved by
\[
  \hat{\rho}(t) = e^{-i\hat{H}t/\hbar}\hat{\rho}(0)e^{i\hat{H}t/\hbar}
  \mbox{ .}
\]

This completes the argument showing that the quantum mechanics of
mixed states may be expressed in terms of the density matrix.

\paragraph*{NB}
The rules of quantum mechanics may
be presented directly in terms of density matrices.
Such a presentation has the advantage of assigning states to all systems.
However, it is traditional to define the rules in terms of pure states
and derive the more general rules as above.

\subsection*{The abstract density matrix}

\begin{thm}
  Let $\hat{\rho}=\sum_i w_i|\psi_i\rangle\langle\psi_i|$ be a density
  matrix for some quantum mechanical system.
  Then
  \begin{itemize}
  \item
    $\hat{\rho}$ is Hermitian
  \item
    $\hat{\rho}$ is positive semi-definite
  \item
    $\Tr\hat{\rho}\M=1$.
  \end{itemize}
\end{thm}

\begin{proof}
  The projection operators $|\psi_i\rangle\langle\psi_i|$ are Hermitian
  and positive semi-definite and hence $\hat{\rho}$ has both these
  properties too.

  For the last part,
  \[
    \Tr\hat{\rho} = \sum w_i \Tr(\hat{P}_{|\psi_i\rangle}) =\sum_iw_i=1
    \mbox{ .}
  \]
\end{proof}

\begin{defn}
  Any linear operator (defined everywhere and) satisfying the three
  conditions in the theorem is called a density matrix.
\end{defn}

Then the theorem simply states that every density matrix of a mixed state
is an abstract density matrix.
The converse is also true.

\begin{thm}
  Every density matrix is the density matrix for some ensemble.
\end{thm}

\begin{proof}
  Let $\hat{\rho}$ be a density matrix.
  It follows from the finite trace of $\hat{\rho}$ that
  $\hat{\rho}$ has a discrete spectrum.
  Let $|1\rangle,|2\rangle,\dots$ be a complete orthonormal sequence
  of eigenvectors with corresponding eigenvalues $\lambda_1,\lambda_2,\dots$
  (which  necessarily sum to $1$).
  Then
  \begin{equation*} 
    \hat{\rho} = \sum_i \lambda_i|i\rangle\langle i|
  \end{equation*}
  so $\hat{\rho}$ is the density matrix of the ensemble with states
  $|1\rangle,|2\rangle,\dots$ in the respective proportions
  $\lambda_1,\lambda_2,\dots$.
\end{proof}

%

\section{The Logico-algebraic approach} \label{s:logicoalgebraic}

\subsection*{Classical assertions}

An assertion about a classical system may be that a particle has
a given position or momentum, or, more generally, that the
position and momentum are in given ranges.
Any such assertion may  be identified with a subset
of phase space.

It may be that not every subset of phase space corresponds to a useful
assertion (eg.\ non-measurable sets).
Instead, the experimental propositions will correspond to some field
of subsets of phase space.

The assertions may be manipulated using classical logic.
This simply corresponds to manipulating the corresponding subsets.
Implication between assertions corresponds to set inclusion,
the negation of an assertion corresponds to the set complement,
disjunction corresponds to set union and conjunction to set intersection.

\subsection*{Quantum assertions}

In discussing quantum mechanics, two changes occur.

Firstly, since position and momentum cannot be measured
simultaneously, assertions can only relate to one or the other
(in practise to the one which is going to be measured).
A statement fixing the particle's position and momentum
cannot be checked and, arguably, is physically meaningless.

If one particular observable is picked, assertions may be made about
the values of that observable and these may be manipulated using classical
logic as above.
However, in general an assertion will be neither true nor false ---
this is the second difference.
Instead, there will be a certain probability of it being true.

Of course, one can make assertions about several commuting observables.
However, this amounts to nothing more than making a single assertion about
some observable with a more refined spectrum
(any commuting self-adjoint operators may be written as a functions of a
single common self-adjoint operator).

This suggests the following definition of an assertion.
\begin{defn}
  Let $\FH$ be a Hilbert space.
  An assertion is a pair $(\hat{A},\Omega)$ where
  $\hat{A}$ is a densely defined self-adjoint operator on $\FH$
  and $\Omega$ is a Borel subset of $\R$.
\end{defn}

Informally, the semantics of such an assertion is that if the
observable $\hat{A}$ is measured then the result
will be in $\Omega$.

If one fixed observable $A$ is considered, classical logic
can easily be applied to all the assertions involving $\hat{A}$.
It is possible to talk about the value of a measurement of $\hat{A}$
lying in $\Omega_1$ {\it and\/} in $\Omega_2$,
in $\Omega_1$ {\it or\/} in $\Omega_2$,
or {\it not\/} lying in $\Omega$.
Finally one can say that if it lies in $\Omega_1$ this {\it implies\/}
that it lies in $\Omega_2$.

As in the classical case, these manipulations amount to nothing
more than set intersection, union,  complement and inclusion.

Von Neumann and Birkhoff \cite{BirkhoffvonNeumann:36} introduced
an elegant abstract representation for these assertions.
The probability of the assertion $(\hat{A},\Omega)$ holding
in a mixed state with density matrix $\hat{\rho}$ is just
\begin{equation} \label{e:projectormeasure}
  P[(\hat{A},\Omega);\hat{\rho}] = \Tr(\hat{\rho}\hat{P}_{\Omega})
\end{equation}
where $\{\hat{P}_{\Omega}\}$ is, of course, the p.v.m.\ associated with
$\hat{A}$.
Therefore the assertion may be represented simply by the projection
operator $\hat{P}_{\Omega}$.  Although this projection may also represent
some other assertion
$(\hat{A}',\Omega')$, the abstract form is sufficient for calculating
the probabilities.

The classical logic operations may also be carried out in terms of
the projections, or, more conveniently, in terms of the spaces on to
which they project.  In terms of these spaces, conjunction corresponds
to intersection, disjunction to the linear span $+$, negation to
orthogonal complement and implication to set inclusion.

Although all the assertions for all the operators are now neatly
represented by a single structure, the lattice of projection operators
on $\FH$, they may not all be treated using classical logic.
Mathematically, this is because the lattice of projectors is not
boolean, ie.\ if we blindly started applying classical logic to
all the projectors, basic identities such as
\[
  p \wedge (q \vee r) = (p \vee q) \wedge (p \vee r)
\]
will fail.

Instead, for any particular experiment, the projectors which arise in
the p.v.m.\ for the relevant operator must be picked out.
These form a boolean sublattice and may be discussed using classical
logic.

\subsection*{Probabilities}

Up to now the assertions have been discussed.
In a given (mixed) state, each assertion, ie.\ each projector,
is associated with a probability
by \eqref{e:projectormeasure}:
\begin{equation} \label{eqn:qmi:prjprb}
  \hat{P} \mapsto \Tr(\hat{\rho}\hat{P})
  \mbox{ .}
\end{equation}
In other words, there is a map from the lattice of projectors to $[0,1]$.
Does this map behave like a probability?

One might expect that if $V$ and $W$ are disjoint linear subspaces
then the corresponding probabilities should be additive, ie.\
\[
  \Tr(\hat{\rho}\hat{P}_{V}) + \Tr(\hat{\rho}\hat{P}_{W})
  =
  \Tr(\hat{\rho}\hat{P}_{V+W})
  \mbox{ .}
\]
This is false.
For example, consider a spin $\half$ system, in a state with $z$-component
of spin up.  The projection associated with $z$-component of spin up
has probability $1$ while the projection associated with $y$-component
of spin up has probability $\half$.  These corresponding spaces
are disjoint and yet the probability associated with the span of the
two spaces is not, of course, $1\half$.

In general, probabilities {\em are\/} additive for mutually exclusive
assertions
which may be tested at one time but {\em not\/} for assertions which cannot
be tested at one time.

Assertions can be tested at one time if their projectors project onto
mutually orthogonal spaces because then these spaces may appear
as eigenspaces for a self-adjoint operator.

Therefore, \eqref{eqn:qmi:prjprb} defines a measure in the sense
that if $V$ and $W$ are {\em mutually orthogonal\/} then
\[
  \Tr(\hat{\rho}\hat{P}_{V}) + \Tr(\hat{\rho}\hat{P}_{W})
  =
  \Tr(\hat{\rho}\hat{P}_{V+W})
  \mbox{ .}
\]
More abstractly, \eqref{eqn:qmi:prjprb} defines a probability
measure on every boolean sublattice of the lattice of projectors.

The map \eqref{eqn:qmi:prjprb} defines such a measure for each mixed state
$\hat{\rho}$.
Gleason's theorem \cite{Gleason:57} is that these are the only measures.

\begin{thm}[Gleason's theorem] \label{t:qmi:Gleason}
  Let $V$ be a separable Hilbert space of dimension more
  than two.
  Let $\mu$ be a map from the set of closed linear subspaces
  of $V$ to $[0,1]$ such that one of the following two (equivalent)
  conditions hold.
  \begin{itemize}
  \item
    $\mu$ is a measure on every maximal boolean sublattice of
    the lattice of subspaces
  \item
    If $U_1,U_2,\dots$ are pairwise orthogonal closed linear subspaces
    then
    \[
      \mu\left(\bigcup_{i\in\N}U_i\right)=\sum_{i\in\N}\mu(U_i)
      \mbox{ .}
    \]
  \end{itemize}
  Then there is some abstract density matrix such that
  \[
     \mu(U) = \Tr(\hat{\rho}\hat{P}_U)
  \]
  for all $U$.
\proofend
\end{thm}

Therefore {\it the state of a quantum mechanical system can be defined
as a probability measure, in the above sense,
on the projection lattice of the associated Hilbert space.}

This may be used to give an alternative formulation of quantum mechanics.
However, the motivation for presenting this material here is different:
the idea of projections as assertions is central to the histories
formalism, considered next, which in turn is central to some modern
interpretations of quantum mechanics.

\section{Spin half}

One particular quantum system will be used in examples, that of a particle
of spin half.  In such a system the $x$-component, $y$-component and
$z$-component of spin are associated respectively with operators
$\hat{\sigma}_x$, $\hat{\sigma}_y$ and $\hat{\sigma}_z$ which
satisfy
\[
  [\hat{\sigma}_x,\hat{\sigma}_y] = i\hat{\sigma}_z
  \quad\quad
  [\hat{\sigma}_y,\hat{\sigma}_z] = i\hat{\sigma}_x
  \quad\quad
  [\hat{\sigma}_z,\hat{\sigma}_x] = i\hat{\sigma}_y
  \mbox{ .}
\]
Each of the operators has eigenvalues $\pm 1$.
Each eigenvalue is associated
with a single eigenvector typically written $|z\!=\uparrow\rangle$,
$|z\!=\downarrow\rangle$, for example,
or, when context allows, just $|\!\uparrow\rangle$ and $|\!\downarrow\rangle$.

Whenever the system is in an eigenstate of one of the components,
the other two components take values up and down with equal probability.

\bib

The rules of quantum mechanics, the Heisenberg picture and
Schr\"odinger representation and the density matrix construction
are discussed in most quantum mechanics textbooks,
eg.\ \cite{dEspagnat:71,LandauLifschitz:77}.

The logico-algebraic approach was introduced in
\cite{BirkhoffvonNeumann:36}.
A concise treatment is given in \cite[\S 1.4]{Redhead:87}.

\mychapter{Histories} \label{c:histories}

A history is a sequence of results from different measurements at
different times.
Histories are here presented as sequences of
measurements which may be analysed using the rules of quantum mechanics.

However, the importance of histories lies elsewhere.
In some interpretations of quantum mechanics, histories are
given a fundamental role and it is the rules of quantum mechanics which are
derived.
This approach is not considered until the end of Part~III.

One of the motivations for discussing histories is the limitations
of ordinary quantum logic (\S\ref{s:logicoalgebraic}).
Since at one time it is possible to measure position or momentum but
not both, classical logic can only be used to discuss either position
or momentum, for example.
But, of course, it is quite possible to measure position
at one time and momentum at another.

\subsection*{Histories}

In this approach, the idea of a projection operator representing
an assertion is generalised to sets of projection operators
which represent assertions about different times.
This is clearly most easily presented in the Heisenberg picture.

Let $\hat{P}$ be a projection operator.
$\hat{P}(t)$ may be thought of as the assertion that $\hat{P}$ holds
at time $t$.  In the Heisenberg picture it may be defined formally
as
\[
  \hat{P}(t) = e^{i\hat{H}t/\hbar} \hat{P} e^{-i\hat{H}t/\hbar}
  \mbox{ .}
\]

\begin{defn}
  An exhaustive set of exclusive assertions at time $t$ is a sequence
  $(\hat{P}_1(t),\hat{P}_2(t),\ldots)$
  such that
  \[
    \sum_{\alpha}\hat{P}_{\alpha}(t) = 1
    \quad\quad
    \hat{P}_{\alpha}\hat{P}_{\beta}=\delta_{\alpha\beta}\hat{P}_{\alpha}
    \mbox{ .}
  \]

  A {\bf set of alternative histories} consists of
  a sequence $t_1<t_2<\cdots$ of times and a corresponding sequence
  \[
    (\hat{P}^1_1(t_1),\hat{P}^1_2(t_2),\ldots),
    (\hat{P}^2_1(t_1),\hat{P}^2_2(t_2),\ldots),
    \ldots
  \]
  of exhaustive sets of exclusive assertions.

  A {\bf history} in a set of alternative histories is a sequence
  \[
    (\hat{P}^1_{\alpha_1}(t_1),\hat{P}^2_{\alpha_2}(t_2),\ldots)
  \]
  of assertions at different times, one for each set of assertions
  in the set of histories.
\end{defn}

\paragraph*{NB}

The times play no essential role in the formalism.
A history may be seen simply as a sequence of projectors.

\subsection*{Fine-graining / course-graining}

One set of alternative histories may be obtained from another
by replacing the alternatives $\hat{P}^i_1(t_i),\hat{P}^i_2(t_i),\ldots$
with a smaller set of alternatives in which projectors
$\hat{P}^i_{r_1}(t_i),\ldots,\hat{P}^i_{r_k}(t_i)$
are replaced by the projector
$\hat{P}^i_{r_1}(t_i)+\cdots+\hat{P}^i_{r_k}(t_i)$.
This process is called {\it course-graining\/}
(although Omn\`es calls it {\it reduction}).

The set of alternative histories obtained by this process (which
may be repeated arbitrarily many times) is called a course-graining
of the original set of alternative histories.
The opposite relation is {\it fine-graining}.

In the case where all the operators at a given time $t_i$ are replaced
by the identity operator, the time $t_i$ may be completely dropped
in the course-grained set of histories.

Although these relations are between sets of alternative histories,
it is often convenient to talk about a course-grained history
meaning a history in a course-grained set of alternative histories.
Similarly a fine-grained history will mean a history in a fine-grained
set of alternative histories.

Each history in a course-graining corresponds, in an obvious way,
to a set of histories in the original set of alternative histories.
Formally, the course-grained history
\[
  \left(\sum_{\alpha\Min S_1} \hat{P}^1_{\alpha}(t_1),\ldots,
        \sum_{\alpha\Min S_n} \hat{P}^n_{\alpha}(t_n) \right)
\]
in which the projectors $\hat{P}^i_{\alpha}(t_i)$ are from the original
fine-grained set of alternative histories, corresponds to the
set
\[
  \{(\hat{P}^1_{\alpha_1}(t_1),\ldots,\hat{P}^n_{\alpha_n}(t_n) )
      \; | \; \alpha_1\Min S_1,\ldots,\alpha_n\Min S_n\}
\]
of fine-grained histories.

The following property of histories will be important.

\begin{lemma} \label{l:sumhists}
  For a set of alternative histories,
  \[
    \sum_{\alpha_1,\ldots,\alpha_n}
    \hat{P}^1_{\alpha_1}(t_1)\cdots\hat{P}^n_{\alpha_n}(t_n)
    =
    \hat{1}
    \mbox{ .}
  \]
\end{lemma}

\begin{proof}
  By induction on $n$.
  \begin{equation*}
   \begin{split}
    \sum_{\alpha_1,\ldots,\alpha_n}
    \hat{P}^1_{\alpha_1}(t_1)\cdots\hat{P}^n_{\alpha_n}(t_n)
    &=
    \left(
    \sum_{\alpha_1}
    \hat{P}^1_{\alpha_1}(t_1)
    \right)
    \left(
    \sum_{\alpha_2,\ldots,\alpha_n}
    \hat{P}^2_{\alpha_2}(t_2)\cdots\hat{P}^n_{\alpha_n}(t_n)
    \right) \\
    &=
    \sum_{\alpha_1,\ldots,\alpha_n}
    \hat{P}^2_{\alpha_2}(t_2)\cdots\hat{P}^n_{\alpha_n}(t_n)
    \cdots
    \mbox{ .}
   \end{split}
  \end{equation*}
\end{proof}

\subsection*{Probabilities}

In the conventional rules, what is the probability of measuring
$P_1,P_2,\ldots,P_n$ and obtaining $1$ for each?
Here it is assumed that these are indeed physical observables
(in practice histories are only useful if they are)
and, more importantly, that they are measured morally.

This probability may be given by
\[
  \Pr(P_1,\ldots,P_n)
  =
  \Pr(P_1) \x \Pr(P_1|P_2) \x \ldots \x \Pr(P_n|P_1,\ldots,P_{n-1})
\]
where
$\Pr(Q|Q_1,\ldots,Q_r)$ means the probability of measuring $1$ for $Q$ given
that
$Q_1,\ldots,Q_r$ have been measured in that order and have all been
found to be equal to $1$.
Now by \eqref{e:qm:dnstycollps}, the right hand side is given by
\[
  \Tr(\hat{P}_1\hat{\rho}\hat{P}_1)
  \x
  \frac{\Tr(\hat{P}_2[\hat{P}_1\hat{\rho}\hat{P}_1]\hat{P}_2)}
       {\Tr(\hat{P}_1\hat{\rho}\hat{P}_1)}
  \x \cdots \x
  \frac{\Tr(\hat{P}_n[\hat{P}_{n-1}\cdots\hat{P}_1\hat{\rho}
              \hat{P}_1\cdots\hat{P}_{n-1}]\hat{P}_n)}
{\Tr(\hat{P}_{n-1}\cdots\hat{P}_1\hat{\rho}\hat{P}_1\cdots\hat{P}_{n-1})}
  \mbox{ .}
\]
Thus
\[
 \boxed{
  \Pr(\hat{P}_1,\ldots,\hat{P}_n)
  =
  \Tr(\hat{P}_n\cdots\hat{P}_1\hat{\rho}\hat{P}_1\cdots\hat{P}_n)
  \mbox{ .}
 }
\]
This formula is used to define probabilities for histories.

Note that it is inevitable that the probabilities depend on the order
of the projectors.  For example, consider a particle of spin $\half$
with the projectors $\hat{P}$ and $\hat{Q}$ corresponding respectively
to $\sigma_x=\hbar/2$ and $\sigma_y=\hbar_2$.
It is clearly more probable to obtain all $1$s when measuring
$P$ several times followed by $Q$ several times rather
than measuring them alternately.

\subsection*{Consistency}

Above it was shown that a course-grained history corresponds to a set
of fine-grained histories.
Can the course-grained history be considered as a disjunction of
the corresponding fine-grained histories?
Does its probability equal the sum of the probabilities of
the fine-grained histories?
This question is now analysed.

Given a fixed set of alternative histories, a history
$(\hat{P}^1_{\alpha_1}(t_1),\ldots,\hat{P}^n_{\alpha_n}(t_n))$
may be denoted simply by the sequence
$(\alpha_1,\ldots,\alpha_n)$
of indices.
The notation $[\boldsymbol\alpha]$ will be used as a shorthand for such a
sequence.

Now given a particular set of alternative histories and an initial
state $\hat{\rho}$, define for any two histories $[\boldsymbol\alpha]$,
$[\boldsymbol\alpha']$,
\[
 \boxed{
  D([\boldsymbol\alpha],[\boldsymbol\alpha']) =
  \Tr[\hat{P}^1_{\alpha_1}(t_1)\cdots\hat{P}^n_{\alpha_n}(t_n)
      \hat{\rho}
      \hat{P}^n_{\alpha_n'}(t_n)\cdots\hat{P}^1_{\alpha_1'}(t_1)]
  \mbox{ .}
 }
\]
$D$ is called the decoherence functional (for state $\hat{\rho}$).
Note that $D([\boldsymbol\alpha],[\boldsymbol\alpha])$ gives the probability of
$[\boldsymbol\alpha]$ and that
\begin{equation*}
  \begin{split}
    D([\boldsymbol\alpha],[\boldsymbol\alpha'])
    & = \Tr[\hat{P}^1_{\alpha_1}(t_1)\cdots\hat{P}^n_{\alpha_n}(t_n)
      \hat{\rho}
      \hat{P}^n_{\alpha_n'}(t_n)\cdots\hat{P}^1_{\alpha_1'}(t_1)] \\
    & = \Tr[\{\hat{P}^1_{\alpha_1}(t_1)\cdots\hat{P}^n_{\alpha_n}(t_n)
      \hat{\rho}
    \hat{P}^n_{\alpha_n'}(t_n)\cdots\hat{P}^1_{\alpha_1'}(t_1)\}^{\dagger}]^*\\
    & = \Tr[\hat{P}^1_{\alpha_1'}(t_1)\cdots\hat{P}^n_{\alpha_n'}(t_n)
      \hat{\rho}
      \hat{P}^n_{\alpha_n}(t_n)\cdots\hat{P}^1_{\alpha_1}(t_1)]^* \\
    & = D([\boldsymbol\alpha'],[\boldsymbol\alpha])^*
  \end{split}
\end{equation*}
using the facts that $\Tr(A^{\dagger})=\Tr(A)^*$
and that $(\hat{A}\hat{B})^{\dagger}=\hat{A}^{\dagger}\hat{B}^{\dagger}$
and that all the operators are self-adjoint.

Now let $[\boldsymbol\alpha]$ be a course-grained history with
its $i$-th projector
$\hat{P}^i_{\alpha_i}(t_i)$ equal to
$\sum_{\beta\Min S_i} \hat{Q}^i_{\beta}(t_i)$.

Then
\begin{equation} \label{e:qmi:Ddecomp}
 \begin{split}
  D([\boldsymbol\alpha],[\boldsymbol\alpha])
  &=
  \sum_{\mbox{\scriptsize $\begin{array}{c}
           [\boldsymbol\beta]  \in S_1\x S_2\x\cdots  \\ {}
           [\boldsymbol\gamma] \in S_1\x S_2\x\cdots
        \end{array} $}
        }
    D([\boldsymbol\beta],[\boldsymbol\gamma])  \\
  &=
  \sum_{[\boldsymbol\beta] \in S_1\x S_2\x\cdots}
    D([\boldsymbol\beta],[\boldsymbol\beta])
  +
  \sum_{\mbox{\scriptsize $\begin{array}{c}
           [\boldsymbol\beta] \in S_1\x S_2\x\cdots  \\ {}
           [\boldsymbol\gamma]\in S_1\x S_2\x\cdots  \\ {}
           [\boldsymbol\beta] \neq [\boldsymbol\gamma]
        \end{array} $}
        }
    D([\boldsymbol\beta],[\boldsymbol\gamma])
  \mbox{ .}
 \end{split}
\end{equation}
If the last term vanished, this result would give the required probabilities
sum rule giving the probability of the course-grained history as the
sum of the probabilities of the corresponding fine-grained histories.
Note that the last term is necessary real because it sums conjugate pairs.
This motivates the following definition.

\begin{defn}
  A set of alternative histories is {\bf weakly decoherent} or
  {\bf consistent} with respect
  to an initial state $\hat{\rho}$ if for every two distinct histories
  $[\boldsymbol\alpha]$ and $[\boldsymbol\alpha']$ in the set,
  $\Real  D([\boldsymbol\alpha],[\boldsymbol\alpha'])\M=0$.
  The set is {\bf decoherent} or {\bf medium decoherent} if
  for every two distinct histories
  $[\boldsymbol\alpha]$ and $[\boldsymbol\alpha']$ in the set,
  $D([\boldsymbol\alpha],[\boldsymbol\alpha'])=0$.
\end{defn}

\eqref{e:qmi:Ddecomp} leads immediately to the following theorem.
\begin{thm}
  If a set of histories is weekly decoherent (with respect to
  initial state $\hat{\rho}$) then the
  probability of a history in any course-graining of this set
  (with respect to $\hat{\rho}$) is the
  sum of the probabilities of the corresponding fine-grained
  histories.
\proofend
\end{thm}

In practice, the approaches to establishing decoherence generally establish
decoherence and not merely weak decoherence.
Therefore weak decoherence, although sufficient for the theorem,
will not be very useful in practice.

Why are consistent sets of histories important?
If the probabilities in a fine-graining do not add up to the probability
of the original course-grained history, then the fact that finer
measurements are being made is affecting the overall chances of the
course-grained history occurring.
To sum up,
\begin{quote}
\em
  In a decoherent set of histories, every fine-grained measurement is
  providing extra information without affecting the overall chance
  a any course-grained history.
\end{quote}
Therefore, if one wants to obtain information about a quantum system
by experiment, without disturbing the results of later experiments,
only consistent histories may be considered!

\subsection*{Decoherence, pure initial states, and records}

Let $\hat{\rho}=|\psi\rangle\langle\psi|$ with $|\psi\rangle$ normalised,
the projector onto the space spanned by $|\psi\rangle$.
Let $[\boldsymbol\alpha]$ range over a set of alternative histories.
Then
\begin{equation*}
 \begin{split}
  D([\boldsymbol\alpha],[\boldsymbol\alpha'])
  &=
  \Tr[\hat{P}^1_{\alpha_1}(t_1)\cdots\hat{P}^n_{\alpha_n}(t_n)
      \hat{\rho}
      \hat{P}^n_{\alpha_n'}(t_n)\cdots\hat{P}^1_{\alpha_1'}(t_1)] \\
  &=
  \Tr[|\psi\rangle\langle\psi|
      (\hat{P}^1_{\alpha_1'}(t_1)\cdots\hat{P}^n_{\alpha_n'}(t_n))^{\dagger}
      \hat{P}^1_{\alpha_1}(t_1)\cdots\hat{P}^n_{\alpha_n}(t_n)
      |\psi\rangle\langle\psi|] \\
  &=
  \langle\psi|[|\psi\rangle\langle\psi|
      (\hat{P}^1_{\alpha_1'}(t_1)\cdots\hat{P}^n_{\alpha_n'}(t_n))^{\dagger}
      \hat{P}^1_{\alpha_1}(t_1)\cdots\hat{P}^n_{\alpha_n}(t_n)
      |\psi\rangle\langle\psi|]|\psi\rangle \\
  &=
\langle\psi\,|(\hat{P}^1_{\alpha_1'}(t_1)\cdots\hat{P}^n_{\alpha_n'}(t_n))^{\dagger}
  \hat{P}^1_{\alpha_1}(t_1)\cdots\hat{P}^n_{\alpha_n}(t_n)\,|\psi\rangle
  \mbox{ .}
 \end{split}
\end{equation*}
Therefore an equivalent condition for the set of histories to decohere
with respect to initial state $|\psi\rangle$ is that
$\hat{P}^1_{\alpha_1}(t_1)\cdots\hat{P}^n_{\alpha_n}(t_n)\,|\psi\rangle$
and
$\hat{P}^1_{\alpha_1'}(t_1)\cdots\hat{P}^n_{\alpha_n'}(t_n)\,|\psi\rangle$
should be orthogonal for all $[\boldsymbol\alpha]\Mneq[\boldsymbol\alpha']$.

If the histories do decohere,
by Lemma~\ref{l:sumhists}, $|\psi\rangle$ may be written as the sum
\[
  \sum_{\alpha_1,\ldots,\alpha_n}
  \hat{P}^1_{\alpha_1}(t_1)\cdots\hat{P}^n_{\alpha_n}(t_n)\,|\psi\rangle
\]
of orthogonal vectors.

This alternative definition of decoherence shows that decoherence is
equivalent to the existence
of {\it records\/}\footnote
{
  Note that  the meaning of this term varies in the literature
  on histories.
}.
Suppose one were to perform all the experiments of a set of
alternative histories and lose the results.  Is it possible  to
reconstruct these by doing a further experiment on the system?

According to the conventional rules, if the experiment gives
results $\alpha_1,\ldots,\alpha_n$ then the system is left in state
$\hat{P}^1_{\alpha_1}(t_1)\cdots\hat{P}^n_{\alpha_n}(t_n)\,|\psi\rangle$.
An experiment to determine which of these states the system is in, is
of course possible {\it iff\/}\footnote
{
  Strictly, the if clause depends on von Neumann's dubious postulate that
  every self-adjoint operator belongs to some physical observable.
  Nevertheless, decoherence implies that different results leave to
  orthogonal states and this seems significant even if there does not
  happen to be an experiment available to distinguish these states.
}
these states are mutually orthogonal.
But is has just been shown that this is the case {\it iff\/} the set
of decoherent histories decoheres (with respect to $|\psi\rangle$).

\subsection*{Prediction and retrodiction}

A history $(\alpha_1,\ldots,\alpha_n)$ may be split into two parts
$(\alpha_{i_1},\ldots,\alpha_{i_k})$ and $(\alpha_{j_1},\ldots,\alpha_{j_l})$
each of which is a history in some (different) coarse-graining.
Given an initial state, define the conditional probability
\[
   p\bra(\alpha_{i_1},\ldots,\alpha_{i_k}) \: | \:
        (\alpha_{j_1},\ldots,\alpha_{j_l})\ket
   =
   \frac{p(\alpha_1,\ldots,\alpha_n)}
        {p(\alpha_{j_1},\ldots,\alpha_{j_k})}
   \mbox{ .}
\]
Consider an experiment in which
$P^1_{\alpha_1}(t_1),\ldots,P^n_{\alpha_n}(t_n)$ are measured.
It is tempting to think of the above probability as the probability
that measurements $\alpha_{i_1},\ldots,\alpha_{i_k}$ give $1$ given that
measurements $\alpha_{j_1},\ldots,\alpha_{j_l}$ give $1$.
But, in general, that probability is given by
\[
   \frac{p(\alpha_1,\ldots,\alpha_n)}
        {\sum_{
           \mbox{\scriptsize$\begin{array}{c}
              \beta_{i_1},\ldots,\beta_{i_k} \\
              \beta_{j_1}=\alpha_{j_1},\ldots,\beta_{j_l}=\alpha_{j_l}
           \end{array}$} }
         p(\beta_1,\ldots,\beta_n)}
   \mbox{ .}
\]
However, if the set of alternative histories decoheres then these {\em are\/}
equal.

\subsection*{Classical logic}

\begin{defn}
  A proposition over a set of alternative histories
  is a subset of the histories in this set.
\end{defn}

Of course, the propositions over a set of alternative histories may be
discussed using classical logic by identifying conjunction, disjunction,
negation and implication with, respectively, set intersection, set union,
set complement and set inclusion.

However, this will not give a useful implication relation.
In a given system we would expect to have $a\Rightarrow b$ even if
$b$ is not a subset of $a$, provided that the histories in $b\M-a$ have
probability zero.
In effect, histories of probability zero can be neglected.

Suppose a system with initial state $\hat{\rho}$ is being considered.
Write $a\equiv b$ if all the histories in $a \Delta b$ (the set-theoretic
symmetric difference of $a$ and $b$) have probability zero.
This relation is clearly reflexive and symmetric and it also transitive
since for any sets $p,q,r$,
\[
  p \Delta r \subseteq (p \Delta q) \cup (q \Delta r)
  \mbox{ .}
\]

In order for the operations $\wedge$, $\vee$ and $\neg$ (identified
with $\cap$, $\cup$ and set complement)
to carry over to equivalence classes, it is necessary to check that
$\equiv$ is a congruence with respect to these operations.
To see that this is true, suppose $a\equiv a'$ and $b\equiv b'$.
Then
\[
  \begin{array}{rcl@{\mbox{ follows from }}rcl}
    a \wedge b & \equiv & a' \wedge b' &
    (p\cap q) \Delta (p'\cap q') & \subseteq & (p \Delta p') \cup (q \Delta
q')\\
    a \vee b & \equiv & a' \vee b' &
    (p\cup q) \Delta (p'\cup q') & \subseteq & (p \Delta p') \cup (q \Delta
q')\\
    \neg b & \equiv & \neg b' &
    (S-p) \Delta (S-p') & = & p \Delta p' \mbox{ if } p,p'\subseteq S
    \mbox{ .}
  \end{array}
\]

Then by a standard result of universal algebra\footnote
{
  See eg.\ \cite{Gratzer:68} or the beginning of \cite{Hennessy:88}.
}
$\wedge$, $\vee$ and $\neg$
may be defined naturally on equivalence classes and, like the original
operations, satisfy the axioms for a complemented boolean lattice
(in the algebraic sense, ie.\ no partial order is assumed).
Implication may then be defined by
\[
  [a] \Rightarrow [b]  \mbox{ \ iff \ } [a] = [a] \vee [b]
\]
(this is the standard way for defining a partial order on an
algebraic lattice)
thus recovering classical logic.

\paragraph*{NB}
The above procedure is different to that presented by Omn\`es
\cite{Omnes:88a,Omnes:90}.
There, $a\Rightarrow b$ means that all the histories in
$a\M-b$ have probability $0$
(although it is not explicitly defined in this form).
$a=b$ is written for $a\Rightarrow b$ and $b\Rightarrow a$ so that
$=$ coincides with the equivalence above.

However, Omn\`es does not then redefine $\Rightarrow$, $\wedge$, $\vee$
and $\neg$ in terms of equivalence classes, a gap in the paper's
presentation.
Omn\`es directly proves the 20 or so rules of abstract logic
(ie.\ properties of a boolean lattice) whereas here we have taken the more
elegant route of inferring these immediately from the lattice structure
of set theory by using a theorem of universal algebra.

Note that the above procedure is valid whether or not the set of alternative
histories weakly decoheres.
Omn\`es writes that conventional
logic applies {\it iff\/} consistency conditions are satisfied.
However,
neither his approach to logic nor the above approach seem to support this
conclusion.
{\it Any\/} set of histories represents a valid programme of experiments
and the results of such a programme can always be discussed in classical
logic.
\proofend

Now that logic has fully been constructed for equivalence classes, what
can be said about the original propositions?
Without attempting to reconstruct all of logic, one can just define
implication between propositions by
\[
  a \Rightarrow b \quad \mbox{\it iff\/} \quad [a] \Rightarrow [b]
  \mbox{ .}
\]
It is this type of implication which is used in physical reasoning.

Now what is the importance of consistent sets of histories?
Suppose $a$, $b$ are propositions in some set of histories and $a'$, $b'$
correspond to the fine-grainings of the histories in $a$ and $b$ in some
fine-grained set of histories.  Now if these sets of histories are
consistent then $a\Rightarrow b$ {\it iff\/} $a'\Rightarrow b'$.
Of course, if $a''$ and $b''$
correspond to the fine-grainings of $a$ and $b$ in some
other consistent fine-grained set of histories then this is also
equivalent to $a''\Rightarrow b''$.

Therefore, {\it any physical implications which may be derived in one
consistent set of histories is also valid in any other consistent
set of histories}.  It is this result which allows histories to
be used consistently in interpretation.

\paragraph*{Bibliography}

Most of the work on decoherent histories is concerned with
the interpretation of quantum mechanics, see
Chapters~\ref{c:omnes}--\ref{c:gmh}.
Many of the papers do include careful descriptions of the histories
formalism and of the associated probabilities, logic and consistency
conditions.  Here, pointers to some of these papers are given.

Histories were introduced by Griffiths \cite{Griffiths:84} in a slightly
different form to the above.
This paper also introduced the ideas of
consistent sets of histories, and the probability measure for histories.

Histories were defined in a form closer to that used above by Omn\`es
\cite{Omnes:88a,Omnes:90,Omnes:92}.
These papers developed the idea of applying classical logic
to decohering sets of histories.

Another variant of the histories formalism plays a central role
in the work of Gell-Mann and Hartle on interpretation
and is described in many of their papers, particularly
\cite{GellMannHartle:90a,GellMannHartle:90b,Hartle:93b}.

Other papers which include a description of the formalism
include \cite{Halliwell:93c,Halliwell:94,DowkerKent:94}.

One aspect of histories not discussed above is the notion of equivalence
between histories.  Discussions are included in
\cite{Hartle:89,GellMannHartle:90b,DowkerKent:94,GellMannHartle:94}.

There have been a small number of papers
concerned with the histories formalism {\it per se}, eg.\
\cite{Diosi:94a,Diosi:94b}.
%
%
\addtocontents{toc}{\protect\thispagestyle{empty}\protect\newpage}
\newpart{III}{Interpretation}
{
  Nine interpretations of quantum mechanics
  are discussed.

  The first
  consists of the rules of quantum mechanics as already presented.
  This approach derives its importance from the fact that most people learn
  quantum mechanics in this form; many never learn any other.
  The chapter points out the ambiguities and difficulties inherent in the
  rules.

  The second chapter discusses Bohr's ideas in which the rules
  are rapped up in an interpretation of sorts.
  The third chapter discusses a variant in which the observer-system
  cut in the rules is fixed between mind and body.

  The fourth chapter discusses hidden variables.  This is not a particular
  interpretation but rather defines a programme for finding an interpretation.

  Chapters five and six discuss two variants of a more radical interpretation
  in which there are many realities.

  Chapter seven discusses Bohm's interpretation.

  Finally, chapters eight and nine present two modern interpretations
  based on histories.

  In each chapter, the interpretation is introduced and formulated.
  Attention is then given to modelling the process of measurement within
  the interpretation.  Then, determinism, locality and any relevant
  philosophical issues are discussed.  Finally, the interpretation is
  criticised and variants of the interpretation are noted.
}
\renewcommand{\thechapter}{III.\arabic{chapter}}
\setcounter{chapter}{0}
\mychapter{The Orthodox interpretation} \label{c:qstate}

\mysection{Introduction}

Most physicists think of quantum mechanics in terms of the
rules presented in Chapter~\ref{c:msrmnt}.
These rules, without any further interpretation, can perhaps be called
the {\it orthodox interpretation\/} of quantum mechanics.

\mysection{Formulation}

The formulation of orthodox quantum mechanics is as in
Chapter~\ref{c:msrmnt}.

\mysection{Measurement}

The rules of quantum mechanics are formulated in terms of {\em external\/}
measurements on a quantum system.
measurement may also be modelled {\em within\/} a system as follows.

Suppose a system has eigenstates $|\phi_1\rangle$, $|\phi_2\rangle$,$\ldots$
corresponding to values $1,2,\ldots$ of some observable.
Suppose a measuring apparatus has initial state $|\chi\rangle$
and states $|\chi_1\rangle$, $|\chi_2\rangle$,$\ldots$ corresponding
to the pointer showing values $1,2,\ldots$.  (Such a description
for a macroscopic pointer is almost obscenely simplified but that
does not matter for now.)

Suppose an interaction may be found in which
\[
  |\chi\rangle |\phi_i\rangle
\]
evolves over a very short time into
\[
  |\chi_i\rangle |\phi_i\rangle
  \mbox{ .}
\]

In such circumstances one might say that the apparatus is measuring
the observable.

\subsubsection*{The measurement problem}

The main difficulty with the orthodox interpretation is that
while it stipulates what happens when external measurements are made
on a quantum system, different results are obtained when the same
process of measurement is treated within the theory as above.
This is called the {\it measurement problem}.

Because of this problem, the theory is ambiguous.
One never knows whether to treat an interaction within the theory or
as an external measurement. Or in other words, one does not know what
is an external measurement --- where the system-observer cut should be
placed.

The following example illustrates the different predictions for a
measurement treated as an interaction and for an external measurement.

\begin{example}

In a benign version of the famous story, Schr\"odinger measures the $x$
component of spin of an electron finding it to be $\!\uparrow$.
He then locks
his cat in a box, telling it to measure the $z$ component of spin.
(Surely any cat belonging to the great physicist would
be up to such a task.)

The cat knows that the electron's state is
\[
  (|z\!=\uparrow\rangle + |z\!=\downarrow\rangle)/\sqrt{2}
  \mbox{ .}
\]
It obtains a result of either $\uparrow$ or $\downarrow$ and writes
it down.
It believes that the electrons state has collapsed into
$|z\!=\uparrow\rangle$ or $|z\!=\downarrow\rangle$ respectively.
To an outside observer, the cat would expect these to appear, together
with his own state, as
\begin{equation} \label{e:qmi:cat}
  |\mbox{cat wrote `up'}\rangle|z\!=\uparrow\rangle
  \mbox{ \ or \ }
  |\mbox{cat wrote `down'}\rangle|z\!=\downarrow\rangle
  \mbox{ .}
\end{equation}

Sitting in his office, Schr\"odinger thinks about his cat and
the electron.  He decides to treat them both using quantum theory.
The system's initial state is
\[
  |\mbox{initial cat}\rangle
  \bra |z\!=\uparrow\rangle + |z\!=\downarrow\rangle \ket /\sqrt{2}
\]
and, by linearity, he expects that the state will now have become
\begin{equation} \label{e:qmi:supervisor}
  \bra |\mbox{cat wrote `up'}\rangle|z\!=\uparrow\rangle +
   |\mbox{cat wrote `down'}\rangle|z\!=\downarrow\rangle \ket /\sqrt{2}
  \mbox{ .}
\end{equation}

If Schr\"odinger opens the box and
has a look at the cat's
result and checks the measurement, he will find the two in agreement.
If he repeats the whole process many times, the result will be up or down
in equal proportions.
This he will readily explain in terms of \eqref{e:qmi:supervisor}.
It is also entirely consistent with \eqref{e:qmi:cat}.

But suppose that instead he carefully checks which of the following
four orthogonal states the cat-electron system is in:
\begin{equation} \label{e:qmi:studentestate}
  \begin{array}{rcl}
  \bra |\mbox{`up'}\rangle|z\!=\uparrow\rangle +
   |\mbox{`down'}\rangle|z\!=\downarrow\rangle \ket /\sqrt{2} \\
  \bra |\mbox{`up'}\rangle|z\!=\uparrow\rangle -
   |\mbox{`down'}\rangle|z\!=\downarrow\rangle \ket /\sqrt{2} \\
  \bra |\mbox{`up'}\rangle|z\!=\downarrow\rangle +
   |\mbox{`down'}\rangle|z\!=\uparrow\rangle \ket /\sqrt{2} \\
  \bra |\mbox{`up'}\rangle|z\!=\downarrow\rangle -
   |\mbox{`down'}\rangle|z\!=\uparrow\rangle \ket /\sqrt{2}
  \mbox{ .}
  \end{array}
\end{equation}
He expects to find the state to be the first of these with certainty.

If, on the other hand, the cat has succeeded in collapsing the
wave function as it believes, the two possible states \eqref{e:qmi:cat}
may be written
\begin{equation*}
 \begin{split}
  \half  \bra |\mbox{`up'}\rangle|z\!=\uparrow\rangle +
   |\mbox{`down'}\rangle|z\!=\downarrow\rangle \ket \\
   \pm
  \half \bra |\mbox{``up''}\rangle|z\!=\uparrow\rangle -
   |\mbox{``down''}\rangle|z\!=\downarrow\rangle \ket
 \end{split}
\end{equation*}
and therefore the first two states above will occur equally often.
\end{example}

\subsubsection*{Decoherence}

The problem above has been partly solved by understanding of a process
called {\it decoherence}.
The point is that it is not feasible to isolate the cat in the box
so perfectly from its environment that there will be no record of whether
it wrote `up' or `down'.
There will be some air particle of photon somewhere in states
$|u\rangle$ or $|d\rangle$ correlated with what the cat wrote.
Then instead of the state \eqref{e:qmi:supervisor}, the system as
viewed by Schr\"odinger has state
\begin{equation}
   \bra |u\rangle \, |\mbox{`up'}\rangle \, |z\!=\uparrow\rangle +
   |d\rangle \, |\mbox{`down'}\rangle \, |z\!=\downarrow\rangle \ket /\sqrt{2}
  \mbox{ .}
\end{equation}
Then it is easy to check that, after all, Schr\"odinger should expect
to obtain either of the first two states in \eqref{e:qmi:studentestate}
with equal probability.

The process of decoherence explains why, in practice, there is no
ambiguity in the predictions of quantum theory when the measuring
apparatus is macroscopic.

To sum up, the orthodox interpretation of quantum mechanics is
ambiguous in that it does not determine the observer-system cut.
If there is no such cut, the theory predicts an absurd universe in which
nothing has a state.

In practice, the environment of a macroscopic system, treated within
quantum mechanics, simulates the effect of an outside observer as
defined in quantum mechanics.  This is decoherence.
The effect means that the ambiguity in quantum theory
gets blurred at larger scales.

The measurement problem is devastating.
The orthodox interpretation of quantum mechanics is practically
useful but conceptually unacceptable.

\mysection{Determinism}

The orthodox interpretation accepts a fundamental non-determinism
in the laws of the microscopic world.

However, this non-determinism only occurs when an external measurement
is made.  It is not at all clear what constitutes an external measurement
and therefore not at all clear when non-deterministic transitions occur.

\mysection{Locality}

The orthodox interpretation of quantum mechanics is nonlocal
in the sense that an action at one point in space may have
repercussions at a distant point without any apparent intervening
mechanism.  This well known result is due to
Einstein, Podolsky and Rosen (EPR) \cite{EinsteinPodolskyRosen:35}.

In the best know version of their thought experiment,
due to Bohm \cite[pp.614--623]{Bohm:51},
a particle of spin zero decays into two particles of spin half which
fly off in opposite directions.
In the basis of eigenstate of $z$-component of spin, the two particle
system is in state
\[
  (| \!\uparrow \rangle | \!\downarrow \rangle +
   | \!\downarrow \rangle | \!\uparrow \rangle)
  \mbox{ .}
\]
Neither particle has a definite $z$-component of spin.

But if the $z$-component of spin of the first particle is measured
then the entire two particle system collapses into the state
$| \!\uparrow \rangle | \!\downarrow \rangle$ or into the state
$| \!\downarrow \rangle | \!\uparrow \rangle$.
In other words, the measurement performed on one particle has caused
the other particle, which may be distant, to obtain a definite
$z$-component of spin (anti-correlated with the first).

However, this nonlocality cannot be used to send signals to a distant
point.  The remote observer cannot tell whether his or her measurement
is itself causing the state to collapse or is revealing an already
collapsed value.

This feature of orthodox quantum mechanics
is, of course, related to {\it non-separability\/}: distant particles
cannot in general be assigned separate states.

\mysection{Philosophy}

The orthodox interpretation attempts, as much as possible, to give
an ontological model of physical systems using the concept
of the {\it quantum state}.

However, the interpretation does not fully succeed in giving such a model
because the formalism cannot sensibly be applied to the entire universe.

One solution to this problem is to abandon any attempt at an ontological
model and to put quantum mechanics on a purely epistemological footing.
This is the approach of Bohr and it is discussed in the next chapter.

However, most physicists simply accept the above presentation.
Since the ambiguity is imperceptible in practice, why worry about it?

\mysection{Criticism}

Even if the interpretation works in practice, the conventional interpretation
cannot provide an ontological model for the universe because of the
measurement problem.
This has been the main motivation in the search for a new interpretation.

The measurement problem manifests itself in other ways.
For example, no one has shown how to recover classical physics
from orthodox quantum mechanics in the large-number-of-bound-particles-%
high-temperature-limit.
The most obvious barrier to achieving this is that if a large
object is treated using orthodox quantum mechanics, its wave
function does not collapse and it will not have a well defined position
let alone fully classical behaviour.

Essentially the same problem may be expressed by saying that orthodox
quantum mechanics cannot be applied to closed systems (such as the universe).

At the same time, nonlocality is considered totally unacceptable by
some physicists.  In particular, Einstein, Podolsky and
Rosen \cite{EinsteinPodolskyRosen:35} do not even consider the
possibility that locality might fail.

Others have found non-determinism unacceptable.
Here again, Einstein's view, that God does not play with dice, is
particularly famous.

\mysection{Variations}

Instead of leaving the observer-system split completely open, it is
sometimes stated that anything macroscopic should be treated
as an observer.
This means that whenever a macroscopic object gets into a superposition
of states, its wave function collapses.

Although very popular, this version suffers from multiple vaguenesses.
How big is macroscopic?  A cat?  A catfish?  A cation?
Also, in which basis does the wave function collapse?
And how closely bound do particles have to be to be considered part of one
macroscopic object?

More seriously, this view may be ruled out by
modern evidence for quantum effects in macroscopic objects:
tunneling in Superconducting Quantum Interference Devices
(SQUIDs) \cite{Leggett:80,Leggett:84,Leggett:85,Leggett:86,Leggett:87},
charge density waves \cite{Bardeen:79,Bardeen:80,GrunerEtAl:81}
and the photon field in a ring laser \cite{LettEtAl:81}.

\mysection{Conclusion}

The orthodox interpretation works well in practice but cannot provide
a model for the universe.

\bib

The ideas of quantum states and of operators corresponding to
observables are due to von Neumann \cite{vonNeumann:55} and
Dirac \cite{Dirac:47}.
More up to date texts include
\cite{dEspagnat:71,Scheibe:73,LandauLifschitz:77}.

Decoherence is investigated in
\cite{FeynmanVernon:63,Zurek:81,Zurek:82,CaldeiraLegget:83,JoosZeh:85,%
Zurek:86,Zurek:91} and \cite{GellMannHartle:90b}.
Note that some of these authors argue that decoherence totally
destroys large-scale quantum effects whereas this chapter has followed
\cite{Bell:75,dEspagnat:90} in arguing that decoherence destroys
quantum effects in practice but not in principle.

\mychapter{Bohr's interpretation} \label{c:bohr}

\aquote{Any ontology whatsoever is ruled out by the very nature
of reality as revealed throughout quantum theory.}{Niels Bohr}

\mysection{Introduction}

The orthodox interpretation is orthodox not only in the sense that most
physicists use it but also in the sense that it aims to
emulate classical physics in assigning states to quantum systems.

In the orthodox interpretation a system always has a well-defined state.
It is only the observables which are not always sharply defined.
In other words, the process of measurement cannot be controlled.

There is a less orthodox approach which was developed by Bohr and his
colleagues in Copenhagen.
It is often called the {\it Copenhagen interpretation\/}\footnote
{
  However, this chapter is dedicated specifically to Bohr's views
  while {\it the Copenhagen interpretation\/} is a more general term
  which includes the subtly different approach of Heisenberg and others.
}.
This approach abandons the idea of talking ontologically about
quantum objects.  Instead, it is decided essentially by fiat that
one should only talk about properties of quantum objects by talking
about the measuring apparatus.  The measuring apparatus are assumed
to behave classically and may be described using classical physics and
everyday language.

It is very difficult to get used to this approach.
Bohr's own presentation style is highly idiosyncratic although this
barrier has been removed by the secondary literature.
Even so, Bohr's ideas are worded vaguely and intentionally lack the customary
mathematical formulation.
More importantly, the very idea of introducing a physical theory
which is epistemological --- describing what knowledge can be obtained
about a system rather that describing the nature of the system --- is
extremely unfamiliar in physics.

Nevertheless, for those who are happy with a physical theory describing
experimental results without giving a model for what is going on,
Bohr's approach has a number of advantages.

\mysection{Formulation}

Bohr accepts that calculations should be performed using the quantum
rules of Chapter~\ref{c:msrmnt}.
Bohr's interpretational ideas are centred around two postulates.
Bohr presents these
in different ways in different papers.  The following wording is
taken from \cite{Scheibe:73a} and based closely on Bohr's own wordings.

\subsection*{Two postulates}

\begin{description}
\item[Quantum postulate]
    Every quantum phenomenon has a feature
    of {\it wholeness\/} or {\it individuality\/} which never occurs in
    classical physics and which is symbolized by the Planck quantum of action.
\item[Buffer postulate]
    The description of the apparatus and of the results of observation,
    which forms part of the description of a quantum mechanical phenomenon,
    must be expressed {\em in the concepts of classical physics\/} (including
    those of ``everyday life''), eliminating the Planck constant of action.
\end{description}
(Note that Bohr, who distrusted technical terminology, did not name the
second postulate; this name is due to \cite{Scheibe:73a}.)

An attempt will now be made to make some sense of these.

\paragraph*{The quantum postulate}

In the quantum postulate, Bohr discusses the quantum {\it phenomenon\/} by
which he means a property of a quantum object attached to a particular
experimental apparatus.  Examples of quantum phenomena would
be `an electron {\em in a double slit experiment\/} behaves like a wave'
or `a proton {\em in a photographic emulsion\/} traces a definite path through
space'.

The `wholeness' or `individuality' of the phenomenon is meant to imply
a ban on discussing properties of the object or apparatus in isolation.
Statements such as `a muon always behaves like a wave' or
`a pion always has a definite position', which are merely false in the orthodox
interpretation, are here excluded as meaningless because they
do not involve a description of the quantum apparatus.

This restriction on our freedom of speech applies to experiments which are
``symbolized by the Planck quantum of action''.  This means that it only
applies to experiments in which one is trying to talk about position
and momentum, for example, to an accuracy of order
$\Delta x \Delta p\approx h$.  When talking about macroscopic phenomena, their
is no such wholeness and one is free to talk about the object and the
experiment independently.

\paragraph*{The buffer postulate}

Since one is not allowed to talk about properties of the quantum
object on its own, one had better be allowed to talk about the
apparatus or else physicists would be doomed to silence.
The buffer postulate permits talk about the apparatus using
everyday language and classical physics.

The buffer postulate refers to any item which is being treated as
an apparatus.  There is therefore flexibility in the object--apparatus
divide.  The buffer postulate talks about eliminating the Planck
constant of action meaning that one is only free to treat something
as an apparatus when one is observing it with an accuracy less than
$\Delta x \Delta p\approx h$.

\subsection*{The interaction}

The quantum postulate implies that there is a non-negligible interaction
between the object and apparatus.  Otherwise, the object and apparatus
could hardly be an unanalysable whole.  Note that this contrasts
with classical physics where the interaction involved in an observation
may be made arbitrarily small and ultimately neglected.

Using the buffer postulate, Bohr argues that the interaction is
{\it not separably accountable}.  For example, one cannot determine how much
momentum is transferred from the screen to the electron in a double
slit experiment.  The reason is that the buffer postulate
insists that effects of order $h$ are neglected in treating the apparatus.
Therefore, the treatment of the apparatus is not sensitive enough
to determine the interaction.

In Bohr's style of language, this is expressed by saying that
the interaction is an integral part of
the phenomenon.  It cannot be analysed on its own.

\subsection*{The object}

Since it is impossible to determine the nature of the interaction in
a particular experiment, it is also impossible to assign independent
properties to a specific quantum object.  There is no concept of the
state of an object.

Properties of an object may only be defined through specific apparatus,
ie.\ in the context of a specific {\it phenomenon}.
For example, for a particle travelling in a photographic emulsion, the
position may be {\em defined\/} as the (newer) edge of the track.
In this phenomenon, momentum will be undefined.

\subsection*{Complementarity}

When different phenomena are
\begin{itemize}
\item
  incompatible in the sense that the
  different apparatus cannot be set up simultaneously and
\item
  mutually complete in the sense that both give important information about
  the object which in classical terms would combine to give complete
  information
\end{itemize}
then Bohr calls the phenomena {\it complementary}.

\mysection{Measurement}

As Bohr insists that the interaction cannot be surveyed, measurement
cannot be modelled in this interpretation.

\mysection{Determinism}

\aquote
{The view-point of complementarity allows us indeed to avoid any futile
discussion about an ultimate determinism or indeterminism of
physical events, by offering a straightforward generalisation of
the very idea of causality.}
{Niels Bohr, \cite[p.25]{Bohr:39}}

In other words,
`futile' discussion about ultimate determinism or indeterminism is
avoided by Bohr's ban on looking for an ontological model for quantum
theory.

Bohr regarded determinism as being complementary to the space-time
description.  Ie.\ a particle in an emulsion traces a definite path
in space-time but behaves non-deterministically while a particle
in a vacuum behaves deterministically but does not trace a definite
path in space-time.
Thus, complementarity generalises causality
in that causality becomes just one possible type of behaviour.

\mysection{Locality}

The formulation of the EPR paradox is not possible in Bohr's view
since one cannot talk about the state of an unobserved physical system.
The only type of statement Bohr would allow about a spin zero particle
decaying into particles of
spin half is that detectors placed at a distance will register
opposite spins.  This is a phenomenon which cannot be ``split''.
Ie.\ one may not talk about the intervening state and the paradox
does not arise.

However, there is still a form of nonlocality in that the phenomenon that
cannot be split spreads over two distant sights \cite[\S7.2]{BohmHiley:93}.

\mysection{Philosophy}

Bohr believed that quantum mechanics forces us into an
uncompromising {\it positivist\/} stance, ie.\ recognising
only empirical facts and denying science
a role in determining the reality behind such facts.
This in particular meant that he rejected any attempt to find
an ontological theory of the quantum world.

\mysection{Criticism}

Most physicists are not positivists and are therefore not comfortable
with Bohr's views.

Bohr argued that the unpredictability and uncontrollability of
quantum effects ruled out the possibility of an underlying ontology.
It has already been pointed out \cite[\S2.5]{BohmHiley:93} that
this  view  is not sustainable; chaotic dynamical systems
are unpredictable and uncontrollable but {\em may\/} be explained in terms of
an underlying deterministic ontology.  In any event, several ontological
models for quantum theory {\em have\/} been suggested.

Bohr believed that the measurement problem was purely a result of futile
talk about states and the like.  It could, however, be argued that
Bohr was covering up this very real problem by placing restrictions on
the type of questions one is allowed to ask.


\mysection{Conclusion}

Bohr's view is historically very important.
It is an epistemological interpretation in good agreement with
experiment.
However, Bohr's arguments against an ontology are unconvincing
and his positivism may be covering up ambiguities rather than resolving them.

\bib

Bohr's works on the interpretation of quantum mechanics include
\cite{Bohr:34,Bohr:35,Bohr:58,Bohr:61,Bohr:63}.
Commentaries by authors regarding themselves as Bohr's disciples include
\cite{Rosenfeld:53,Petersen:63,Rosenfeld:63,MeyerAbich:65,Petersen:68}.

The account above was based on \cite{Scheibe:73a}.  Other useful
discussions of Bohr's views include
\cite[Ch.21]{dEspagnat:71},\cite[\S2.2]{BohmHiley:93}.

It is interesting to note that despite the advent of
ontological interpretations of quantum mechanics, some experts still feel that
the epistemological approach is preferable and perhaps even
inevitable \cite{PeresZurek:82}.
\mychapter{Mind causes collapse}

\mysection{Introduction}

Schr\"odinger's equation cannot be accepted on its own
because our minds perceive macroscopic objects as always having definite
states, not linear superpositions of definite states.

It is therefore possible to assume that the unitary mechanics applies
to the entire physical universe and that wave function collapse
occurs at the last possible moment, in the mind itself.
This, of course, assumes a non-physical mind.

This interpretation was hinted at by Von Neumann \cite[\S VI.1]{vonNeumann:55}
and later advocated in
\cite[\S11]{LondonBauer:39}, \cite{Wigner:67}.
It was at one time known as the {\it standard interpretation}.

\mysection{Formulation}

The idea that consciousness causes the wave function to collapse
may be formulated as follows.
\begin{quote}
  The rules of quantum mechanics are correct but
  there is only one system which may be treated with quantum mechanics,
  namely the entire material world.
  There exist external observers which cannot be treated within
  quantum mechanics, namely human (and perhaps animal) minds,
  which perform measurements on the brain causing wave function
  collapse.
\end{quote}

Of course, the state of particles in a persons brain
will be correlated with the
state of particles outside the person's brain so the collapse will
have far reaching consequences.
For example, assuming that Schr\"odinger's cat is not itself conscious,
its fate will be finally decided after Schr\"odinger first takes a look at it,
when the information enters his mind.

This interpretation makes a prediction that is, in principle,
experimentally testable, namely that some particles in the human
brain do not obey Schr\"odinger's equation.
For example, if a person's mind measures say the position of a particle in the
person's brain at time $t$, this will have an effect
which may be observed by measuring the momentum of the same particle
at times $t\M-\epsilon$ and $t\M+\epsilon$.

\mysection{Measurement}

In this interpretation, the only true measurement is the mind measuring
the brain.

\mysection{Determinism}

This interpretation accepts that the universe is inherently non-deterministic.

\mysection{Locality}

This interpretation accepts that the universe is nonlocal;
when the mind measures the brain it causes wave-function collapse
which may have consequences far beyond the brain.
In the EPR situation, for example, a conscious observation of one
particle's $z$-component of spin causes the distant particle to
obtain a definite $z$-component of spin.

\mysection{Philosophy}

This interpretation depends on a particular ontological view of the
mind-body question.  Many physicists have criticised the interpretation
because it does not accord with their own understanding of the mind-body
question without in any way making clear that this is the reason.

Therefore, in order to put the debate about this interpretation
into its correct philosophical context, a brief
description of the main philosophical approaches to the mind is
required here.

\begin{figure}[t]
%
\epsfbox{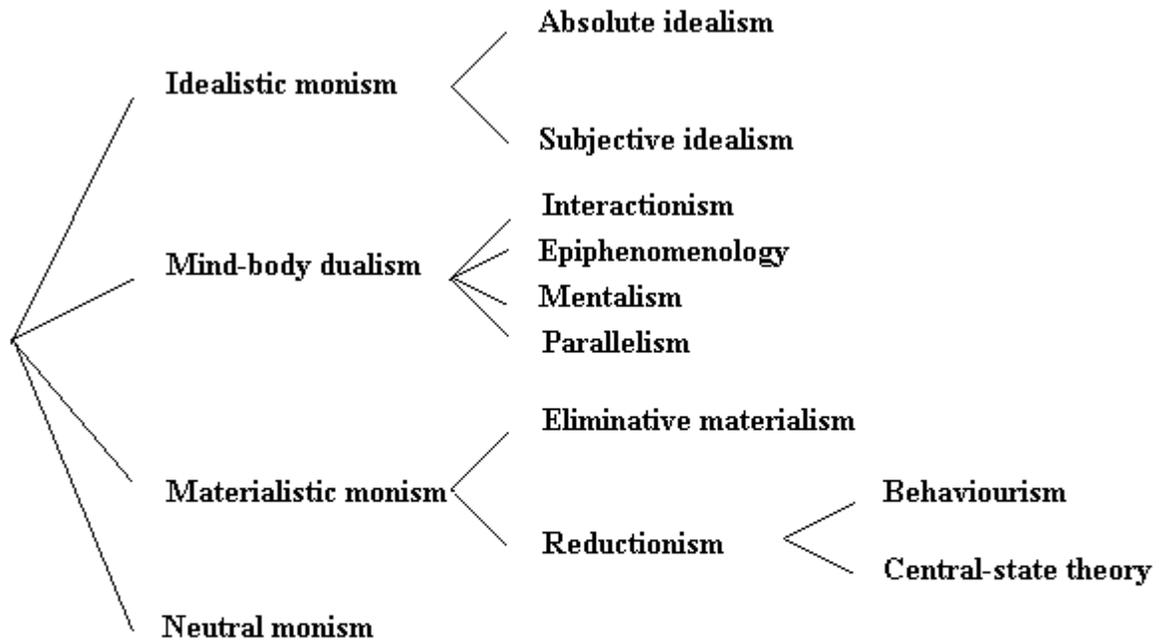}
\caption{Some philosophies of the mind}
\end{figure}
\subsubsection*{Idealism}

Russell uses the term {\it idealism\/}
for ``{\it the doctrine that whatever exists,
or at any rate whatever can be known to exist,
must be in some sense mental.}''.
It is a monistic theory as it recognises only one type of entity.
It may be traced back to the Irish bishop George Berkeley
\cite{Berkeley:1710,Berkeley:1713}
This belief has been very popular with philosophers over the ages.

The main argument against idealism is that elements of the physical
world evolve with time in a way which is often predictable and which
does not depend on whether or not I am looking at them.
Even if this does not {\it prove\/} that these items have an independent
existence, it makes it a lot {\it simpler\/} to assume that they do.
And there is, after all, no evidence whatsoever that they do not.

(One interpretation of quantum mechanics, the many-minds interpretation
[Chapter~\ref{c:mmi}] is arguably idealistic.)

\subsubsection*{Materialism}

At the other extreme is materialism, another monistic theory,
the doctrine that the physical universe is all that exists.
It may be traced back to the early Greeks.

How does this doctrine explain sensations and emotions?
In one version it is believed that these words should be {\it eliminated\/}
altogether from the philosopher's vocabulary, at least for the purpose of
describing facts.
In the other version, {\it reductionism},
sensations and emotions may be explained as complex physical
phenomena.
That is, the study of the mind can in principle be reduced to the study
of chemistry and physics.
They may either be explained as nothing more that types of behaviour
({\it behaviourism\/}) or as particular states of neurons in the brain
({\it central-state theory\/} or the {\it identity thesis\/}).

Materialism gains credence from recent progress in understanding the brain
(eg.\ \cite{Konishi:93,LeDoux:94}), work which has not found any need for
a mind.

Materialism is very much in vogue at the moment following the proclamation
by Francis Crick and Chrostof Koch (the former of double-helix fame)
in 1990 that the time is ripe for science to tackle consciousness
(see \cite{Crick:94}).  This has helped to inspire ``an intellectual
stampede'' of `scientific research' into consciousness
\cite{Horgan:94}.

This doctrine, by definition, precludes consciousness from having a
privileged role in quantum mechanics.

\subsubsection*{Dualism}

{\it Mind-body dualism},
or {\it psychoneural dualism},
is the doctrine that the mind
or consciousness exists as an entity separate to the physical universe.
In this model, there is an objective physical universe and, in addition,
associated with humans and perhaps some other beings is an extra-physical
mind which observes (and possibly influences) the universe.
This view may be traced back to Plato although it is
Rene Descartes \cite{Descartes:1642} who gave the best known formulation.

Advocates of this view may accept that the behaviour of a fellow human can
be understood without assuming that they have an extra-physical mind.
But they feel that their {\it own\/} sensations must be extra-physical.
After all, one could imagine a human-body obeying the law of physics
without having any sensations.  The advocate of dualism believes
that he or she must comprise more than ordinary matter and,
for philosophical or emotional
reasons, believes that other humans, and perhaps animals,
must be conscious in the same way.

Do the body and mind interact?
Descartes thought that they do.
The mind observes the world but also influences it; for example,
one makes a conscious decision to move one's arm.
This is called {\it interactionism}.

Others have suggested that the decision to move an arm is made by the
physical brain and the mind can do nothing but observe.
This is {\it epiphenomenology}.
Still others have suggested that the mind can affect the body but is
not affected by it.
This is {\it mentalism}.
Finally, in a view that goes back to Leibniz, it has been suggested
that the mind and brain do not interact at all but work in parallel
maintaining
corresponding states, rather like synchronised clocks.
This view is called {\it parallelism\/} or, more fully,
{\it psycho-physical parallelism}.

\subsubsection*{Neutral monism}

There is one more form of monism which warrants a mention,
{\it neutral monism}.
In this doctrine there is only one type of entity which is neither
purely physical nor purely mental.
This doctrine was first proposed by Benedict de Spinoza \cite{Spinoza:1677}
and various other forms have been proposed since.

\subsubsection*{Science and philosophy}

Scientists may, for the most part, ignore the multitude of theories
of the mind.  Their job is to model the physical world and they have
been able to do so, up to know, without assuming any kind of mind
whatsoever.
Almost without exception, scientists believe, perhaps implicitly,
that their success in modelling the physical world without reference
to the mind, vindicates the idea that the physical world exists in its
own right.
In terms of philosophical models, this means that scientists tend to
be materialists or dualists.

Many scientists believe that it is possible to model the physical
universe completely without reference to the mind, ie.\ they believe that
the mind does not affect the physical world and so
reject any form of interactionism or mentalism.

In fact, many scientists are materialists, ie.\ they reject
the existence of anything outside the physical universe.
(In a cynical assessment of this, Wigner \cite{Wigner:67} writes
``The reason is probably that it as emotional necessity to exalt
the problem to which one wants to devote a lifetime.''.)

In this context, it is easy to see why the present interpretation,
which stipulated a form of interactionism\footnote
{
  Contrary to this, von Neumann \cite[Ch.VI]{vonNeumann:55},
  who apparently invented the ``standard'' interpretation,
  argues that it is consistent with psycho-physical parallelism.
  But in fact he merely shows that the effects of the mind making
  quantum measurements on the brain could not be observed outside the body.
  However, true psycho-physical parallelism demands that these effects
  be undetectable even to an observer with perfect access to the brain.
  It seems that von Neumann used the term {\it parallelism\/} not in the
  strict philosophical sense but rather to describe the phenomenon that
  in practice the effects of someone else's consciousness are not apparent.
},
is often rejected out of hand.

\mysection{Criticism}

One important difficulty with this interpretation
is that it assumes a very specific model of the mind, a
dualistic interactionist model.
However, even if this model is accepted, the interpretation is
problematic.

Firstly, it is often criticised for giving
predictions which are not clear-cut in that they depend on which animals
are conscious\footnote
{
  Theories vary from solipsism, the doctrine that only Zvi Schreiber
  is conscious, to views ascribing consciousness to all kinds of animals,
  plants and computers.
}
and on the precise nature of the measurements performed by the mind.
However, this criticism is unscathing as
these open questions could, in principle, be settled by  experiment.
It is true that even with the arsenal of modern invasive and non-invasive
techniques for observing the brain, the above experiments will not be
feasible for the foreseeable future.  But the interpretation is a good theory
in that it has testable consequences and in that it prescribes a
scientific programme for filling in the unspecified parameters.

A more serious criticism is hinted at by Bohm and Hiley
\cite[\S 2.4]{BohmHiley:93}.
They write that ``it is difficult to believe that the evolution of the
universe before the appearance of human beings depended fundamentally on
the human mind''.
This criticism can be expanded as follows.
Assuming that the only minds belong to humans and to certain animals,
the universe in this interpretation would initially undergo
no wave function collapse.  If that were true, the universe's wave function
would become a linear superposition of many different possibilities
and human beings or animals would not come into existence at any well-defined
moment.  It is therefore difficult to see at what point minds could
start to observe the universe.

Bohm and Hiley write ``Of course one could avoid this difficulty
by assuming a universal mind.  But if we know little about the human mind,
we know a great deal less about a universal mind.  Such an
assumption replaces one mystery by an even greater one.''.
Indeed, if the interpretation were correct and if
there were minds observing points outside human and animal brains,
we would not know were
to start looking for the quantum effects they must be causing!

\mysection{Variations}

Some have suggested that the mind not only observers the universe causing
a random collapse of the wave function but that rather it {\em chooses\/}
between different quantum alternatives
\cite{Margenau:67,Margenau:68,Walker:74,WalkerHerbert:77,Margenau:84}
This is seen as a way of attributing free will to the mind.

In the normal version of the interpretation,
it was shown that the effects of the mind
could be detected by sensitively analysing the brain.
But if the mind can actually choose between alternatives, its effects
could be detected outside the body.  (Unless the mind is in some way
forced to obey Copenhagen statistics in the long run.)

For example, if Schr\"odinger had many cats each sharing a box with a bomb
in the normal way then when he observed the cats his brain would
be in a linear superposition of seeing them alive and seeing them dead
and his mind could choose to make them all alive thereby causing
a blatant deviation from the predictions of quantum theory.
(There have been claims that the mind can
be seen to exert just such an influence on macroscopic objects in the lab
\cite{Forwald:69}!)

\mysection{Conclusion}

The interpretation ties down the Copenhagen model by saying
that the only true measuring apparatus are non-physical minds.
Unlike the previous two interpretations, it is ontological and
is not inherently ambiguous.

For those who accept the philosophical arguments for dualism and are
not scared of a little interactionism, this interpretation seems
attractive.
Even so, it does not allow sensible discussion of the big-bang and
evolution.

Scientists must hope for a different interpretation of quantum mechanics.

\bib

This idea is advocated by \cite{vonNeumann:55} (especially \S VI.1),
\cite{LondonBauer:39} (especially \S11), \cite{Wigner:67}
and criticised all over the place.

An excellent concise overview of the various philosophical
approaches to the mind-body question may be found in \cite{Shaffer:76}.
A well known known philosophical discussion of the relevant issues is
\cite{Russell:12}.
A classification of the various views may be found in
\cite{Bunge:77}.
\mychapter{Hidden variables}

\aquote{I$\ldots$ believe$\ldots$ that the flight into statistics
is to be regarded only as a temporary expedient that bypasses
the fundamentals.}{Albert Einstein, \cite{Einstein:38}}

\mysection{Introduction}

The statistical results of quantum mechanics seem at first sight
reminiscent of the non-deterministic results of statistical mechanics.
The latter result from underlying laws which are deterministic.
Could the same be true of quantum mechanics?
Could quantum mechanics be incomplete?

The hidden variables programme suggests that in the complete theory,
every quantum system is always in a state in which every observable
has a definite value.
The predictions of quantum mechanics reflect
the fact that there is no mechanism for completely observing or controlling
every aspect of a system's state.
Ie.\ there is no way to {\em prepare\/} an ensemble of systems in which
every system has the same position and the same momentum even though such
an ensemble may exist.
In fact, the maximal knowledge we can ever obtain may be summed up by a vector
in Hilbert space.
It is precisely for this reason that the specification of the state beyond
the state vector is called {\it hidden}.

There exist no hidden variables theories at least in the broad sense
described above.  Work on hidden variables has in fact often concentrated
on proving that such theories cannot exist.
This chapter therefore does not describe a theory.  Instead, it
outlines what would constitute a hidden variables theory and discusses
the proofs that no such theories exist.

\mysection{Formulation}

The following might be given as the definition of a completion of a quantum
mechanical system.

%

\begin{defn}[Completion]  \label{d:qmcompletion}
  Let a quantum mechanical system have states given by the Hilbert space $\FH$
and have
  Hamiltonian $\hat{H}$.
  A {\bf completion} of this system is a tuple $(\Omega,U,V,\mu)$ where
  \begin{itemize}
  \item
    $\Omega$ is a set (the hidden states)
  \item
    for every state $s\Min \Omega$ and densely define self-adjoint
    operator\footnote
    {
      $V$ could be defined only for those operators which belong
      to physical observables.  However, historically it was assumed
      that {\em all\/} densely defined self-adjoint operators belong
      to physical observables.  Without this assumption much of
      the work on impossibility proofs makes little sense.
    }
    $\hat{Q}$,
    $V(s,\hat{Q})$ is a value in the spectrum of $\hat{Q}$ (so in each state
    each physical quantity has a definite value --- this is often expressed
    by saying that the states in $\Omega$ are  `dispersion free')
  \item
    for every $t\M>0$, $U_t:\Omega\rightarrow \Omega$ is a map
    which has a measurable inverse
    (giving the time evolution over a time interval of length $t$)
    where the $\sigma$-field of subsets of $\Omega$ generated by the subsets
    \[
      \cbra s\Min\Omega \:|\: V(s,\hat{Q})\Min (a_1,a_2] \cket
    \]
    are called measurable
  \item
    for every $|\psi\rangle\Min\FH$, $\mu_{|\psi\rangle}$ is a probability
    measure on $\Omega$ defined on some $\sigma$-field including
    the sets called measurable above (giving the ratios of systems
    with different states in $\Omega$ in an ensemble described by
    $|\psi\rangle$)
  \end{itemize}
  such that
  \begin{Itemize}
  \item[{\bf C.1}]
    the values of observables are consistent with relationships between
    the observables
    \[
      V(s,f(\hat{Q})) = f(V(s,\hat{Q}))
      \quad (\forall s\Min\Omega, \hat{Q},
      f\M:\R\rightarrow\R \mbox{ Borel measurable})
    \]
  \item[{\bf C.2}]
    the ensemble defined by $\mu_{|\psi\rangle}$ gives Copenhagen statistics:
    \[
      \mu_{|\psi\rangle} \bra\{s\Min\Omega \:|\: V(s,\hat{Q})\Min B\} \ket
      =
      \Pr \bra Q\Min B \,;\,|\psi\rangle \, \ket
      \quad (\forall Q,\psi,B \mbox{ Borel})
    \]
  \item[{\bf C.3}]
    the composure of the ensemble associated with a quantum system
    does not change
    with time (although its constituent systems do, of course, evolve with
    time)
    \[
      \mu_{e^{-i\hat{H}t/\hbar}|\psi\rangle}(U_t(K))
      =
      \mu_{|\psi\rangle}(K)
      \quad (\forall t\M>0,\psi,K\subseteq\Omega \mbox{ measurable})
    \]
  \item[{\bf C.4}]
    it is consistent to assume that the wave function collapses
    \[
      \frac
      {\mu_{|\psi\rangle}
        \bra K \cap \{s\Min\Omega \:|\: V(s,\hat{Q})\Min B\} \ket}
      {\mu_{|\psi\rangle}
        \bra \{s\Min\Omega \:|\: V(s,\hat{Q})\Min B\} \ket}
      =
      \mu_{\hat{P}|\psi\rangle}(K)
      \quad (\forall K\subseteq\Omega,Q,\psi,B \mbox{ Borel})
    \]
    where $\hat{P}$ projects onto eigenvectors of $\hat{Q}$ with
    eigenvalues in $B$
    (this represents the case when $Q$ is measured in an ensemble in state
    $|\psi\rangle$ and the systems with result in $B$ are selected).
  \end{Itemize}
\end{defn}

A completion in the above sense would provide a deterministic
theory in which all the predictions of quantum mechanics hold.

In fact, the programme of completing quantum mechanics runs into difficulties
long before the details of ensembles of dispersion free state are
considered.
The debate about hidden variables has centred on the feasibility
of finding {\em any\/} dispersion free states $s$ with sensible associated
values $V(s,\hat{P})$.
In order to concentrate on this problem, a dispersion free state
may be viewed more abstractly as a map from operators to $\R$.

\begin{defn}[Dispersion free state] \label{d:qmi:dsprnfree}
  A dispersion free state is a map $V$ from self-adjoint operators to $\R$
  which maps every observable $P$ to a value in the spectrum of
  $\hat{P}$ such that
  \begin{Itemize}
  \item[{\bf C.1'}]
    the values of observables are consistent with relationships between
    observables\footnote
    {
      Again, notice the perhaps unreasonable assumption that
      $V$ is defined for
      all densely defined self-adjoint operators.
    }
    \[
      V(f(\hat{Q})) = f(V(\hat{Q}))
      \quad (\forall \hat{Q}, f\M:\R\rightarrow\R)
      \mbox{ .}
    \]
  \end{Itemize}
\end{defn}

The trouble is that no such states exist.  This is shown
under {\it Criticism}.

\mysection{Measurement}

It is not immediately apparent whether a hidden variables theory
would facilitate the modelling of measurement within the theory.
After all, in the above formulation the world is still split into
system and observer.  Whether the observed can be brought within the
system more successfully that in ordinary quantum mechanics is an open
question.

There is, however, room for optimism on this point.
In quantum mechanics one could not describe the whole universe
within the formalism because doing so led to the strange result that
nothing ever has a definite state.
On the other hand with hidden variables, the hidden variables may
provide this definiteness.
Just such an approach is taken in Bohm's interpretation,
Chapter~\ref{c:Bohm}.

\mysection{Determinism}

A hidden variables theory is completely deterministic.
It is of course this which motivates the programme.

\mysection{Locality}

One of the original motivations of hidden variables theories was to eliminate
nonlocality from quantum theory.
If the spin in an EPR experiment is determined by hidden variables,
each particle
will have a definite $z$-component of spin all along and measuring the spin
of one merely reveals the spin of the other without influencing it.

However, a famous paper by Bell established that {\it nonlocality
is implied not only by the Hilbert space formalism but by the
statistical predictions of quantum mechanics}.
Any hidden variables theory which reproduces the statistical predictions
of quantum mechanics would necessarily be nonlocal.
A far simpler version of Bell's argument is reproduced
here from \cite{Mermin:90a,Mermin:90b,Mermin:90c}.

\subsection*{Proof of nonlocality}

Consider three particles numbered $1$, $2$ and $3$ of spin $\half$.
Write eg.\ ${1x}$ for the $x$
component of spin of particle $1$ and $\hat{\sigma}_{1x}$ for the
associated operator.
These satisfy the commutation relations
\[
  [\hat{\sigma}_{nx},\hat{\sigma}_{ny}]=i\hat{\sigma}_{nz},
  \quad
  [\hat{\sigma}_{ny},\hat{\sigma}_{nz}]=i\hat{\sigma}_{nx},
  \quad
  [\hat{\sigma}_{nz},\hat{\sigma}_{nx}]=i\hat{\sigma}_{ny}
\]
with other commutators vanishing.

It is an easy matter to check that the three operators
$\hat{\sigma}_{1x}\hat{\sigma}_{2y}\hat{\sigma}_{3y}$,
$\hat{\sigma}_{1y}\hat{\sigma}_{2x}\hat{\sigma}_{3y}$
and $\hat{\sigma}_{1y}\hat{\sigma}_{2y}\hat{\sigma}_{3x}$
commute.
Suppose the particles are in a common eigenstate of these with
each having value $1$.

The product of the three operators is
$-\hat{\sigma}_{1x}\hat{\sigma}_{2x}\hat{\sigma}_{3x}$ and so the
system will also be in an eigenstate of this operator with eigenvalue
$1$.

But now assume hidden variables and locality.
This means that each particular spin component has a definite value
(which is revealed by the measurement) and that this value, for
a particular particle,
cannot depend on which component of spin is measured for a different
particle.
Let $m_{1x}$ represent
the hidden value of $1x$, etc.

Now in this particular state it must be that
$m_{1x}m_{2y}m_{3y}\M=1$ etc.\footnote
{
  To see this, suppose $1x$, $2y$ and $3y$ were measured giving results
  $m_{1x}$, $m_{2y}$ and $m_{3y}$ respectively.
  Since $\hat{\sigma}_{1x}$, $\hat{\sigma}_{2y}$ and $\hat{\sigma}_{3y}$
  each commute with
  $\hat{\sigma}_{1x}\hat{\sigma}_{2y}\hat{\sigma}_{3y}$,
  at the end of the measurement the system is still in an eigenstate of
  $\hat{\sigma}_{1x}\hat{\sigma}_{2y}\hat{\sigma}_{3y}$ with
  eigenvalue $1$.
  But now the system is also in an eigenstate of {\em each\/} of
  $\hat{\sigma}_{1x}$, $\hat{\sigma}_{2y}$ and $\hat{\sigma}_{3y}$
  with eigenvalues $m_{1x}$, $m_{2y}$ and $m_{3y}$ respectively
  and it follows that $m_{1x}m_{2y}m_{3y}\M=1$.
}
Therefore
\[
  1
  =
  m_{1x}^2 m_{2x}^2 m_{3x}^2
  =
  (m_{1x}m_{2y}m_{3y})(m_{1y}m_{2x}m_{3y})
  (m_{1y}m_{2y}m_{3x})(m_{1x}m_{2x}m_{3x})
  =
  -1
\]
giving the required contradiction.

To make the implications clear, suppose that a triplet of particles
in the above state is repeatedly sent out with each particle going
in a different direction.  Suppose that at three distant sights sit
three observers and each of them on each occasion measures some component of
spin.  They later meet and compare results.

Now suppose that the observers assume that on each occasion
the $m$ values are well defined, ie.\ the result of their experiment
was predetermined and depended only on their own experimental set up,
as is the case in hidden variable theories.
Looking back on their data they will all agree that an {\it invariant\/}
property of these $m$ values is that
\[
  m_{1x}m_{2y}m_{3y} = m_{1y}m_{2x}m_{3y} = m_{1y}m_{2y}m_{3x}
  = -m_{1x}m_{2x}m_{3x} = 1
\]
since each of these identities held true in every experiment which happened
to test it.  But the observers then find that no six integers
satisfy these equations!

The result of each individual experiment cannot be predetermined.
If one wants to insist that the world is deterministic, the only was
is to assume that the result of the three experiments is predetermined
only in terms of the experimental apparatus of all three observers.
In other words, the result of one experiment depends nonlocally on the
apparatus set up at the other sights.

In conclusion, from the {\em predictions\/} of quantum mechanics
(as opposed to the formalism) it follows that if the world is
deterministic it is nonlocal.

\mysection{Philosophy}

It is a little futile to discuss in any detail the philosophy
behind a non-existent theory.
Suffice to reword the discussion  under {\it Measurement\/} and to say
that although one might initially be looking for an epistemological
hidden variables theory, one might expect such a theory to easily
yield an underlying ontological theory.

\mysection{Criticism}

\aquote{What is proved by impossibility proofs is lack of imagination.}
{A `suspicion' aired by John S.\ Bell, \cite{Bell:82}}

Consider a quantum mechanical system with states given by the
Hilbert space $\FH$ and with Hamiltonian $\hat{H}$.

Do there exist dispersion free states?
\subsubsection*{The proof of von Neumann}

This theorem is mainly of historical interest.
As will be seen, Gleason's theorem leads to a much  more general result.

\begin{thm}
  There are no dispersion-free states
  with the following additional property.
  \begin{Itemize}
  \item[{\bf C.1+}]
    The values of (even non-commutative) operators are additive.
    \[
      V(\hat{P}\M+\hat{Q}) = V(\hat{P}) + V(\hat{Q})
      \quad (\forall s\Min\Omega,\hat{P},\hat{Q})
      \mbox{ .}
    \]
  \end{Itemize}
\proofend
\end{thm}

The imposition of {\bf C.1+} is dubious.
If $\hat{P}$ and $\hat{Q}$ do not commute then
measuring $P$, $Q$ and $P\M+Q$ involves three
totally different experiments.  $P\M+Q$ is the sum of $P$ and $Q$
only in the formal mathematical sense that the operator associated
with $P\M+Q$ is $\hat{P}\M+\hat{Q}$.

Therefore, although it happens to be a fact that
\[
  E(\hat{P}+\hat{Q}) = E(\hat{P}) + E(\hat{Q})
\]
for the expectation values in quantum mechanics,
there is no reason to expect this to hold for arbitrary states
in some completion of quantum mechanics (although it must hold
for quantum ensembles).
The postulate described by von Neumann as `very general and plausible'
is `absurd' \cite{Bell:82}!

This point may be emphasised by constructing a time-independent completion
for a two-dimensional system, eg.\ for a particle of spin $\half$
(\cite[\S 2]{Bell:66}, \cite[\S VI]{KochenSpecker:67}) in which,
of course, {\bf C.1+} is violated.

\subsubsection*{The proof of Jauch and Piron}

The same conclusion is proved by
Jauch and Piron \cite{JauchPiron:63}, starting from the following
weaker premises.
\begin{Itemize}
\item[{\bf C.1+'}]
  If $\hat{P}_V$ and $\hat{P}_W$ are projections and \
  $V(\hat{P}_V)=V(\hat{P}_W)=1$ \ then \
  $V(\hat{P}_{V\cap W})\M=1$.
\end{Itemize}
It is easy to see that this follows from {\bf C.1+}.

Bell \cite{Bell:66} however argues that the property in {\bf C.1+}
is once again peculiar to quantum mechanics and that
there is no reason to expect this to hold in some new theory.
It is true that every element in an ensemble with state in $V\cap W$
will have to have this property.
But there may be other hidden states which do not appear in such
ensembles and which violate this property.

\subsubsection*{Implications of Gleason's theorem}

\begin{thm}
  Let $\FH$ be a separable Hilbert space of dimension more than two.
  Then there are no dispersion free states for $\FH$.
\end{thm}

\begin{proof}
Suppose $V$ were a dispersion free state.
Define a map $\mu$ from the lattice of closed linear subspaces of $\FH$
to $\R^+$  by letting $\mu(U)\M=V(\hat{P}_U)$
(in fact $\mu$ must take the values $0$ or $1$).
It will be shown that this map satisfies the conditions of
Gleason's theorem (Theorem~\ref{t:qmi:Gleason}).

Let $U_1,\ldots,U_n$ be orthogonal linear subspaces of $\FH$.
Let
\[
  \hat{A} = \sum_{i=1}^n \hat{P}_{U_i} = \hat{P}_{\sum U_i}
  \mbox{ .}
\]
Now $\hat{A}$ and the $\hat{P}_{U_i}$ all commute so,
by a well-known theorem,
each is a function of some common
operator $\hat{Q}$.  {\bf C.1'} will therefore imply that
\[
  V(\hat{A}) = V(\hat{P}_{\sum U_i}) = \sum_i V(\hat{P}_{U_i})
  \mbox{ .}
\]
Thus, $\mu(\sum U_i)\M=\sum\mu(U_i)$.

Then Gleason's theorem implies
that there is some density matrix such that
\[
  \mu(U) = \Tr(\hat{\rho}\hat{P}_U)
\]
for all $U$.
This in turn implies that the state is not dispersion free after
all because there will always be some assertions which are neither
true or false (eg.\ either a projector tying down position or one
tying down momentum).
\end{proof}

Is this the final nail in the coffin of hidden variables?
Bell \cite{Bell:66} argues that even the condition {\bf C.1'}
(Definition~\ref{d:qmi:dsprnfree}) is
unreasonable and that hidden variables which do not have property
{\bf C.1'} are therefore feasible.
He writes as follows.
\begin{quote}
\it
  The result of an observation may reasonably depend not only on the
  state of the system (including hidden variables) but also on the
  complete disposition of the apparatus.
\end{quote}

As an example, suppose $\hat{A}=f(\hat{P})$ and $\hat{A}=g(\hat{Q})$.
Then it so happens that in quantum mechanics one can measure $A$
by measuring $P$ and applying $f$ or by measuring $Q$
and applying $g$.  In each case one is obtaining different side
information; the first type of measurement reveals the value of any
$\hat{B}=f'(\hat{P})$ while the second reveals the value of any
$\hat{B}=g'(\hat{Q})$. However, since in quantum mechanics the
statistical predictions for $A$ are the same in both kind of
measurement, one becomes accustomed to thinking of $A$ as a physical
observable and of the two experiments as different ways of measuring $A$.

But perhaps deep down there is no physical observable called $A$ and
the two experiments are revealing different properties which happen
to have the same statistical distributions in the ensembles allowed by
quantum theory!

Thus, the whole definition of completions and dispersion free states
has been coloured by some deep rooted but unfounded assumptions about
what constitutes a physical observable.

So what Gleason's theorem tells us is that in any hidden variables
theory, some quantities which appeared as coherent physical observables in
quantum theory will manifest themselves as a group of
distinct
observables, which correspond to different experimental set ups\footnote
{
  At first sight, even this conclusion is avoidable if we reject the
  idea that all densely defined self-adjoint operators correspond to
  observables.  Perhaps for those particular operators that do
  correspond to observables, condition {\bf C.1'} can be satisfied.
  However, see the comments below about Bell's theorem.
}.
Such observables are called {\it context dependent observables\/} although this
slightly obscures the point that they are {\em not really observables}.
As Bell \cite{Bell:82} points out, all of this is in accord
with Bohr's philosophy that the properties of a system are inextricably
tied up with the set-up of the experiment.

So it turns out that the proofs against hidden variables are
disappointing.  They only tell us that some quantities which seem
like observables in quantum mechanics
will turn out to depend on the experimental set up
in a hidden variables theory.  And this is not new!  Bell's theorem
has already established that some results will have to depend on the
experimental set up, in fact, on the experimental set up at remote sights!


\bib

Von Neumann's proof is contained in \cite[\S IV.1--2]{vonNeumann:55}.
It is criticised in eg.\ \cite[pp.99--104]{Hermann:35},
\cite{Bell:66,Bell:71}.

The implications of Gleason's theorem were pointed out by Jauch.
The relevant corollary of Gleason's theorem is established using elementary
arguments in \cite{Bell:66}.
Kochen and Specker \cite{KochenSpecker:67} proved the same conclusion
without starting from Gleason's theorem.
A simpler proof, for the case of two particles of spin $\half$
was given by Peres \cite{Peres:90}.
A proof for more than two particles is given in
\cite{GreenbergHorneShimonyZeilinger:90}.
These proofs are summarised concisely in \cite{Mermin:90c}.

A concise account of the proofs of von Neumann, Jauch and Piron and of the
consequences of Gleason's theorem is given in \cite{Bell:66}.
A more general survey is \cite{Belinfante:73}.

Bell's theorem is presented in \cite{Bell:64} and again, in a slightly
more general form, in \cite{Bell:71}.
The theorem establishes that the {\em statistical\/} correlations in a certain
experiment on two particles of spin one half are inconsistent with
hidden variables and locality.
The relevant predictions of quantum mechanics have been checked
experimentally \cite{AspectGrangierRoger:81,AspectDalibardRoger:82}.
The same result may be established by considering angular momentum
\cite{Mermin:80}.

In an alternative experimental arrangement with three particles of
spin one half, there are {\em perfect\/}
correlations which are inconsistent with hidden variables and locality
\cite{GreenbergHorneZeilinger:89,GreenbergHorneShimonyZeilinger:90}.
The proof presented above is a simplified version of this
due to Mermin \cite{Mermin:90a,Mermin:90b,Mermin:90c}.
The implications of these results are discussed carefully in \cite{Stapp:93}.
The same result may be achieved considering two particles \cite{Hardy:92}
or one particle \cite{Czchor:94}.

Einstein was the best know advocate of hidden variables and a nice
book on his views is \cite{Fine:86}.
\mychapter{Many-worlds interpretation} \label{c:mwi}

\aquote{Every quantum transition taking place on every star, in every
galaxy, in every remote corner of the universe is splitting our
local world on earth into myriads of copies of itself.}
{B.\ S.\ DeWitt, \cite{DeWitt:71}}

\mysection{Introduction}

At the heart of the measurement problem is the contrast between the
coherent state \eqref{e:qmi:supervisor}:
\begin{equation*}
  (|\mbox{``up''}\rangle|\uparrow\rangle +
   |\mbox{``down''}\rangle|\downarrow\rangle)/\sqrt{2}
  \mbox{ .}
\end{equation*}
predicted by Schr\"odinger's equation, on the one hand, and the two
possible outcomes \eqref{e:qmi:cat}:
\[
  |\mbox{``up''}\rangle|\uparrow\rangle
  \mbox{ \ or \ }
  |\mbox{``down''}\rangle|\downarrow\rangle
\]
predicted by the wave function collapse, on the other.

The many-worlds interpretation suggests that these are one and the same!

If it were possible to imagine that the two summands
$|\mbox{``up''}\rangle|\uparrow\rangle$
and
$|\mbox{``down''}\rangle|\downarrow\rangle$
represents equally real but separate realities then the wave function
collapse would actually follow from the Schr\"odinger equation!

In the many-worlds interpretation, this is achieved by the bold suggestion
that the summands in the coherent state represent separate worlds,
existing in parallel and equally real, one in state
$|\mbox{``up''}\rangle|\uparrow\rangle$
and the other in state
$|\mbox{``down''}\rangle|\downarrow\rangle$.

The idea of decomposing the state into orthogonal summands
to show that the unitary mechanics
and wave function collapse are one and the same is highly seductive.
Here, an extreme view will be taken: {\em this observation is entirely
irrelevant to the interpretation of quantum mechanics}.

Why? The measurement problem arises because in studying {\em our world\/}
it is found that both the Schr\"odinger equation and wave function collapse
are needed.  The challenge is therefore to find a way to understand
both the Schr\"odinger equation and the wave function collapse
within {\em our world}.  There is no virtue in finding a way to apply
the Schr\"odinger equation
elsewhere, ie.\ in some collection of worlds.

The many-worlds interpretation  has always been presented in a vague way
along with exaggerated claims (such as ``{\it The mathematical formalism
of the quantum theory is capable of yielding its own interpretation\/}''
\cite{DeWitt:71}) and overlooked technical problems.
Nonetheless, it has received broad attention and is therefore
considered here.

\mysection{Formulation}

\subsubsection*{Rules}

The universe may be associated with a
separable Hilbert space $\FH$;  a densely defined self-adjoint
operator $\hat{H}$
on $\FH$ called the Hamiltonian; and an element $|\psi\rangle\Min\FH$
called the initial state,
in such a way that the following hold.
\begin{itemize}
\item
  At any time,
  the universe comprises a countable number of worlds each of which
  has a space-time like our own world
  and each of which may be described by a state in
  $\FH$.  The worlds have a branching structure\footnote
  {
    Formally, one might say that there are countably many branches
    $b_i(t)\Min\FH$.  Then for any two branches $b_i$ and $b_j$,
    $t_{ij}$ could give the time at which these two branches split
    from each other.
    One would then impose that $t_{ik}$ cannot precede both
    $t_{ij}$ and $t_{jk}$ (a form of transitivity) and that
    $b_i(t)\M=b_j(t)$ for $t\M\leq t_{ij}$.
  },
  ie.\ sometimes one world splits into countably many worlds.
\item
  Parallel worlds at one time have orthogonal states.
\item
  The total state of any subset of the branches
  evolves according to Schr\"odinger's equation
  with Hamiltonian $\hat{H}$.
\item
  If one assigns a measure $\langle\psi|\psi\rangle$ to a world in
  state $|\psi\rangle$ (the previous rule implies that this measure
  is conserved across a split) then
  one should only expect to reproduce the predictions of
  quantum mechanics in {\it almost all\/} worlds
  (with respect to this measure).
\item
  The way in which vectors in $\FH$ describe the world is understood
  using orthodox quantum mechanics, ie.\ self-adjoint operators and the like.
\end{itemize}

\begin{figure}[t]
%
\epsfbox{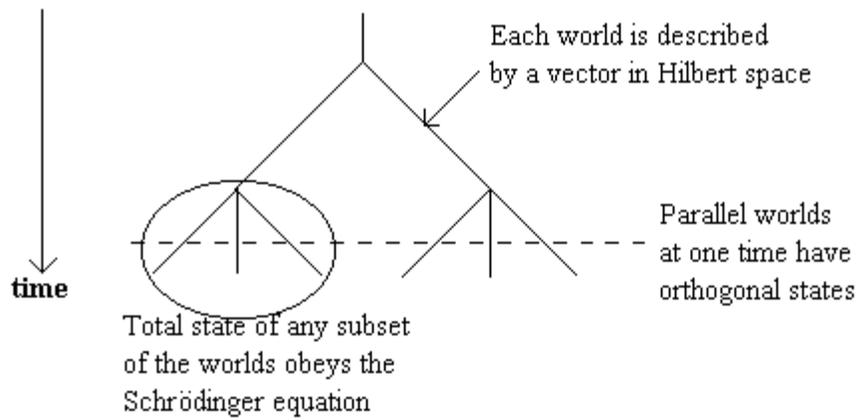}
\caption{The many-worlds ontology\label{f:mwi}}
\end{figure}

The ontology is sketched in Figure~\ref{f:mwi}.

\subsubsection*{Discussion}

At any one time the states of parallel worlds are orthogonal so these
states may be normalised and extended (perhaps in many ways) to a
complete orthonormal set of vectors.  Such a set is called a
{\it preferred basis}.

Conversely, given that
$|\psi_1\rangle,|\psi_2\rangle,\ldots$ is a preferred basis at some time,
the global state may be decomposed into the form $\sum_i a_i|\psi_i\rangle$
giving  the states of the different worlds at that time:
$a_1|\psi_1\rangle,a_2|\psi_2\rangle,\ldots$ (ignoring zeroes).

The interpretation is still incomplete in that it does not
give details of the branching structure or of the preferred basis and
its time evolution.

It is necessary to check that the rules {\em can\/} be satisfied.
Suppose that an arbitrary branching structure is fixed.
At any time when no splitting occurs, the state of each world must
evolve according to the Schr\"odinger equation.
(This means that the elements of the `preferred basis' are also evolving
according to the Schr\"odinger equation.  One cannot have a fixed
preferred basis.)

What happens at a split?  Of course, the total state must not change
discontinuously. What happens is that a world in state $|\psi\rangle$
splits into worlds with orthogonal states $|\phi_1\rangle$,
$|\phi_2\rangle$,$\ldots$, where
$|\psi\rangle=\sum_i|\phi_i\rangle$.
(This corresponds to a discontinuous change in the preferred basis.)

This is sufficient to ensure that all the rules are satisfied.

\subsubsection*{Achievements}

What has been achieved?
Consider one branch.
At times the state of that branch evolves in accordance with
the Schr\"odinger equation.  At other times (at the splits) it
undergoes a wave function collapse.

This, of course, is a possible way to solve the measurement problem.
The world obeys the Schr\"odinger equation except that the wave
function sometimes collapses ensuring that things have some sort of
definite state.
Of course, the big question is when and how this collapse occurs
and the many-worlds interpretation has left this completely open.

So what has been gained by assuming many-worlds?
Merely that it might be possible for the entire collection of worlds
to be described by Schr\"odinger's equation.  This, as already argued,
is no gain at all.

\mysection{Measurement}

There is one thing that works in the many-worlds interpretation, namely
that the Copenhagen statistics for a measurement are reproduced.
This does assume that things have been sorted out so that the
preferred basis at the end of the measurement is
precisely a basis consisting of the eigenstates of the system and
apparatus (tensor producted with some states for the rest of the world).
(Of course, the central difficulty with this type of interpretation is how
to sort things out so that this always holds.  This in turn requires
a rigorous definition of a measurement.)

Consider a system with normalised eigenstates
$|\chi_1\rangle,|\chi_2\rangle,\dots$
coupled with an observer in initial normalised
state $|\phi\rangle$ in such a way that
over a specified time the initial state
$|\phi\rangle\otimes|\chi_i\rangle$
evolves into the normalised state
$|\phi_i\rangle\otimes|\chi_i\rangle$
in which the observer has completed the measurement and has a record of
the result.

Let $|R\rangle$ be the state of the rest of the world
and assume that
$|R\rangle\otimes|\phi_i\rangle\otimes|\chi_i\rangle$
is a part of a preferred basis at this time.

What is the result of performing a measurement on the normalised state
$\sum_i a_i|\chi_i\rangle$~?
Of course, the state
$|R\rangle\otimes|\phi\rangle\otimes\sum_i a_i|\chi_i\rangle$
evolves to the state
$|R\rangle\otimes\sum_i a_i (|\phi_i\rangle\otimes|\chi_i\rangle)$
in which there are a number of worlds in each of which a specific
value has been measured.

In the world where the observer measured value $i$ (that is where the observer
is in state $|\phi_i\rangle$),  the state of the system is $|\chi_i\rangle$
after the measurement, ie.\ each observer observes
the wave function collapse predicted in the Copenhagen interpretation.

To check that the Copenhagen statistics are observed, it is necessary
to consider the same experiment on an ensemble.

Consider an ensemble of $n$ copies of the system and
a team of $n$ observers
with states
\[
  |\boldsymbol\phi\rangle = |\phi\rangle\cdots|\phi\rangle
  \quad\quad
  |\boldsymbol\phi_{i_1,\dots,i_n}\rangle
  = |\phi_{i_1}\rangle\cdots|\phi_{i_n}\rangle
  \mbox{ .}
\]
Then the initial state
\[
  |R\rangle\otimes
  |\boldsymbol\phi\rangle \otimes
  \left[\left(\sum_i a_i|\chi_i\rangle\right)
  \otimes \cdots \otimes
  \left(\sum_i a_i|\chi_i\rangle\right)\right]
\]
evolves into the state
\[
  |R\rangle\otimes
  \sum_{i_1,\dots,i_n} a_{i_1}\cdots a_{i_n}
  |\phi_{i_1,\dots,i_n}\rangle \otimes
  |\chi_{i_1}\rangle\otimes\cdots\otimes|\chi_{i_n}\rangle
  \mbox{ .}
\]
It must now be assumed that each of these summands is in a preferred
basis.

The measure of the observer who observes the sequence $i_1,\dots,i_n$
is $|a_1|^2\cdots |a_n|^2$.
It follows that as $n\rightarrow\infty$,
the measure of observers who observe the proportion of results $i$ to be
within $\epsilon$ of $|a_i|^2$ tends to one\footnote
{
  Note that the measure of the observer who observes the sequence
  $i_1,\dots,i_n$ is $|a_1|^2\cdots |a_n|^2$ which
  is precisely equal to the probability that a sequence of $n$
  integers in which each given integer equals $i$ with probability
  $|a_i|^2$ is in fact $i_1,\dots,i_n$.

  Now as $n\rightarrow\infty$, the probability that the proportion of entries
  in such a sequence with value $i$ is within $\epsilon$ of $|a_i|^2$
  (for any $\epsilon\M>0$) tends to one.  The result follows.
}.
In the limit, {\em almost all observers obtain the probabilities
predicted by the Copenhagen interpretation}.

Note that it is also easy to check that almost all observers will
obtain the results predicted by Schr\"odinger's equation if they
wait before performing the measurement.

\mysection{Determinism}

The interpretation is deterministic.
This does not, however, allow definite predictions to be made because
a typical prediction will be wrong in most futures.

\mysection{Locality}

In the many-world interpretation an experiment performed at one point
in the universe causes a split at all points in the universe.
In this sense, the interpretation cannot be considered local.

\mysection{Philosophy}

The interpretation presents a bizarre realistic ontology.

\mysection{Criticism}

The central difficulty with the many-worlds interpretation is
that it is formulated so
vaguely as to hardly warrant the title {\it interpretation}.
With no indication of how to sort out the preferred basis and
branching structure, the interpretation has achieved nothing.

The other disturbing feature of this interpretation
is the ontological extravagance of an infinity
of parallel worlds.  (These are intended to recover the Schr\"odinger
equation.)

Another criticism of the interpretation which is sometimes raised
(eg.\ \cite[p.115]{Albert:92})
is what sense can be made, in the interpretation,
of the statement that an event will happen with probability $p$?
However, sense {\em can\/} be made of this by saying that (in almost all
futures) if the experiment is repeated many times then the given result will
occur with frequency $p$.

\mysection{Variations}

Some people try to avoid the extravagance of many-worlds by talking about
one world with many stories (eg.\ \cite{Lockwood:89}).
However, it is not at all clear what this really means.

It is possible to avoid the extra worlds by assuming that only one survives.
Explicitly, one may assume that whenever the world branches,
only one world is retained, it being chosen probabilistically with
probability proportional to the branch's measure in the ordinary many-worlds
interpretation.  This has been called the one-world version of the many-worlds
interpretation \cite{Healy:84}.  It is the version which Bell considered
most defensible.  It is, of course, not a theory with many-worlds although
there are still many possible future worlds of which one will be realised.

It seems that in the many-worlds interpretation one can choose between
a deterministic universe with many worlds and a stochasticly governed
universe with one world.

\mysection{Conclusion}

The many-worlds interpretation makes no progress in helping us to
understand the wave function collapse.  It introduces extra worlds to allow
Schr\"odinger's equation to be recovered but this is pointless.

\bib

The many-worlds interpretation was introduced as the
``relative state'' formulation of quantum mechanics by Hugh Everett III
in his PhD dissertation \cite{Everett:57a}.
A fuller account is \cite{Everett:57b}.
The book \cite{DeWittGraham:73} reproduced these and some other papers
explaining the many-worlds interpretation.

Everett did not explicitly say that the different summands represented
different {\it worlds\/} but only that at each observation ``the
observer state `branches' into a number of different states'' and
that ``all branches exist simultaneously''.
De Witt \cite{DeWitt:70,DeWitt:71} and others added the idea of many-worlds
(mistakenly attributing this idea to Everett).

A thorough discussion of the development of the interpretation is included in
\cite{BenDov:87}.
Other discussions of many-worlds include
\cite{CooperVechten:69,Zeh:71,Primas:81,Healy:84,Geroch:84,Stein:84,AlbertLoewer:89},
\cite[Ch.20]{dEspagnat:71},\cite[Ch.13]{BohmHiley:93}.
A typical philosophical reaction to the interpretation is \cite{Gibbens:87}.

The interpretation is criticised in several papers of Bell
(eg.\ \cite{Bell:76,Bell:81,Bell:86}) for the vagueness regarding
the preferred basis problem, for simplistically treating instruments
as single particles, and for extravagance.
Bell and others \cite{Ballentine:73} also criticised the exaggerated
claims such as quantum mechanics `yields its own interpretation'.
The interpretation is also criticised in \cite{Barrett:92}.

\cite{Albert:90} points out the surprising result that
one observer can prove to another that the latter exists in two
universes.

Astoundingly, presumably for lack of choice, the interpretation is
accepted by many cosmologists.
Some have gone so far as to consider travelling to parallel universes
(eg.\ \cite{DeutschLockwood:94} for a non-technical discussion).

\mychapter{Many-minds interpretation} \label{c:mmi}

\aquote{We have little doubt that the MMV (many-minds view)\\
will invoke in most readers an ``incredulous stare''.}
{David Albert and BarryLoewer \cite{AlbertLoewer:89}}

\mysection{Introduction}

The many-minds interpretation accepts Everett's idea that all results
of an experiment are equally real but shies away from the idea that
this involves making an infinite number of copies of the universe.
Instead, different minds perceive the different results.

\mysection{Formulation}

\subsubsection*{Rules}

The universe may be associated with a
separable Hilbert space $\FH$;  a densely defined self-adjoint
operator $\hat{H}$
on $\FH$ called the Hamiltonian; and an element $|\psi\rangle\Min\FH$
called the initial state,
in such a way that the following hold.
\begin{itemize}
\item
  The state of the universe may be represented by a vector in
  Hilbert space which evolves according to Schr\"odinger's equation
  with Hamiltonian $\hat{H}$.
\item
  Each conscious brain in the universe is associated with many
  non-physical minds.
  There is a specific
  set of normalised orthogonal states of a brain written
  $|B_1\rangle,|B_2\rangle,\ldots$
  for each of which the attached mind is in a given state, ie.\ these
  brain states correspond to particular sensations.
  (States of the mind may therefore be labelled
  $B_1,B_2,\ldots$.)
\item
  Suppose the universe is in state $\sum |B_i\rangle |R_i\rangle$
  where the $|R_i\rangle$ are normalised states of the rest of the world.
  Suppose that $|B_i\rangle|R_i\rangle$ evolves over time to the
  state $\sum_jb_j|B_j\rangle|R_j'\rangle$ with the $|R_j'\rangle$ normalised.
  Then the probability that any one mind initially in state $B_i$
  ends up in state $B_j$ is $|b_j|^2$.
\end{itemize}

There is an immediate difficulty with this formulation, similar to
the vagueness regarding the branching structure in the many-worlds
interpretation.

In order to decide the probability for a particular mind to be
in a particular state $B_i$ at time $t$, one may look at its state
at some earlier time and the above rule will give the probability.
But for {\it each\/} earlier time there will be such a  probability!
In general these probabilities will all be different.

Of course, it is always possible that with a large enough set of
minds the allocation could be done very cleverly so that {\it all\/}
the rules of probabilities are satisfied.  It is just that nobody has
ever been clever enough to propose such a scheme\footnote
{
  Albert \cite[p.130]{Albert:92} has suggested that
  in the state $\sum_jb_j|B_j\rangle|R_j\rangle$,
  the proportion of minds in state $B_i$ is $b_i$.
  Far from making a start at solving the above problem, this
  statement actually contradicts the probability
  rule above as a bit of algebra shows.
}.

\mysection{Measurement}

In this interpretation, a measurement occurs when a conscious brain
becomes aware of the result.  For example, let $\hat{A}$ be
an observable on some system with spectrum $|A_1\rangle,|A_2\rangle,\ldots$
and let $|``A_i\mbox{''}\rangle$ (equal to some $|B_j\rangle$)
represent the state of a brain which has noted
the value of $A$ to be $i$.

A measurement consists of a chain of interactions under which any
state $|?\rangle|A_i\rangle$ evolves into the state
$|``A_i\mbox{''}\rangle|A_i\rangle$.

To see that the orthodox statistics are reproduced, let a system
be in state $|A_i\rangle$.  Someone measures the state leading to
a state $|``A_i\mbox{''}\rangle|A_i\rangle$.  Now suppose the two systems
evolve independently for a time into state
$|?\rangle(\sum_i a_i |A_i\rangle)$. Another measurement changes this
into state $\sum_i a_i |``A_i\mbox{''}\rangle|A_i\rangle$.
Of the minds which originally perceived $``A_i\mbox{''}$ the rules dictate
that a proportion $|a_j|^2$ now perceive $``A_j\mbox{''}$ in accordance
with the orthodox statistics.

\mysection{Determinism}

The state of the universe evolves deterministically but the state
of a given mind evolves probabilistically.

\mysection{Locality}

The inventors of the interpretation claim that it is local.
In the EPR situation, when each observer measures the $z$-component of
spin, half of their minds perceive it to be up and the other half of their
minds perceive it to be down.  It makes no difference in what order
the experiments are performed.  (Even so, if both observers write
down the result then {\em all\/} minds will perceive the results to
be anti-correlated.)

More generally, it is easy to check that the probabilistic evolution
of the state of a mind does not depend on what happens in a part
of the universe which is isolated from the associated brain or on what
happens to a mind attached to a distant brain.

Nevertheless, there is a sense in which the interpretation is not local.
Considering again the EPR situation, when a particular mind becomes committed
to seeing the local particle's $z$-component of spin as being up or down,
it is also being committed to seeing the distant particle's $z$ component
as being, respectively, down or up.  So, as in orthodox quantum mechanics,
the local experiment causes the distant particle to take on a definite
$z$-component of spin, at least
as far as the local mind is concerned (even though
the distant minds will be unaffected).

This ambiguity in the understanding of locality is due to the fact
that the interpretation is mainly concerned with describing minds
(ie.\ it is idealistic, see {\it Philosophy\/} below). As such,
it is not terribly clear that the concept of locality is even relevant.

\mysection{Philosophy}

The interpretation is most easily understood as being idealistic.
In other words, it assumes that no universe exists independently of
minds and that Schr\"odinger's equation may be used to predict
the time evolution of minds.  There may be many conscious entities
each associated with infinitely many minds and their sensations
are governed by a single equation so are correlated.

\mysection{Criticism}

It is not clear that the probability rule can be satisfied.
If it can, this interpretation suffers from the same difficulty as
the interpretation in which the mind causes collapse; it cannot
explain the evolution of the universe prior
to the appearance of (animals and) man.

\mysection{Variations}

The interpretation was originally formulated in terms of sets of minds
so that to any one mind attached to one person's brain their
corresponded a particular mind attached to another person's brain
\cite{AlbertLoewer:89}.
The idea was that, for example, in an EPR type situation it is this
pair of minds which would be anti-correlated in their perceptions.
If they ever met they could confirm this.
But since minds cannot meet, this idea seems a bit pointless and
indeed it was dropped \cite{Albert:92}.

\mysection{Conclusion}

The fact that this interpretation is extremely bizarre would be
more palletable if the technical stuff actually worked.

\bib

The interpretation is presented in
\cite{AlbertLoewer:88,Albert:88,AlbertLoewer:89}
and \cite[esp.\ pp.130--133]{Albert:92}.

\mychapter{Bohm's interpretation} \label{c:Bohm}

\aquote{The God of the BB (Bohm-Bell) view doesn't play dice\\
  but he has a malicious sense of humor.}
{David Albert and Barry Loewer, \cite{AlbertLoewer:89}}

\mysection{Introduction}

Bohm's interpretation of quantum mechanics came about not so much as an
attempt to solve the measurement problem but more as an attempt to explain
the wave-particle duality.
In quantum mechanics electrons, for example, display the
interference properties of waves while, at the same time,
coming in discrete multiples of some basic mass like particles.

In 1927 de Broglie \cite{deBroglie:27}
suggested that the duality could be resolved by saying
that an electron was a wave {\it and\/} a particle, ie.\ the wave function
represents a physical field and in it moves a particle with well defined
position and momentum.  This theory was retracted in 1930
\cite{deBroglie:30}.

The idea was taken up (independently) by Bohm in 1952 \cite{Bohm:52}.
Here it was cast as a hidden variables theory in which the position
of the particle is the hidden variable.

Later Bohm stopped using the term `hidden variables' and emphasised
the fact that his interpretation promotes quantum mechanics from
epistemology to ontology.

Here, the interpretation is first presented as a hidden variables theory.
Its application to closed systems is then discussed.

\mysection{Formulation}

\subsection*{As a hidden variables theory}

Bohm's interpretation gets around the impossibility proofs with the
following bold conjecture:  the only true observable
is position.

The interpretation may be formulated for a spinless system
with $n$ generalised coordinates as follows.

\subsubsection*{Rules}

  The system may be associated with a self-adjoint Hamiltonian $\hat{H}$
  densely defined on $L^2(\R^n)$, with a
  wave function $\psi(t,q_1,\ldots,q_n)$
  (which may be seen as an objective physical field) and with
  $n$ sharply defined coordinates following
  continuous paths $r_1(t),\ldots,r_n(t)$,
  in such a way that the following hold.
\begin{itemize}
\item
  The wave function evolves according to Schr\"odinger's equation
  with Hamiltonian $\hat{H}$.
\item
  The rate of change $\dot{r}_i(t)$ of the $i$th coordinate at time $t$
  is equal to
  \begin{equation} \label{e:bohmvelocity}
   \boxed{
    \dot{r}_i(t)
    =
    \frac{\Bj_i(t,r_1(t),\ldots,r_n(t))}{|\psi(t,r_1(t),\ldots,r_n(t))|^2}
   }
  \end{equation}
  where $\Bj$ is the ($n$-dimensional) probability density current for
  the wave function $\psi$.
\item
  A measurement which tests in which of a number of (Borel) sets $B_i$
  the $j$th coordinate lies at time $t$ will yield the
  set $B_k$ containing $r_j(t)$.  It will also cause the wave function
  to collapse in such a way that  $\psi(t,q_1,\ldots,q_n)$  will be
  zero for $q_j\not\in B_k$.
\item
  In any ensemble of systems with wave function $\psi$, the number
  of systems with coordinates in the Borel set $B$ is proportional to
  \(
    \int_B |\psi(\Br)|^2 d\Br
    \mbox{ .}
  \)
\end{itemize}

\subsubsection*{Discussion}

The idea is that the sharp coordinates $r_i(t)$ are hidden variables
determining the result of any measurement of a coordinate.
A quantum mechanical ensemble with wave function $\psi$ really consists
of systems all of which have wave function $\psi$ and in which the
proportion of systems with coordinate values $r_1,\ldots,r_n$ at time
$t$ is proportional to $|\psi(t,r_1,\ldots,r_n)|^2$.
If this is true at time $t$, the definition of the probability density
current ensures that it stays true at all later times.

It is possible to check that this is a completion of quantum mechanics in
the formal sense of Definition~\ref{d:qmcompletion} if
``every self-adjoint operator $\hat{Q}$'' is replaced in the definition
by ``each of the operators $\hat{q}_1,\ldots,\hat{q}_n$''.
The set of states $\Omega=L^2(\R^n) \x \R^n$;
a typical state may be written $(\psi,\Br)$
(strictly one should call $\psi$ and $k\psi$ equivalent and
consider equivalence classes).
$Q((\psi,\Br),\hat{q}_i)$ is of course just $r_i$.
The time evolution $U_t$ evolves the wave function according to
Schr\"odinger's equation and the coordinates according to
\eqref{e:bohmvelocity}.

For every Borel set $B \subseteq \R^n$, one may define
$\mu_{\psi}(\{\phi\}\x B) = \delta_{\psi\phi} \int_B |\psi(\Bx)|^2 \,d\Bx$.

It may be checked that C.1--C.4 hold.

\subsection*{For closed systems}

The same formalism precisely may be applied to closed systems, in particular
to the entire universe.  Again, the $n$-dimensional
spinless case is considered.

\subsubsection*{Rules}

  The closed
  system may be associated with a self-adjoint Hamiltonian $\hat{H}$
  densely defined on $L^2(\R^n)$, with a
  wave function $\psi(t,q_1,\ldots,q_n)$
  (which may be seen as an objective physical field) and with
  $n$ sharply defined coordinates following
  continuous paths $r_1(t),\ldots,r_n(t)$,
  in such a way that the following hold.
\begin{itemize}
\item
  The wave function evolves according to Schr\"odinger's equation
  with Hamiltonian $\hat{H}$.
\item
  The rate of change $\dot{r}_i(t)$ of the $i$th coordinate at time $t$
  is equal to
  \begin{equation*} 
    \dot{r}_i(t)
    =
    \frac{\Bj_i(t,r_1(t),\ldots,r_n(t))}{|\psi(t,r_1(t),\ldots,r_n(t))|^2}
  \end{equation*}
  where $\Bj$ is the ($n$-dimensional) probability density current for
  the wave function $\psi$.
\item
  Human experience results from a perception of the ``hidden'' coordinates
  (not from a perception of the wave function).
\end{itemize}

\subsubsection*{Discussion}

It is now necessary to establish that the two presentations are
mutually consistent.  One must show that in a closed system one may treat
any subsystem (which does not contain conscious entities) using
the hidden variables formulation.

Suppose the subsystem comprises coordinate $1,\ldots,m$.
Then the wave function of this subsystem may be obtained from the total
wave function by
\begin{equation} \label{e:bohmreduced}
  \psi_0 (t,r_1,\ldots,r_m)
  =
  \int_{\R} \psi (t,r_1,\ldots,r_n) dr_{m+1}\cdots dr_n
  \mbox{ .}
\end{equation}
The hidden coordinates of the subsystem are $r_1(t),\ldots,r_m(t)$
as in the total system.

The fact that an ensemble of such subsystems will tend to have a
$|\psi|^2$ distribution is proved in \cite{Valentini:91a}.
Note that once this distribution is obtained it will be maintained
(this follows from the definition of $\Bj$).

When there is no interaction, the wave function and the coordinates
will have the correct time evolution.  When a measurement of position
takes place,
this must be modelled in the total system and it must be shown
that the interaction preduces the result predetermined by the
$r_1(t),\ldots,r_m(t)$ and that it causes the reduced wave function
\eqref{e:bohmreduced} to collapse.  This will be shown under
{\it Measurement}.

When Bohm's interpretation is applied to the whole universe, the
particle coordinates, far from being hidden, are the only things which
we perceive. This important assumption is sometimes swept under the carpet
in discussions of Bohm's interpretation.
But if our perceptions are determined by the wave function, the coordinates
would play no role whatsoever.
Note, however, that when considering a subset of the universe which
includes no conscious brain, the particle coordinate {\em are\/} hidden.

\subsection*{Comparison with many-worlds interpretation}

The
many-worlds  interpretation seeks to explain the definite position
of macroscopic objects by writing the wave function as a sum of
different vectors and saying that different vectors are experienced
in different branches of the universe.
Bohm's interpretation achieves the same thing using the ``hidden''
coordinates and it is instructive to cast this in the language of many
worlds.

Let $\psi(t,\Bx_1,\Bx_2,\ldots)$ be the wave function of the universe.
At time $t$, this may be written as the sum
\[
  \psi(t,\Bx_1,\Bx_2,\ldots)
  =
  \int \psi(t,\Br_1,\Br_2,\ldots)
    \delta(\Bx_1\M-\Br_1)\delta(\Bx_2\M-\Br_2)\cdots
    \, d\Br_1 \, d\Br_2 \cdots
  \mbox{ .}
\]
This may be interpreted as a continuum of universes.
A typical world has wave function
$\delta(\Bx_1\M-\Br_1)\delta(\Bx_2\M-\Br_2)\cdots$
and measure $|\psi(t,\Br_1,\Br_2,\ldots)|^2$.
In each world every particle has a definite position.

A branch may be said to have the defining property that every particle
moves continuously in that branch.

Now the ``hidden'' coordinates of Bohm's interpretation clearly define
a branch in this many-worlds model.  These are a very special class
of branches in which the laws of quantum mechanics are observed to hold.

Clearly, Bohm's interpretation is not a many-worlds theory.
In the first place, the above account uses uncountably many worlds.
It has not been shown how to write the wave function in terms of the
branches.  And, most importantly, only a very special class of
branches is picked out.

Nevertheless, it is interesting to see the relationship between
these seemingly unrelated interpretations.

\subsection*{Spin}

There has been an attempt to model spin in Bohm's interpretation
as the physical spin of the particle \cite{BohmSchillerTiomno:55}.
However this fails where there is more than one particle.

Instead, spin should be considered as a property of the wave function only.
It is then only necessary to recast \eqref{e:bohmvelocity} in terms
of the spinor wave function and this can be done in two ways
(compare \cite{Bell:84} and \cite[\S10.4--10.5]{BohmHiley:93}).

\mysection{Measurement}

\subsection*{Of position}

The aim now must be to recover the hidden variables version of the
theory from the theory for closed systems.
In other words, if the measured particle and measured apparatus are
all modelled within the theory, one must derive the result that the
measured position will correspond to the coordinates of the measured
particles and that wave function will appear to collapse.

Consider a simple impulsive measurement of position in which a
macroscopic pointer initially has wave function $\psi_{ini}$ and after the
measurement
has position $y$ corresponding to the measured position of the particle.
This is only intended as an example and so the mathematically dubious
Dirac delta functions will be used here.

During the measurement, an initial wave function $\delta(x-r)\psi_{ini}(y)$
corresponding to a particle in position $r$
evolves into the wave function $\delta(x-r)\delta(y-r)$ in which both
the particle and pointer have position $r$.
In this case the {\em hidden\/} coordinate of the particle
must initially be $r$ and both hidden coordinates end up as $r$.

If, more generally, the particle initially has wave function
\[
  \psi(x)\psi_{ini}(y) = \int \psi(r) \delta(x-r)\psi_{ini}(y)\, dr
\]
then, by linearity, it evolves into the wave function
\begin{equation} \label{e:bohm:msrmnt:rslt}
  \int \psi(r) \delta(x-r)\delta(y-r)\, dr = \psi(x)\delta(y-x)
\end{equation}
in which the coordinate $x$ and $y$ are perfectly correlated.

What about the hidden coordinates?
If the initial coordinate was $q$ for the particle
then in the sharp case of an initial wave function
$\delta(x-q)\psi_{ini}(y)$ the coordinates end up as $q$ for both the
particle and pointer.  Here we have a superposition of wave packets
$\int \psi(r) \delta(x-r)\delta(y)\, dr$ but the wave packets never
overlap (because of the $\delta(x-r)$ factor which does not change
with time) so just one of them affects the hidden coordinates and therefore
the coordinates vary in exactly the same way ending up
as $q$ for both the position and particle.

Thus the hidden variables result is recovered.
The final coordinate of the pointer, which is perceived by us, corresponds
accurately to the hidden coordinate of the particle.

What about wave function collapse?
In any further experiments on the particle, the perceived results will
again be based entirely on the ``hidden'' coordinate of the pointer.
This coordinate will evolve according to the probability density current
of the wave function \eqref{e:bohm:msrmnt:rslt}.
But here again, the packets do not overlap and only the packet
$\delta(x-q)\delta(y-q)$ will contribute to the behaviour of the pointer
coordinate.   This shows that for the purposes of predicting further
results the wave function of the particle must be taken to be the
collapsed form $\delta(x-q)$.

\subsection*{Of other observables}

Measurements of other observables will not lead to definite answers.

Consider what happens if the above argument is repeated for momentum.
An initial wave function $e^{ipx/\hbar}\psi_{ini}(y)$ evolves
into $e^{ipx}e^{ipy}$.  The more general wave function
\[
  \int \phi(k) e^{ikx} \psi_{ini} dk
\]
evolves, by linearity, into
\begin{equation} \label{e:bohmmomentum}
  \int \phi(k) e^{ikx} e^{iky} dk
  \mbox{ .}
\end{equation}

But here all the wave packets will affect the hidden coordinates,
(not just the one with momentum equal to the initial momentum of the
coordinate, for example).  So the hidden coordinates will not
predetermine the result.  In fact, no definite result will
emerge since the momentum of the hidden coordinates will
not settle down to a fixed value.

In doing further calculations, all the wave packets in
\eqref{e:bohmmomentum} will influence the hidden coordinates and no
wave function collapse will be observed.

In conclusion, {\it there is no measurement until there is a measurement
of position}.

\mysection{Determinism}

The theory is deterministic.

\mysection{Locality}

The theory is explicitly nonlocal.
The velocity \eqref{e:bohmvelocity} of a particle depends on the
multidimensional wave function at the point given by all the other
particles.  Therefore any change in the position of one particle
instantly changes the velocity of the other particles.

It is reassuring, however, that if the wave function is separable
(ie.\ if two systems
have not interacted in the past) then the systems do not interact nonlocally.
This is easily checked.

In any event, the nonlocal effect on a distant system may effect that
system but will not have a perceptible effect on a $|\psi|^2$ ensemble of such
systems since such an ensemble obeys the rules of quantum mechanics.
This makes nonlocal communication rather difficult.

Even so, there is currently no proof that nonlocal communication is impossible
in Bohm's interpretation because there may be situations in
which there is no $|\psi|^2$ ensemble.
(This necessarily entails signal nonlocality \cite{Valentini:91b}.)

\mysection{Philosophy}

The theory gives a realistic ontological model of the universe.

It gives a special role to position.
The term {\it positionism\/} was coined in \cite{AlbertLoewer:89} for
this philosophy!

The particles in the theory are influenced by the wave but do not influence it
(epiphenomenalism) a position which offends some people's sense of justice.

\mysection{Criticism}

Nonlocality has often been sighted as the interpretations least attractive
feature.  There are other difficulties too.

\subsection*{The privileged role of position}

The interpretation gives position a unique role which it does not
obviously deserve.  It is true that most experiments on quantum system
go through a stage of recording positions eg.\ in a photographic plate
or bubble or spark chamber.  But is this necessarily the case or is
it just convenient in practice?

It is also no use having a measurement of position somewhere along the line
--- the final measurement must be of position.
This really means that the information presented by the brain to
the mind (be that a component of the brain or a separate entity) had
better be presented using position.
Otherwise, humans would not perceive the world in a definite state.

This important implication of Bohm's theory has not been made
clear up to know.  Indeed one of the main attractions of the theory
has been its apparent independence from any considerations of the mind.

\subsection*{The mysterious quantum potential}

Consider a particle moving through space in Bohm's interpretation with
a wave function $\psi$.
It may be checked that the fact that its velocity is given by
\eqref{e:bohmvelocity} implies that the particle is subject to a
potential
\[
  -\frac{\hbar^2}{2m} \frac{\nabla^2 |\psi|}{|\psi|}
  \mbox{ .}
\]

Bohm himself felt that this potential is ``rather strange and arbitrary''
and ``has no visible source''.
At that time Bohm accepted this as grounds for being dissatisfied with
the interpretation.

Later, however,
Bohm and Hiley \cite[p.37]{BohmHiley:93} suggested that the wave function
should not be seen as a potential pushing the particle around.
Instead, the wave function acts like a map (they call
this {\it active information\/}) guiding the particle which
reads the wave function and then adjusts its own speed.

They write:
``The fact that the particle is moving under its own energy,
but being guided by the information in the quantum field, suggests
that an electron or any other elementary particle has a complex and subtle
inner structure (eg.\ perhaps even comparable to that of a radio).''

This suggestion seems strange.  There are many ways to reassemble the parts
of a radio and obtain something else.
There is no
evidence that the same may be done with an electron.

However, this problem and the awkward solution proposed by Bohm and Hiley
only come about from trying to cast the interpretation in the familiar
language of potentials.
But here a new and fundamental theory of the universe is being proposed
and there is no reason to expect it to conform to this model.
Instead, the formula \eqref{e:bohmvelocity} for velocity, which seems
neither strange nor arbitrary, may be taken as a new fundamental law of
physics.

\mysection{Variations}

In one variation it is assumed that there are sub-quantum fluctuations.
These ensure that every ensemble rapidly reaches a $|\psi|^2$
distribution \cite{BohmVigier:54,Nelson:66,BohmHiley:89}.

Bohm's interpretation does not extend naturally to quantum field
theories because such theories lack a position observable.
However,  versions of the interpretation do exist for
the electromagnetic field \cite[Appendix]{Bohm:52}, the Fermi
field \cite{Bell:84} and for Boson fields \cite[Ch.11]{BohmHiley:93}.

\mysection{Conclusion}

\bib

The idea of the wave function guiding a particle was originally presented
in \cite{deBroglie:27} for a one-body system.
In this presentation it was suggested that a non-linear modification of the
Schr\"odinger equation might give rise to a stable singularity or pulse
(corresponding to Bohm's particle) with values elsewhere agreeing
with the ordinary wave function.
This idea was called {\it double solution}.
This was criticised at the 1927 Solvay congress particularly by
Pauli \cite{Pauli:28}.
The most important criticism was that the theory did not explain two-particle
experiments.  Indeed, the double solution has never been extended to
many particles.

The social reasons for the theory's rejection are discussed in
\cite{Pinch:77,Cushing:}.
The theory was retracted \cite{deBroglie:30}.

The theory was again and independently proposed, in great detail,
by Bohm \cite{Bohm:52}.  Bohm encouraged de Broglie to take it up
again \cite{deBroglie:56a,deBroglie:56b}.
Interestingly,

Einstein found the theory ``too cheap''
\cite{Born:71}\footnote
{
  See page 192 and letters 81,84,86,88,97,99,103,106,108,110,115,116.
}
although, as Born put it, ``it was quite in line with his own ideas''.

De Broglie himself did not like it very much \cite{deBroglie:70}.
(These comments were collected by Bell \cite{Bell:82}.)

Bohm \cite{Bohm:80} accepted that the form of the quantum potential is
``rather strange and arbitrary'' and ``has no visible source'' but
later Bohm and Hiley \cite[p.37]{BohmHiley:93} devised the concept
of active information to help explain the potential.

Other references including a discussion of Bohm's interpretation include
\cite{Bell:71,Bell:76,Bell:81,Bell:82,AlbertLoewer:89,Bell:90}
\cite[Ch.7]{Albert:92}.  A less technical discussion is \cite{Albert:94}.

Examples of how Bohm's interpretation works in different cases include
\cite{DewdneyHiley:82,DewdneyKyprianidisVigier:84,Dewdney:85,%
DewdneyHollandKyprianidis:86,DewdneyHollandKyprianidis:87,%
VigierDewdneyHollandKyprianidis:87,DewdneyLam:90,DewdneyMalik:93},
\cite[Ch.5]{BohmHiley:93}.

The proof that any ensemble tends to a $|\psi|^2$ distribution
was presented in \cite{BohmVigier:54,BohmHiley:89}, by assuming
the existence of subquantum fluctuations.
It was proved in \cite{Valentini:91a} for the original Bohm theory.

\cite{Valentini:91b} proves that the theory exhibits signal-locality
{\it iff\/} the ensemble has $|\psi|^2$ distribution.
The uncertainty principle is also shown to be peculiar to this case.
\mychapter{Decoherent histories (Ontology)} \label{c:gmh}


\mysection{Introduction}

A consistent set of histories (Chapter~\ref{c:histories})
gives a whole experimental plan which may be carried out
on a quantum system without disturbing the system (at least in the sense
that later results will be unaffected by the fact that
earlier experiments took place).

Thinking purely in the orthodox interpretation, consider a quantum system
and a set of histories consistent with respect to the system's Hamiltonian
and initial state.
Left to its own devices, the quantum system will evolve into a state in
which nothing has a definite value.

Now one could actually carry out the experiments in the set of histories.
This would ensure that various useful observables do take on definite
values.  And, most importantly, this would be achieved without in anyway
disturbing the predictions of the Schr\"odinger equation.
(It should be emphasised that the predictions are not disturbed
in the sense that these predictions may be checked against the available
data, whether in their fine-grained form or in any coarse-graining.  Of
course, these are the only ways in which the predictions can be checked
so this statement is quite strong.)

How is this relevant to interpretation?
If the Schr\"odinger equation is applied to the whole universe, one could
understand why observables take definite values simply by assuming that
a set of external experiments is being performed on the universe.
At first this seems no different to the idea that the mind, for example,
is performing
experiments on the universe.  However, the big difference is that,
assuming the set of histories is consistent, in this model
the predictions of the Schr\"odinger equation are perfectly valid
while effects of the mind would lead to violation of the Schr\"odinger
equation in the brain.

The usefulness of this idea depends, of course, on there existing useful
consistent sets of histories.  In fact, sets of histories involving
macroscopic observables (such as the approximate position or momentum of a
large object) do tend to be consistent as a result of the phenomenon of
decoherence (see bibliography).

This idea leads in an obvious way to an ontological view of a universe
which is experimented on in accordance with a consistent set of histories
causing random wave-function collapse, and which, between experiments,
evolves in accordance with Schr\"odinger's equation.
Below,  this is slightly abstracted.  The interpretation is formulated
in terms of the results of the experiment and the quantum state is
done away with.

It should however be noted immediately that there is a more subtle approach.
If the experimental programme defined by a consistent set of histories
does not interfere with a quantum system (in some sense), perhaps it is
not necessary to perform the experiments at all.  Perhaps it is not
necessary to think about a quantum state at all.  Perhaps it is possible
to talk about the system's properties directly by talking about the
histories, their logical relations and their probabilities.
This abstract idea does not lead to an ontological model but does lead
to a powerful epistemology in which one can make predictions and
retrodictions.  This approach is discussed in the next chapter.

\mysection{Formulation}

\subsubsection*{Rules}

\begin{itemize}
\item
  The universe may be associated with a separable Hilbert space $\FH$;
  an initial state
  consisting of a density matrix $\hat{\rho}$ on $\FH$;
  a Hamiltonian consisting of a densely defined self-adjoint operator
  $\hat{H}$ on $\FH$; and a set of histories $\CS$ over $\FH$ consistent
  with respect to $\hat{\rho}$ and $\hat{H}$;
\item
  facts about the macroscopic world may be associated with
  projectors on $\FH$
\end{itemize}
in such a way that the following hold.
\begin{itemize}
\item
  The projectors associated with the same
  fact at different times are related by the Heisenberg equation.
\item
  The history of the universe is non-deterministic and
  corresponds to some history in $\CS$ with probability given by
  the standard histories formalism.
\end{itemize}

\subsubsection*{Achievements}

Note that in this formulation it is assumed that it is possible
to construct a consistent set of histories in which every possible
fact which humans might directly perceive appears as a projector.
In particular, the approximate positions and momenta of macroscopic
objects would appear at very frequent intervals in this set of histories.

Assuming that this is possible, this interpretation represents an
impressive success.  On the one hand, in this interpretation
Schr\"odinger's equation applies perfectly in the sense that the
relationship between any two facts that humans can perceive will be
perfectly (probabilistically) explained by Schr\"odinger's equation.
This equation may also be used for prediction.

On the other hand, the definite state of the world around us is also
explained.

\subsubsection*{Relationship with orthodox interpretation}

This interpretation has a strange relationship with orthodox quantum
mechanics.  Firstly, in order to complete the interpretation into a
theory of quantum cosmology one needs to find $\hat{\rho}$, $\hat{H}$
and to find a way of associating projectors with macroscopic facts.
Only orthodox quantum mechanics can guide this programme.

Further, in this interpretation one must rely on orthodox quantum mechanics
in order to reconstruct our physical intuition.
How so?  The interpretation does not give any meaning to the existence
of microscopic particles.  Of course, it does predict the {\em effects\/}
of these particles, ie.\ the appearance of tracks in bubble chambers\footnote
{
  All of this is very much in line with Bohr's ideas!
}.
But if one wants to talk about these particles, one must, in practice,
use orthodox quantum mechanics to read the existence of microscopic
particles into the Hamiltonian $\FH$.

In other words, orthodox quantum mechanics may be derived from the
interpretation.

\subsubsection*{Choice of $\boldsymbol\CS$}

Physicists have some ideas about choosing $\hat{H}$ (in terms of the
four forces of nature) and $\hat{\rho}$.  Is there a natural choice
of $\CS$?

Essentially this point has been discussed at length in the work
of Gell-Mann and Hartle and in the secondary literature.  They suggest
looking for the set of histories which is most quasiclassical or for one of the
sets of histories which is most quasiclassical, in some sense
to be made precise.  The idea is that a quasiclassical `domain'
should be a consistent set of histories\footnote
{
  This description is really rather obscure.
  Is every history supposed to behave classically?
  Surely not.  There will always be some histories where the laws
  of classical physics are grossly violated.

  Are just the histories with high probability supposed to behave
  classically?
  This may work but it implies the disturbing idea that there are
  other domains with classical variables with completely non-classical
  behaviour.

  A preferable approach is to look for the set of histories which best
  captures
  the type of classical {\em facts\/} in which one is interested.
  Then the histories formalism may be used to prove that
  the histories with classical dynamics have high probability.
  A sketch of such a proof has already been carried out by Omn\`es
  \cite[\S 16]{Omnes:92}
  using the mathematics of micro-local analysis.
}
\begin{quote}
  ``maximally refined consistent with decoherence, with its
  individual histories exhibiting as much as possible patterns of
  classical correlation in time.'' \cite[\S VI]{GellMannHartle:90b}
\end{quote}

\subsubsection*{Branching histories}

The histories formalism may actually be inadequate for the present
interpretation.  It may be necessary to use a slightly more general
formalism in which the set of projectors at a later time may depend
on which projector materialised at an earlier time.
This would allow $\CS$ to contain projectors referring, say, to the
weather today in Paris only in branches in which the Earth formed
and Paris was built. (The importance of this is that histories
which describe the weather of non-existent cities may not
be consistent.)

\mysection{Measurement}

In this interpretation there is no fundamental measurement process.
Of course, if $\CS$ contains a projector at time $t_1$ indicating the
existence of a bubble chamber and projectors at time $t_2$ indicating
the existence of the same chamber with or without a track then the
formalism leads to appropriate probabilities for the chamber being found
with or without a track. A perfect illusion of measurement is created.

\mysection{Determinism}

The interpretation describes a non-deterministic universe.

\mysection{Locality}

The present interpretation sketches what one might call a {\it universal\/}
theory.  There is no way {\it per se\/} to talk about
some subset of the universe in isolation, over time.  Of course,
individual projectors in $\CS$ at a given time may relate to this
part of the universe or that.

Thus, although the term {\it local\/} has not really been defined for
this type of theory, the very formulation of the theory is in some
sense contrary to locality.

As usual, the discussion can be made more concrete by considering the
EPR situation.  Suppose some projector in $\CS$ at time $t_1$ stipulates that
at location $C$ there exists a macroscopic apparatus $S$.
Orthodox quantum mechanics might tell us that this
$S$ will emit a particle of spin zero which will decay
into two particles of spin half.

At time $t_2$, some projectors in $\CS$ may talk about detectors at
two distant points, $A$ and $B$, centred at $C$ and about whether
each of these registers up or down.  Suppose orthodox quantum mechanics
relates this to a measurement of spin.

The interpretation predicts that at time $t_2$ the results on $A$ and $B$
will be anti-correlated.  Since there was nothing within the formalism
(say at $C$)
corresponding to these values being predetermined, this correlation can only
be seen as nonlocal.

\mysection{Philosophy}

The interpretation sketches a realistic ontology.

\mysection{Criticism}

The idea of having to select $\CS$ in order to make everything work
may seem disturbing.  Nevertheless, most physicists accept that
$\hat{H}$ and $\hat{\rho}$ have to be chosen, not derived.
One can not therefore reject this interpretation simply because
in introduces another parameter.

Nevertheless, there is an important difference between the parameters.
One can give clear objective criteria for selecting $\hat{H}$,
namely that it should be such that, when plugged into Schr\"odinger's
equation, it  predicts all the correlations observed in the lab.
$\hat{\rho}$ should be such as to lead to the presently observed cosmology.
The criteria for selecting $\CS$ are, for now, more vague and subjective.
They have to do with capturing the types of facts which humans perceive.

\mysection{Variations}

\subsubsection*{Approximate decoherence}

It is sometimes assumed merely that the set $\CS$ should be approximately
consistent. However, Dowker and Kent \cite[\S4]{DowkerKent:94} have shown
that it is plausible that every approximately consistent set is close,
in some sense, to a consistent set.  It is therefore always possible
to use a consistent set and only these should be given a fundamental
role in interpretation.

\subsubsection*{Universal state}

Above, the idea of a quantum state has been dispensed with leaving
only the histories.  The interpretation may be cast
in a more familiar form by saying that the universe has a state
in $\FH$ which evolves according to Schr\"odinger's equation.
Then, at each time $t$ appearing in $\CS$, say with projectors
$(\hat{P}_1(t),\hat{P}_2(t),\ldots)$, the state $|\psi\rangle$
collapses into one of
$\hat{P}_1(t)|\psi\rangle$, $\hat{P}_2(t)|\psi\rangle$,$\ldots$ with
appropriate probabilities.

This presentation is certainly more familiar\footnote
{
  It is telling, for example, that Omn\`es \cite[\S2]{Omnes:88a},
  \cite[\S2]{Omnes:90} introduces a rule for the time evolution of
  a state vector even though the notion of a state vector plays
  no role whatsoever in his interpretation (although projectors
  do).  The idea of the state vector is very entrenched.
}
especially in that it assigns
a state to the universe at all times, not just at the discrete times in
$\CS$.  However, this only serves to obscure the point that everything
one might want to say about the universe should, in fact, be said
using a projector in $\CS$.

\subsubsection*{Many-histories}

In the {\it many-histories interpretation\/}
it is assumed that there
is {\em one\/} world as described above for every possible set of histories
$\CS$ consistent with respect to $\hat{H}$ and $\hat{\rho}$.
(There is no multiplicity of worlds corresponding to
different histories in one set.)
This is supposed to solve the problem of selecting $\CS$.

However, this solution is inadequate.  It does allows one to apply the
anthropic
principle to explain why humans exist in a world with quasiclassical
behaviour.  But it can give us no hope that the universe will still
be quasiclassical tomorrow!

\subsubsection*{Many-worlds}

Whether there is one or many sets of histories, the idea that one
history from the set is realised may be replaced by the idea
that all the histories are realised in parallel worlds.
In fact, one may imagine one initial world splitting at each time in $\CS$
giving a perfectly rigorous version of the many-worlds interpretation
(with or without the extra multiplicity of many-histories).

Note however that this is {\em not\/} a faithful realisation of
Everett's idea because the total state of the universe does not obey
Schr\"odinger's equation at the times in $\CS$.

As in the ordinary many-worlds interpretation, some might complain that
it is difficult to talk about the probability of future events in this
version, although in fact this is possible.

The choice of many worlds or one comes down to a choice between
determinism and economy.

\mysection{Conclusion}

This interpretation seems to work!  The fact that it is very unfamiliar
need not be a problem.  In addition to the existing search for
$\hat{H}$ and $\hat{\rho}$, the interpretation necessitates a search
for $\CS$.  The interpretation will only be successful if some natural
criteria for selecting $\CS$ emerge.

\bib

The idea of using consistent histories in the interpretation of quantum
mechanics was suggested by Griffiths
\cite{Griffiths:84} although Griffiths did not explicitly suggest
an interpretation.

The work which most closely resembles the above presentation is
that of Gell-Mann and Hartle
\cite{Hartle:89,GellMannHartle:90a,GellMannHartle:90b,GellMannHartle:93,%
Hartle:93a,Hartle:93b,Hartle:93c,Hartle:93d,Hartle:93e,GellMannHartle:94}.
Of these, the best introductions are \cite{GellMannHartle:90b,Hartle:93b}.
See also
\cite{Halliwell:93a,Halliwell:93b,Halliwell:93c,Halliwell:94}.

Gell-Mann and Hartle do not propose
a specific model of the universe although they hint at various
possibilities.
For example, they write that their work aims at a
``extension, clarification, and completion of the Everett interpretation''
\cite[\S III]{GellMannHartle:90b} and that consistent sets of histories
``give a definite meaning to Everett's `branches'\,''
\cite[\S IV.E]{GellMannHartle:90b}.
This is presumably the first reference to the many-world version
mentioned above.  However, Gell-Mann and Hartle do not take this
literally.  They prefer the term `many histories' to `many worlds'
suggesting a many-stories type of many worlds but reject even that.
Instead, there are many worlds only in the sense that ``quantum mechanics
prefers none over the another except via probabilities''
\cite[\S XIII]{GellMannHartle:90b}.
This seems to suggest the primary interpretation above.
They seem to prefer to think about the choice of branch being made at
each time in $\CS$, not in one go.

Gell-Mann and Hartle do talk about other sets of histories
(which they call other domains).  However, they do not mean it in
the sense of the many-histories interpretation above since they
envisage the possibility of communication between the domains
\cite[\S IV]{GellMannHartle:94}.  This type of situation does not
seem to fit in with any of the ontological models proposed above.
It may make sense in the context of the epistemological models of the
next chapter.

It must be emphasised that Gell-Mann and Hartle did not explicitly
introduce any of the interpretation in this chapter and it is doubtful
that they are committed to any of them.  The important point is that
inspiration for all these interpretations may be found in their work.

The idea of many-histories is discussed in \cite[\S5.4]{DowkerKent:94}.
They do not say who invented it but do say that
the idea was mentioned to them by Griffiths.

The idea of branch-dependent sets of histories is discussed in
\cite[\S X]{GellMannHartle:90b}.

The phenomenon of decoherence was briefly discussed in
Chapter~\ref{c:qstate}, see references there.
The  relevance of this to finding consistent histories is discussed
in the work of Gell-Mann and Hartle, see above.

A particularly insightful discussion on many of the technical and
philosophical issues surrounding histories interpretations may be
found in \cite{DowkerKent:94}.
\mychapter{Decoherent histories (Epistemology)} \label{c:omnes}


\mysection{Introduction}

The present interpretation says: why not do away altogether with the
quantum formalism and use the logic of consistent histories as a way
of talking about the world!

This rather abstract idea turns out to work rather well.
The ideas are mainly due Griffiths and to Omn\`es who calls the interpretation
the {\it logical interpretation}.

\mysection{Formulation}

\subsubsection*{Rules}

\begin{itemize}
\item
  Every quantum system may be associated with a separable Hilbert space $\FH$;
  an initial state
  consisting of a density matrix $\hat{\rho}$ on $\FH$; and
  a Hamiltonian consisting of a densely defined self-adjoint operator
  $\hat{H}$ on $\FH$
\item
  any fact about any system may be associated with
  a projector on the associated $\FH$
\end{itemize}
in such a way that the following hold.
\begin{itemize}
\item
  The projectors associated with the same
  fact at different times are related by the Heisenberg equation.
\item
  One may reason about a system, in particular deriving the probabilities
  for unknown past and future facts from known past facts, such reasoning
  always to be carried out within a consistent set of histories.
\end{itemize}

\subsubsection*{Discussion}

This interpretation is purely epistemological.  It accepts that we have
knowledge of the present and past and tells us what can be derived about
the future and about unknown aspects of the past.

In contrast to the interpretation of the previous chapter, the
present interpretation allows us immediately to reason about all systems
using all types of facts, notably, microscopic facts.  As such,
it directly formalises physical intuition on all scales.

The price to be paid is in the lack of an ontology.
The interpretation does not tell us how to describe the state of a system.
It does not tell us how the potential facts about which one may reason
become actual facts.
(In fact, Omn\`es has gone so far as to suggest that the latter question
will never be answered\footnote
{
  although he may have changed his mind \cite{Omnes:94}.
}
\cite[\S5]{Omnes:91}.)

\subsubsection*{Details of reasoning}

Such reasoning is carried out by writing down all known present and
past facts.  One then finds the consistent histories containing all
such facts.

At this point there are two approaches.
One may look for facts which appear in all such histories.
Following \cite[\S6]{Omnes:91} such facts will be called
{\it true facts\/} and, following \cite[\S5.2]{DowkerKent:94},
those which have probability $1$ in all the histories will be called
{\it definite\/} while those which have probability $p$ in all the
histories will be called {\it probabilistic}.

Alternatively, one may look for facts which appear in any one of the
histories.  Following the same authors these may be called
definite or probabilistic {\it reliable facts}.

Although Omn\`es \cite{Omnes:91} attempted to base his interpretation
on the derivation of true facts, it turns out, not surprisingly,
that true facts are rather hard to come by \cite[\S5.2]{DowkerKent:94}.

If one wants to be able to make predictions, one must rely on reliable
facts.
This is the approach taken by Griffiths \cite{Griffiths:93a}.
This has some seemingly strange consequences
(see {\it Locality\/} and {\it Criticism\/}) but it all works in the end.

(Omn\`es has apparently not yet set out his position on true and
reliable facts in the wake of the attack on the practicability
of true facts in \cite[\S5.2]{DowkerKent:94}.)

\subsubsection*{Comparison with ontological version and Copenhagen
interpretation}

Note that in the ontological interpretation of the previous chapter
one is also free to perform the type of reasoning suggested here.
However, there such reasoning should in the first instance be limited to
the macroscopic facts one can directly perceive and these are all
included in the single consistent set of histories $\CS$.
Therefore, all fundamental reasoning is limited to this one set of
histories.  If one wanted to use other reasoning, say about
a microscopic system, one would have to prove that this is valid.

Omn\`es points out the similarities between his
interpretation and Bohr's, despite the different formulations
\cite[\S29]{Omnes:90}.
This is really a reflection of the fact that
both take an epistemological approach.
The profound advantage of Omn\`es' interpretation lies in the fact
that, far from assuming classical physics, it actually implies
classical physics.  In this sense it is manifestly internally consistent
while the Copenhagen interpretation is not.

\mysection{Measurement}

Using micro-local analysis, Omn\`es has been able to outline a proof
of all the measurement axioms of quantum mechanics in terms of the
simple rules above \cite[\S III]{Omnes:90}.

\mysection{Determinism}

The theory is non-deterministic.

\mysection{Locality}

This question has caused some confusion.
Omn\`es states that in the EPR type of situation,
the measurement on one particle entails
as a reliable fact the collapse of the wave function of the other, a form
of nonlocality.
Omn\`es takes comfort in that this is not a true fact
\cite[\S7,8]{Omnes:91} (although it would now seem that almost nothing is).

Griffiths, on the other hand, uses reliable facts but still argues that
the interpretation is local \cite{Griffiths:94b}.

The confusion is due to the fact that the interpretation is purely
epistemological (compare with Bohr's interpretation, Chapter~\ref{c:bohr},
where locality also causes confusion) whereas locality is an ontological
concept.  One can say in the interpretation that when the local
particle is examined it allows the remote spin to be predicted.  But
one cannot say whether this is cause and effect or whether a predetermined
value is being revealed.

\mysection{Philosophy}

The interpretation is epistemological.
It does not lead to an obvious ontological theory.
(In fact the only candidate so far proposed
for such a theory is extremely awkward
involving an interaction between space-time and matter inducing
a Brownian motion which causes potential facts to become reality
\cite{Omnes:94}.)

\mysection{Criticism}

Several criticisms of this interpretation are made by d'Espagnat
\cite{dEspagnat:89} although none stand up to closer examination
\cite{Griffiths:93a}.

The first criticism is that the future influences the past.
Consider a particle which has been prepared
in stationary state $(|u\rangle+|v\rangle)/\sqrt{2}$.
Tomorrow the particle will be measured either in the basis
$|u\rangle,|v\rangle$ or in the basis $(|u\rangle\pm|v\rangle)$.
In the former case, it is consistent to talk about whether its state
now is $|u\rangle$ or $|v\rangle$ while in the latter case it is not.

However, the future is not influencing the past.
Suppose an experiment is performed now measuring the state either
(A) in the basis $|u\rangle,|v\rangle$ or (B) in the basis
$(|u\rangle\pm|v\rangle)/\sqrt{2}$.
An experimenter performing a measurement tomorrow in the basis
$|u\rangle,|v\rangle$ will be able to say ``if (A) was performed
yesterday then I am certain that the result was such and such''.
An experimenter performing an experiment tomorrow in the basis
$(|u\rangle\pm|v\rangle)/\sqrt{2}$ will be able to say no such thing.
This is an indisputable property of the world which is captured
by the histories formalism.
Is does not mean that the future influences the past but merely
that in the future one may choose which aspect of the past to
reconstruct.

The second criticism is that logic fails.
In the same example, if tomorrow the experimenter measures the state in the
basis $|u\rangle,|v\rangle$ and obtains a result of $|u\rangle$
the histories formalism allows him or her to conclude that the state today was
$|u\rangle$.  Equally, a different set of histories gives, from the very same
facts, the conclusion that the state today was $(|u\rangle+|v\rangle)/\sqrt{2}$
(following from the initial state).  Nevertheless, it is not possible
to conclude the nonsensical fact that the state today is
$|u\rangle$ {\em and\/} $(|u\rangle+|v\rangle)/\sqrt{2}$.

Again, it is reality which gives this result, not the histories formalism.
In this situation, if someone had performed either experiment (A) or (B)
the experimenter tomorrow could with certainty say what the result
was in either case.  It does not make sense to infer the conjunction of
the two results because the two experiments cannot be performed
simultaneously.  The situation is familiar from quantum logic
(\S\ref{s:logicoalgebraic}).

\mysection{Variations}

\paragraph*{Choice of quasiclassical domain}

There may be more than one quasiclassical domain.
If so, Gell-Mann and Hartle suggest that {\it Information Gathering
and Utilising Systems\/} (IGUSs) such as humans may be able to subjectively
choose a quasiclassical domain.  Alternatively, these systems may
evolve to exploit a particular quasiclassical domain.

Both ideas seem strange.  By what mechanism does the IGUS choose a domain?
In the alternative, what quasiclassical domain should be used to study
evolution?

\mysection{Conclusion}

The interpretation provides an elegant epistemology.
Ultimately, it would be preferable to have an ontological theory, perhaps
along the lines of the previous chapter.  Then, the present interpretation
may be introduced simply as a convenient way for discussing
microscopic systems.

\bib

Histories were introduced by Griffiths \cite{Griffiths:84}.
This paper suggested their use in interpretation but was somewhat
vague about what the new interpretation should be.
The application of these ideas to the EPR experiment is presented
in \cite{Griffiths:87a}.  The interpretational and logical
aspects are further developed
in \cite{Griffiths:93a,Griffiths:94a,Griffiths:94b}.

In a series of papers, Omn\`es showed how decoherent histories
may be analysed with classical logic
\cite{Omnes:88a}, how histories explain
interference effects and the EPR experiment \cite{Omnes:88b} and
how the classical world is recovered \cite{Omnes:88c}.
The semiclassical world is then investigated in \cite{Omnes:89}.
These ideas are repeated in \cite{Omnes:90,Omnes:92}.
True and reliable facts are investigated in \cite{Omnes:91}.

%
%
\newpart{IV}{Conclusion}
{Nine interpretations have been described.

What does each imply for the cat?

Which are the most promising?

What does the future hold?
}
\renewcommand{\thechapter}{IV.\arabic{chapter}}
\setcounter{chapter}{0}
\mychapter{Conclusion}

\section*{The nine lives}

In its first life, Schr\"odinger's cat may be viewed in two ways.
Either it is part of a quantum system and it exists in a superposition
of being dead and alive until someone checks.  Here curiousity kills
the cat.
Or, perhaps the cat should be seen as an observer and it causes the
trigger into a definite state of fired or not.  Thus, the cat kills
itself.  The two predictions are contradictory but it is practically
impossible to tell them apart.

In its second life, there is no such thing as a cat.
If one tries to analyse a cat with accuracy of order $h$,
a cat may only be defined as part of an integral phenomenon, eg.\
in terms of milk being consumed and hairs being left around the place.
The concepts of dead and alive are not even defined until one tries
to see whether any milk is consumed!

In its third life the cat is in a superposition of dead and alive until
a conscious being checks up on it.

In its fourth life, the cat is either dead or alive.
However, in any ensemble there will always turn out to be some dead and
some alive.  It is impossible to predict which will live and which
will die.

In its fifth life, the cat causes the entire world to split into two.
In one world it lives, in the other it dies.

In its sixth life, the cat is in a linear superposition of being alive
and dead.  Half of Schr\"odinger's minds perceive it to be
definitely alive
while the other half perceive it to be definitely dead.
In a sense, all are wrong.

In its seventh life, the cat is either alive or dead, assuming that these
concepts may be defined purely in terms of the positions of the
cat's constituent particles.  If so, the cat's fate actually follows
deterministically from a full specification of the initial state.
(Otherwise, it will be in a superposition of dead and alive until
someone correlates these with some position.)

In its eighth life, the cat will  either live or die according
to a random transition.
The microscopic trigger doesn't exist except by virtue of its effect
on the bomb and cat.

In its ninth life, one can predict that the cat will be alive or dead
with equal probabilities.  If it is found to be dead or alive one may
conclude that the bomb respectively had triggered or had not.
But one cannot talk about whether the cat is dead or alive.

\section*{Discussion}

The orthodox interpretation works well in practice but
is ambiguous and therefore unacceptable.
Bohr's epistemological views are also somewhat vague and require
classical physics to be assumed.

For those who are satisfied with epistemology, the decoherent histories
approach admirably ties down any ambiguity or vagueness.  It may
be used to recover classical physics.

Those who would like to see an ontological model of the world must
look elsewhere.  Neither many-world nor many-minds provide such models.
The idea that the mind causes collapse does but the model is
problematic.

Bohm's interpretation does seem to work although it is rather awkward,
especially in the context of quantum field theory.

The best hope at the moment for an ontological theory is that the world's
state collapses according to some fixed
set of consistent histories.
However, no clear criterion have emerged for the appropriate set of
histories.

Finally, there are approaches which involve not interpretation of
quantum mechanics but modification.  These approaches have not
been considered here.
The most successful of these
is the stochastic scheme of `GRW' \cite{GhirardiRiminiWeber:86}
(this scheme was put in a very nice form by Bell \cite{Bell:87a})
although it is not without its problems
(eg.\ \cite[pp.92--111]{Albert:92}).
Several other attempts have involved introducing non-linear terms
in the Schr\"odinger equation but none of these seems to work.

\section*{Final world}

Quantum mechanics is a theory lacking an ontological picture of the
world.  The search for such an ontology  has been long, hard and appallingly
haphazard. It is time that the entire programme was defined and
analysed in a systematic and uniform mathematical way.  Strange as this idea
may seem, I am convinced that it is possible.
When this programme is completed, theoretical physicists will finally be
able to put the cat out and take a well earned rest.

\subsection*{Acknowledgements}
I am extremely grateful to Jonathan Halliwell, Jim Hartle, Chris Isham
and Tom Kibble for some useful conversations and e-mail exchanges.

Special thanks to my wife, Rina.
The writing of this thesis was interrupted by the arrival of our son,
Eitan, and by our family moving to Jerusalem.
Without Rina's love and encouragement, the thesis would have got
buried under a pile of dirty nappies and Israeli immigration documents.
\end{document}